\documentstyle[aps,eqsecnum]{revtex}
\begin{document}
%%%
%%%Title page%%%%%%%%%%%%%%%%%%%%%%%%%%%%%%%%%%%%%%%%%%%%%%%%%%
\title{\bf Nonperturbative QCD: A Weak-Coupling Treatment on the
		Light Front}
%%%
\author{\bf Kenneth G. Wilson$^a$, Timothy S. Walhout$^a$,
Avaroth Harindranath$^a$, \\
Wei-Min Zhang\thanks{Present Address: Institute of Physics, Academia
Sinica, Taipei 11529, Taiwan, ROC.}$^a$, Robert J. Perry$^a$,
and Stanis{\l}aw D. G{\l}azek$^b$}
%%%
\address{$^a$Department of Physics, The Ohio State University,
Columbus, Ohio 43210-1106, USA\\
$^b$Institute of Theoretical Physics, Warsaw University, ul. Ho{\.z}a 69,
00-861 Warsaw, Poland }
%%%
\date{\today}
\maketitle
\mediumtext
\widetext
%%%
\begin{abstract}
\begin{quotation}

In this work the determination of low-energy bound states in Quantum
Chromodynamics is recast so that it is linked to a weak-coupling
problem. This allows one to approach the solution with the same
techniques which solve Quantum Electrodynamics: namely, a combination
of weak-coupling diagrams and many-body quantum mechanics. The key to
eliminating necessarily nonperturbative effects is the use of a bare
Hamiltonian in which quarks and gluons have nonzero constituent masses
rather than the zero masses of the current picture.  The use of
constituent masses cuts off the growth of the running coupling
constant and makes it possible that the running coupling never leaves
the perturbative domain.  For stabilization purposes an artificial
potential is added to the Hamiltonian, but with a coefficient that
vanishes at the physical value of the coupling constant.  The
weak-coupling approach potentially reconciles the simplicity of the
Constituent Quark Model with the complexities of Quantum
Chromodynamics. The penalty for achieving this perturbative picture is
the necessity of formulating the dynamics of QCD in light-front
coordinates and of dealing with the complexities of renormalization
which such a formulation entails.  We describe the renormalization
process first using a qualitative phase space cell analysis, and we
then set up a precise similarity renormalization scheme with cutoffs
on constituent momenta and exhibit calculations to second order. We
outline further computations that remain to be carried out.  There is an
initial nonperturbative but nonrelativistic calculation of the hadronic
masses that determines the artificial potential, with binding energies
required to be fourth order in the coupling as in QED.  Next there is a
calculation of the leading radiative corrections to these masses, which
requires our renormalization program.
Then the real struggle of finding the right
extensions to perturbation theory to study the strong-coupling
behavior of bound states can begin.

\end{quotation}
\vskip .25in
\end{abstract}
%%%
%%%
%%%%%%%%%%%%%%%%%%%%%%%%%%%%%%%%%%%%%%%%%%%%%%%%%%%%%%%%%%%%%
%%%%%%%  intro.tex  (input to QCD paper)         %%%%%%%%%%%%
%%%%%%%%%%%%%%%%%%%%%%%%%%%%%%%%%%%%%%%%%%%%%%%%%%%%%%%%%%%%%
%%%%%%%  last modified 1/23/94    %%%%%%%%%%%%%%%%%%%%%%%%%%
%%%%%%%%%%%%%%%%%%%%%%%%%%%%%%%%%%%%%%%%%%%%%%%%%%%%%%%%%%%%%

%%%%%%%%%%%%%%%%%%%%%%%%%%%%%%%%%%%%%%%%%%%%%%%%%%%%%%%%%%%%%
\section{Introduction}
%%%%%%%%%%%%%%%%%%%%%%%%%%%%%%%%%%%%%%%%%%%%%%%%%%%%%%%%%%%%%

The only truly successful approach to bound states in field theory has
been Quantum Electrodynamics (QED), with its combination of
nonrelativistic quantum mechanics to handle bound states and
perturbation theory to handle relativistic effects.  Lattice Gauge
Theory is maturing but has yet to rival QED's comprehensive success.
There are four barriers which prohibit an approach to Quantum
Chromodynamics (QCD) that is analogous to QED.  The barriers are: (1)
the unlimited growth of the running coupling constant $g$ in the
infrared region, which invalidates perturbation theory; (2)
confinement, which requires potentials that diverge at long distances
as opposed to the Coulombic potentials of perturbation theory; (3)
spontaneous chiral symmetry breaking, which does not occur in
perturbation theory; and (4) the nonperturbative structure of the QCD
vacuum. Contrasting the gloomy picture of the strong interaction in
QCD, however, is that of the Constituent Quark Model (CQM), where only
the minimum number of constituents required by the symmetries are used
to build each hadron and where Zweig's rule leaves little role for
production of extra constituents.  Instead, rearrangement of
pre-existing constituents dominates the physics.  Yet the CQM has
never been reconciled with QCD --- not even qualitatively.  Ever since
the work of Feynman, though, it has been clear that the best hope of
reconciliation is offered by infinite momentum frame (IMF) dynamics.

In this paper, a framework closely related to the IMF will be
employed, namely, light-front quantization.  The purpose of this paper
is to provide arguments that all the barriers to a perturbative
starting point for solving QCD can be overcome, at least in principle,
when a light-front framework is used. We present the basic formulation
for such an approach. Coupled to the choice of coordinates are the
introduction of nonzero masses for both quarks and gluons, the use of
cutoffs on constituent momenta which eliminate vacuum degrees of
freedom, and the addition of an artificial stabilizing and confining
potential which vanishes at the relativistic value of the renormalized
coupling but nowhere else.  One result of these unconventional
modifications is a theory with a trivial vacuum. However, the theory
has extra terms in the Hamiltonian which are induced by the
elimination of vacuum degrees of freedom, which account for
spontaneous chiral symmetry breaking, yet can be treated
perturbatively. It is a theory where a second-order treatment of
renormalization effects should closely resemble the phenomenological
CQM, while a fourth-order treatment --- if all goes well --- should
begin to replace phenomenology by true results of QCD.  We have not
carried out these computations but will outline them to show that they
are indeed perturbative ones.

Our basic aim, then, is to tailor an approach to QCD which is based on
the phenomenologically successful approach of the CQM. Central to our
formulation is to start with nonzero masses for both quarks and gluons
and to then consider the case of an arbitrary coupling constant $g$
which is small even at the quark-gluon mass scale --- a running
coupling which is then small everywhere because below this scale it
cannot run anymore. This means sacrificing manifest gauge invariance
and Lorentz covariance, with these symmetries only being implicitly
restored (if at all) when the renormalized coupling is increased to
its relativistic value, which we call $g_s$. The value $g_s$ is a
fixed number because it is measured at the hadron mass scale, and by
asymptotic freedom arguments such a coupling has a fixed value. For
smaller values of $g$ our theory lacks full covariance and is not
expected to have the predictive power of QCD, but it allows
phenomenology to guide renormalization and is defined to maximize the
ease of perturbative computations and extrapolation to $g_s$.

Some key new ideas in this paper have previously been reported only in
unpublished notes\cite{Wil1,Wil2,Wil3,Wil4,P 91}. Some have been
presented in a condensed form in Ref. \cite{Wal1}. These ideas underly
two main parts of this paper. The first part is a qualitative power
counting analysis of divergences in light-front QCD (LFQCD), which
provides a vital basis for the new cutoff scheme and renormalization
framework we develop in the second part. More broadly, this paper
draws on a range of previous research efforts by the authors and
colleagues: the QCD calculations presented in this paper are largely
based on the standard gauge-fixed LFQCD Hamiltonian\cite{QCD2}, and
the specific diagrammatic rules used here are defined with examples in
Refs. \cite{Zh 93a,Zh 93b,Ha 93a} following earlier work\cite{QCD2s}.
This paper follows recent work on QCD \cite{Zh 93a,Zh 93b,Ha 93a,Pe
93a,Zh 93} and the similarity renormalization scheme
\cite{GW93a,GW93b}; and it is also the outgrowth of a line of research
that covers the Tamm-Dancoff approximation \cite{TD}, light-front QED
\cite{LFQED}, and light-front renormalization \cite{REN}. The relation
of this paper to the entire field of light-front field theory will be
outlined in section II.

The plan of the paper is as follows. As our formulation is quite
different not only from previous approaches to QCD but also to
previous studies of light-front dynamics, we start in Section II with
a general overview and motivation of our approach.  We discuss the
apparent contradictions between QCD and its precursor, the CQM. Then
we outline why a light-front approach can bridge the gap between QCD
and the CQM and discuss in turn how each of the equal-time barriers to
a weak-coupling treatment of QCD may be overcome.  In Sections III-V,
the first major part of the paper, we motivate and discuss the
light-front power-counting analysis, describe the canonical LFQCD
Hamiltonian, and use a qualitative phase space cell analysis to
classify the possible counterterms for ultraviolet and infrared
divergences in LFQCD needed for renormalization. Light-front power
counting can determine the operator structure of the LFQCD Hamiltonian
\cite{Wil2}, and we rely on the derivation from the QCD Lagrangian to
determine details such as color factors. This canonical LFQCD
Hamiltonian will be treated as a Hamiltonian in the many-body space of
finite numbers of quark and gluon constituents, a Hamiltonian which
with regularization gives finite answers in this many-body space.
Three terms in the canonical Hamiltonian {\it itself} require infrared
counterterms because of linear infrared divergences, and we proceed to
construct them. We explain the absence of finite parts to many of
these counterterms because kinematical longitudinal boost invariance
is a scale invariance. Because of this kinematical constraint, only
counterterms for logarithmic infrared divergences are allowed to
contain finite parts, which as part of the renormalization process
must be tuned to reproduce physical results.   Counterterms to
canonical terms are prime candidates for bringing in phenomena
associated with true confinement and with the spontaneously broken
vacuum of normal rest frames.  We distinguish true confinement in the
exact theory from the artificial confinement which we introduce by
hand in the weak-coupling starting point.

This leads to the second major part of the paper, where in Sections
VI-X we set up an explicit quantitative scheme for renormalization of
the LFQCD Hamiltonian.  First, a momentum space cutoff procedure is
introduced which regulates the Hamiltonian itself rather than the
individual terms of perturbation theory \cite{Wil3}.  This cutoff
scheme depends on the use of massive constituents and is chosen to
ensure that each state has only a finite number of constituents. Both
gauge invariance and Lorentz covariance are violated by these cutoffs,
and a range of counterterms is needed to enable a finite limit as the
cutoffs are removed. Restoration of the violated symmetries can only
be established by examining the solution to the theory at the
relativistic limit $g_s$ (or good approximations to this solution).
This is the price we must pay for achieving the new framework. We
proceed to the construction of the effective Hamiltonian. We begin by
discussing a novel perturbation theory formalism due to G{\l}azek and
Wilson \cite{GW93a,GW93b} designed to transform a cutoff Hamiltonian
to band-diagonal form while avoiding small energy denominators that
plague other approaches \cite{P93}. This is called the ``similarity
renormalization scheme.''  The end product of the scheme is a
band-diagonal effective Hamiltonian in which dependence on the
original cutoffs has been removed to any desired order of perturbation
theory. As an example we determine in detail the gluon mass
counterterms necessary to remove second-order divergences generated by
two-gluon intermediate states. We then proceed to construct order
$g^2$ light-front infrared counterterms, and we identify finite
counterterms that may be necessary to compensate for the removal of
zero longitudinal momentum modes from the cutoff theory.  The next
step is to analyze the role of an artificial potential in the model.
We use it to maintain the qualitative structure of bound states for
small $g$, except for an overall scale factor $g^4$.  We argue our
theory is closely analogous to a CQM if the band-diagonal Hamiltonian
is computed only to order $g^2$.  Our hope is that higher-order
computations for the effective Hamiltonian will be invaluable
preparation for study of the limit $g\to g_s$.

%%%%%%%%%%%%%%%%%%%%%%%%%%%%%%%%%%%%%%%%%%%%%%%%%%%%%%%%%%%%%
%%%%%%%  prelim.tex  (input to QCD paper)        %%%%%%%%%%%%
%%%%%%%%%%%%%%%%%%%%%%%%%%%%%%%%%%%%%%%%%%%%%%%%%%%%%%%%%%%%%
%%%%%%%    last modified 1/23/94                %%%%%%%%%%%%
%%%%%%%%%%%%%%%%%%%%%%%%%%%%%%%%%%%%%%%%%%%%%%%%%%%%%%%%%%%%%

%%%%%%%%%%%%%%%%%%%%%%%%%%%%%%%%%%%%%%%%%%%%%%%%%%%%%%%%%%%%%
\section{Motivation and Overview}
%%%%%%%%%%%%%%%%%%%%%%%%%%%%%%%%%%%%%%%%%%%%%%%%%%%%%%%%%%%%%

A key ingredient in the constituent picture of hadrons, starting with
the parton model and then moving to the quark parton model with
constituent quarks and gluons, is the infinite momentum frame. The
field theoretic realization of the intuitive ideas from the IMF is
provided by the light-front dynamics of field theories. Recently there
has been renewed interest in LFQCD because of its potential advantages
over the normal equal-time formulation, especially due to the
triviality of the light-front vacuum in the cutoff theory. However, a
basic puzzle has remained --- namely, how do confinement, chiral
symmetry breaking, and other nonperturbative aspects of QCD emerge
from LFQCD?

Since Dirac's formulation of Hamiltonian systems in light-front
coordinates \cite{Di 49} and the development of the infinite momentum
frame limit of equal-time field theory \cite{Wein66}, there has been
intermittent progress in this area, initially driven by the
recognition of its importance for current algebra \cite{CURR} and the
parton model \cite{PART}, but later slowed by the many renormalization
problems inherent to Hamiltonian field theory. After early progress
towards understanding light-front field theory \cite{PERT}, theorists
began to consider light-front QCD \cite{QCD2}.  No clearly successful
approach to nonperturbative QCD emerged, so some theorists tried to
make progress on technically simpler problems such as 1+1 dimensional
field theory \cite{1+1} and bound states in QED and the Yukawa model
in 3+1 dimensions \cite{QED2}.  More recently there has also been much
work on the light-front zero mode problem \cite{ZERO}, primarily by
theorists who advocate a different approach to the vacuum problem than
that developed in this paper --- the relationship between most of this
work and ours is at present obscure. Unfortunately, the barriers that
are discussed in this paper have hindered theorists from attacking
nonperturbative QCD in 3+1 dimensions; and so recent work on
light-front field theory has not focused on the light-front problem
that is of greatest practical interest.

In this paper we establish a new framework for studying QCD in
light-front coordinates by building from an unconventional choice of
the bare cutoff Hamiltonian. The basic question we try to answer is
the following: can one set up QCD to be renormalized and solved by the
same techniques that solve QED; namely, a combination of weak-coupling
perturbation theory and many-body quantum mechanics? The changes from
the standard approach in QED we introduce are: a) the use of
light-front dynamics; b) the use of nonzero masses for both quarks and
gluons; c) the need to handle relativistic effects which give rise to,
for example, asymptotic freedom in QCD, which in turn leads to a
fairly strong renormalized coupling constant and hence relativistic
binding energies; d) the presence of artificial stabilizing and
confining potentials which vanish at relativistic values of the
coupling constant but nowhere else; and e) the concerns about
light-front longitudinal infrared divergences, which cancel
perturbatively in QED because of gauge invariance.  Our theory is not
gauge-invariant order by order because we use a non-zero gluon mass.
Moreover, we propose that these non-cancelling divergences are
necessarily the sources of true confining potentials and chiral
symmetry breaking in QCD. With the cutoff Hamiltonian we have a
trivial vacuum. Furthermore, the free part of the cutoff Hamiltonian
exhibits exact chiral symmetry despite the existence of a quark mass.
(In the canonical light-front Hamiltonian, chiral symmetry is
explicitly broken only by the helicity flip part of the quark-gluon
interaction.  The free Hamiltonian itself only breaks chiral symmetry
if zero mode quarks are included.) We expect light-front infrared
divergences to be sources of confinement and chiral symmetry breaking
because these are both vacuum effects in QCD, and we show that vacuum
effects can enter the light-front theory through light-front
longitudinal infrared effects. Because of the unconventional scaling
properties of the light-front Hamiltonian, these effects include
renormalization counterterms with whole functions to be determined by
the renormalization process.

The basic motivation for our approach is physical rather than
mathematical.  Physically, one's unperturbed or starting Hamiltonian
is supposed to model the physics one is after, at least roughly.  A
Hamiltonian with nonzero (constituent) quark and gluon masses and
confining potentials is closer to the physics of strong interactions
than a Hamiltonian with zero mass constituents and no confining
potentials.  Then, in the spirit of QED, we analyze renormalization
effects with the confining potential itself treated perturbatively,
but only to generate an effective few-body Hamiltonian which can be
solved nonperturbatively.

%%%%%%%%%%%%%%%%%%%%%%%%%%%%%%%%%%%%%%%%%%%%%%%%%%%%%
\subsection{Constituent quarks versus QCD}
%%%%%%%%%%%%%%%%%%%%%%%%%%%%%%%%%%%%%%%%%%%%%%%%%%%%%

Prior to the establishment of QCD as the underlying theory of strong
interactions, there arose the Constituent Quark Model\cite{QM} and
Feynman's Parton Model \cite{PART}. The CQM provides an intuitive
understanding of many low-energy observables. The Parton Model
provides an intuitive understanding of many high-energy phenomena.
Then QCD came along, and along with it came the commonly accepted
notion that the vacuum of QCD is a very complicated medium.
Unfortunately, this is the source of a contradiction between the
constituent quark picture and QCD; and twenty years of study of QCD
have done little to weaken that contradiction.

The complicated vacuum of QCD plays a crucial role in invalidating any
perturbative picture of isolated quarks and gluons. One important
function of the vacuum is to produce confining interactions among
quarks and gluons at large distances, thus overturning the
non-confining gluon exchange potential of perturbation theory. Another
important function of the vacuum is to provide the spontaneous
breaking of chiral symmetry.  The equal-time QCD vacuum is an infinite
sea of quarks and gluons, and baryons and mesons arise as excitations
on this sea. Unfortunately, individual quarks and gluons are lost in
this infinite sea.

According to the CQM, however, a meson is a simple quark-antiquark
bound state and a baryon is a bound state of three quarks; and Zweig's
rule suppresses particle production in favor of rearrangement of
constituents. But how can the hadrons be simply a quark-antiquark or a
three quark bound state if they are excitations over a complicated
vacuum state? One may hope that the effects of quarks and gluons in
the vacuum may be treated via weak-coupling perturbation theory.
Unfortunately, the coupling of quarks and gluons in the vacuum grows
in strength as the average momentum of these constituents decreases as
a result of asymptotic freedom. Thus one expects the density of
low-momentum constituents in the vacuum to be quite large, thereby
invalidating any perturbative treatment of vacuum effects.

The CQM, as we conceive it, requires that both quarks and gluons have
sizable masses. For gluons, this violates the gauge invariance of QCD.
For quarks, this violates the rule that chiral symmetry is not
explicitly broken only for massless quarks.

On the other hand, many high-energy phenomena are most naturally
described in the language of the Parton Model. The constituent picture
and the probabilistic interpretation of distribution functions are
essential for the validity of the Parton Model.  But it is not at all
easy to reconcile the probabilistic picture with the notion of a
nontrivial vacuum in the equal-time framework. Thus both the
Constituent Quark Model and the Parton Model are put in peril by QCD
with a complicated vacuum structure.

At first glance, the blame for the contradiction rests with the
naiv\`et\`e of the constituent picture.  Equal-time QCD seems to
clearly indicate that any few-body, perturbative approach to hadron
bound states is unfounded.  However, in light-front dynamics we argue
that the CQM and QCD can be reconciled so that the apparent
contradictions disappear.

%%%%%%%%%%%%%%%%%%%%%%%%%%%%%%%%%%%%%%%%%%%%%%%%%%%%%%%%%%%%%%%%
\subsection{Why the light front and massive constituents?}
%%%%%%%%%%%%%%%%%%%%%%%%%%%%%%%%%%%%%%%%%%%%%%%%%%%%%%%%%%%%%%%%

A constituent picture of hadrons is certainly very natural in a
nonrelativistic context. However, particle creation and destruction
need to be a part of any realistic picture of relativistic bound
states. Is it possible to build a relativistic constituent picture of
hadrons based on nonrelativistic, few-body intuitions?

As pointed out long ago, the most serious obstacles to this goal are
overcome by changing to light-front coordinates (the ``front-form'' in
Dirac's original work \cite{Di 49}), or moving to the
IMF\cite{PART,PERT,Wein66}. In light-front quantization one quantizes
on a surface at fixed light-front time, $x^+=t+z$ (see Fig. 1), and
evolves the system using a light-front Hamiltonian $P^-$, which is the
momentum conjugate to $x^+$.  A longitudinal spatial coordinate,
$x^-=t-z$, arises, with its conjugate longitudinal momentum being
$P^+$.

The possibility of building a constituent picture in light-front field
theory rests on a simple observation. All physical trajectories lie in
or on the forward light cone.  This means that all physical
trajectories lie in the first quadrant of the light-front coordinate
system, so that all longitudinal momenta satisfy the constraint,
	\begin{eqnarray} k^+ \ge 0. \end{eqnarray}
By implementing any cutoff that removes degrees of freedom with
identically zero longitudinal momentum, one forces the vacuum to be
trivial because it can carry no longitudinal momentum.

For a free massive particle on shell ($k^2 = m^2$),
the light-front energy is
\begin{eqnarray} k^- = {k_\perp^2 + m^2 \over k^+}, \label{dis}\end{eqnarray}
where $k_\perp$ is the transverse momentum.  This means that the
zero-momentum states we must remove to create a trivial vacuum in
theories with positive $m^2$,
have infinite energy unless $k_\perp=0$ and $m=0$.  This makes
it sensible to replace the zero-momentum modes with effective
interactions, since this is exactly the strategy used when
renormalizing divergences from high energy degrees of freedom in
equal-time field theory.  However, such a starting point may be far
removed from the canonical field theory.

When field theories simpler than QCD are analyzed in light-front
coordinates, it becomes apparent that the assumption of a trivial
vacuum can be misleading.  If $m^2 < 0$ because of spontaneous or
dynamical symmetry breaking,
constituents may have exactly zero longitudinal momentum (known as
``zero modes'') and still have finite energy.
In this case the physical
vacuum is free to contain an arbitrarily large number of zero
longitudinal momentum constituents.  The importance of zero modes is
most simply illustrated in models with spontaneous symmetry breaking
such as the sigma model. In the case of the sigma model,
however, using power-counting arguments and demanding current
conservation (see Appendix A) one can easily determine the
counterterms needed to preserve the physics --- at least on the
canonical or tree level --- for a theory in which zero modes are
removed. We compare and contrast the sigma model with and without
the zero modes removed.  Both are reasonable theories, but the
phenomena of the two theories must be described with different
languages. The discussion of Appendix A indicates that an alternative
way to realize the dynamics of spontaneous symmetry breaking on the
light-front is to force the vacuum to be trivial and to include
counterterms in the Hamiltonian which are based on power counting and
explicitly break chiral symmetry. In Appendix A it is shown that
current conservation can be used to fix these counterterms.  It has
also been shown in Ref. \cite{PW93} that in some cases one can use the
renormalization group and coupling coherence, which is discussed
below, to fix such counterterms.

The analysis of QCD is far more complicated than that of the sigma
model. In QCD, the signal of spontaneous symmetry breaking is still a
vanishing pion mass, but the pion is now a composite state.  For the
weak-coupling starting point, we require all hadrons including the
pion to have masses that are close to the sum of their constituent quark
masses.  One
might imagine other limiting procedures, but we require our starting
point to be perturbative and the pion cannot be massless initially
with this criterion.  As the coupling increases,
the pion mass must decrease toward zero if spontaneous chiral symmetry
breaking is to be recovered as $g \rightarrow g_s$.  The
pion mass should be a continuous function of the coupling, so it
cannot reach zero until the coupling reaches its physical value $g_s$;
and it does so only after the inclusion of renormalization effects. In
contrast, the sigma model illustrates spontaneous symmetry breaking
even for arbitrarily small coupling. Thus the complete determination
of the terms necessary to counter the elimination of zero modes in QCD
will not be simple.

But the major problem in LFQCD is not the question of zero modes. To
even address the role of zero modes we need a reliable, practical
calculational framework. LFQCD has severe infrared divergences arising
from small longitudinal momenta when we eliminate the zero modes
\cite{QCD2,Pe 93a,Zh 93a,Zh 93b,Ha 93a}. These infrared divergences
are separate from and in addition to the infrared problems of
equal-time QCD. In equal-time QCD, infrared problems arise due to both
zero quark and gluon masses and to the growth of the running coupling
constant in the infrared domain. In LFQCD the same infrared problems
also exist, but they are divergences associated with small transverse
momenta, which are the only momenta that combine directly with masses.
The longitudinal infrared divergences are special to LFQCD and for
this reason it is normally presumed that they will cancel out if
treated properly. An essential part of this supposed cancellation is
the maintenance of gauge invariance. To preserve manifest gauge
invariance in QCD in perturbation theory one needs massless gluons and
carefully chosen regularization schemes. With massless gluons,
however, the running coupling constant increases without bound when
the energy scale is of the order of hadronic bound state energies.
Thus one needs all orders of perturbation theory to compute
observables in the hadronic bound state range. But all orders of
perturbation theory involve arbitrary numbers of quarks and gluons as
intermediate states, thus contradicting the notion that hadrons are
mostly a quark--antiquark or three-quark bound state.  It is this
problem that has caused us to abandon manifest gauge invariance in
favor of a weak-coupling picture in which gluons have mass.

Suppose we consider a light-front Hamiltonian whose free part
corresponds to that of massive quarks and gluons. What is the
justification for taking massive gluons in the free part of the
Hamiltonian? As is always the case, the division of the Hamiltonian
into a free part and an interaction part is arbitrary; however, it is
also true that the convergence of a perturbative expansion depends
crucially on how this choice is made. We are free to take advantage of
this arbitrariness; we choose the free gluon part of the bare cutoff
Hamiltonian to be that of massive gluons. (In QED, in contrast, where
electrons and photons appear as free particles in asymptotic
scattering states, it would be more difficult to exercise this
freedom.)  Furthermore, as we show in Section XIII, the cutoffs we
employ inevitably introduce quadratic mass renormalization for quarks
and gluons; thus the bare quark and gluon masses are tunable
parameters. Of course, the crucial question is what the renormalized
masses used for the bound state computations will be: these might be
of the order that phenomenology assigns to constituent quarks and
gluons.  (In QED, one would tune the bare masses to reproduce physical
masses; in QCD, we must tune to fit bound state properties.)

The fact that we are giving the gluon a mass should not create any
contradiction with asymptotic freedom when $g$ achieves its
relativistic value $g_s$.  The reason is that $g_s$ is a running
coupling constant, $g_\Lambda$, at a scale where $\Lambda$ is of the same
magnitude as hadronic masses, $\Lambda \sim \Lambda_{QCD}$.  $g_\Lambda$
is small at
extremely large momentum scales, and the running scale is $\Lambda
\approx \Lambda_{QCD}e^{c/g_\Lambda^2}$ for small running $g_\Lambda$,
where $c$ is a positive constant.  But changes to amplitudes due to
masses can be treated perturbatively at such scales and behave as
powers of $\Lambda_{QCD}/\Lambda \approx e^{-c/g_\Lambda^2}$, which
vanishes to all orders in a perturbative expansion in powers of
$g_\Lambda$.

Once we assume the free part to consist of massive gluons, what are
the consequences? A gluon mass automatically prevents unbounded growth
of the running coupling constant below the gluon mass scale and
provides kinematic barriers to unlimited gluon emission. It eliminates
any equal-time type infrared problems. With nonzero quark and gluon
masses it is also possible to develop a cutoff procedure for the
Hamiltonian such that if the cutoffs are imposed in a specific frame,
a large number of states (the upper limit of whose invariant masses
are guaranteed to be above a large cutoff) are still available for
study even in boosted frames. Another consequence of nonzero gluon
mass is that the long-range gluon exchange potential between a pair of
quarks is too small in transverse directions, falling off
exponentially.  Hence an artificial potential must be added to provide
a full strength potential to yield realistic bound states for small
$g$.

%%%%%%%%%%%%%%%%%%%%%%%%%%%%%%%%%%%%%%%%%%%%%%%%%%%%%%%%%%
\subsection{Light-front infrared divergences}
%%%%%%%%%%%%%%%%%%%%%%%%%%%%%%%%%%%%%%%%%%%%%%%%%%%%%%%%%%

Our basic objective is to establish a weak-coupling framework for
studying QCD bound states, so that one can smoothly approach the
strong-coupling limit and use bound state phenomenology to guide
renormalization.  We have argued that the choice of light-front
dynamics, massive quarks and gluons, and a particular cutoff scheme
eliminates the traditional barriers to a perturbative treatment of
QCD. The first step, then, is the construction of a bare Hamiltonian
which incorporates confining potentials, massive quarks and gluons,
and a trivial vacuum.  The cutoffs will violate Lorentz and gauge
symmetries, forcing the bare Hamiltonian to contain a larger than
normal suite of counterterms to enable a finite limit as the cutoffs
are removed.

Now in the equal-time theory, the QCD vacuum is thought to be a
complicated medium which presumably provides both confinement and the
spontaneous breaking of chiral symmetry. But in the light-front theory
with a suitably cutoff Hamiltonian, we have asserted that the vacuum
is trivial. So we have to find other sources for confinement and
spontaneous chiral symmetry breaking in the cutoff theory.  A natural
place to look for them is in the divergences associated with
light-front infrared singularities.

Explicitly, suppose we regulate infrared divergences (where $k^+_i\to
0$) by cutting off the longitudinal momentum of each constituent $i$
so that $k^+_i > \epsilon$, with $\epsilon$ a small but finite
positive constant.  (We will also need to regulate ultraviolet
divergences by cutting off large transverse momenta, for example via
$k_\perp^2 < \Lambda^2$, but this is not important for the present
discussion.) Because the total longitudinal momentum $P^+$ of a state
is just the sum of the longitudinal momenta of its constituents, $P^+
= \sum_i k_i^+$, it follows immediately that the vacuum of the cutoff
theory (for which $P^+=0$) contains no constituents; that is, the
vacuum is trivial.  Moreover, any state with finite $P^+$ can contain
at most $P^+/\epsilon$ constituents. Therefore, effects which in an
equal-time formulation are due to infinite numbers of constituents ---
in particular, confinement and chiral symmetry breaking --- must have
other sources in the cutoff theory on the light-front.  The obvious
candidates are the counterterms which must be introduced in the
effective Hamiltonian in order to eliminate the dependence of
observables on the infrared cutoff $\epsilon$. Of special interest are
counterterms that reflect consequences of zero modes (namely, modes
with $k^+=0$) in the full theory (see discussions in Sections IV.B and
XI).

Now in the canonical Hamiltonian, one particular term of interest is
the instantaneous interaction in the longitudinal direction between
color charge densities, which provides a potential which is linear in
the longitudinal separation between two constituents that have the
same transverse positions.  In the absence of a gluon mass term, this
interaction is precisely cancelled by the emission and absorption of
longitudinal gluons. The presence of a gluon mass term means that this
cancellation becomes incomplete. But what is still lacking is the
source of transverse confining interactions.

According to power-counting arguments, the counterterms for
longitudinal light-front infrared divergences may contain functions of
transverse momenta \cite{Wil2}; and there exists the possibility that
the {\it a priori} unknown functions in the finite parts of these
counterterms will include confining interactions in the transverse
direction. The $g\to g_s$ limit may then be smooth if such confining
functions are actually required to restore full covariance to the
theory.  In the absence of a gluon mass term, the light-front
singularities are supposed to cancel among each other in physical
amplitudes. This has been verified to order $g^2$ explicitly in
perturbative amplitudes for quarks and gluons rather than physical
states, to order $g^3$ for the quark-gluon vertex
\cite{Pe 93a,Ha 93a}, and to order $g^4$ in the gluon four-point function
\cite{Thorn}. A gluon mass term in the free part of the Hamiltonian,
however, spoils this cancellation. What are the consequences?

The light-front infrared singularities give rise to both linear and
logarithmic divergences. The linear divergences, however, contain the
inverse of the longitudinal cutoff $\frac{1}{\epsilon}$, which
violates longitudinal boost invariance; and hence the infinite parts
of the counterterms for these divergences also violate longitudinal
boost invariance. Thus finite parts for these counterterms are
prohibited by longitudinal boost invariance, which is a kinematical
symmetry. So we have to analyze logarithmic infrared divergences in
order to get candidates for transverse confinement.

There is a specific problem that the complexity of the counterterms
creates. In covariant perturbation theory every coupling or mass
introduced by renormalization becomes an independent parameter. One
might worry that the appearance of whole functions as counterterms
could destroy the predictive power of the theory because functions
include an infinite number of parameters and may seem to destroy the
renormalizability of the theory.  We discuss the resolution of this
problem at the end of Section V.

%%%%%%%%%%%%%%%%%%%%%%%%%%%%%%%%%%%%%%%%%%%%%%
\subsection{Zero modes and chiral symmetry breaking}
%%%%%%%%%%%%%%%%%%%%%%%%%%%%%%%%%%%%%%%%%%%%%%

How does chiral symmetry breaking arise in the light-front theory with
the zero modes removed?  The removal of the zero modes has two
important consequences for chiral symmetry in LFQCD --- namely, the
vacuum is trivial, and chiral symmetry is exact for free quarks of any
mass.  The consequence of a trivial vacuum is that, as with the
sigma model, the mechanism for effects associated with spontaneous
chiral symmetry breaking in the equal-time theory will be far
different in LFQCD.  The consequence of chiral symmetry being exact
for massive constituents means that the mechanism for effects
associated with explicit chiral symmetry breaking in equal-time will
also be quite different in LFQCD. We discuss these points further in
this subsection but leave most details of the light-front chiral
transformation to Appendix B.  We just note here for the discussion
which follows that the fermion field naturally separates into
two-component fields $\psi = \psi_+ + \psi_-$ on the light-front,
where $\psi_+$ is dynamical and $\psi_-$ is constrained.  The
light-front chiral transformation applies freely only to the
two-component field $\psi_+$ \cite{Mus92}, because the constraint
equations are inconsistent with the chiral transformation rules when
explicit breaking is present.

In light-front dynamics chiral symmetry is exact for free quarks of
any mass once zero modes have been removed. This is because chiral
charge conservation is simply the conservation of light-front
helicity, which is a fundamental property of free quarks in light
front dynamics in the absence of zero modes (see Appendix B). Despite
the conservation of chiral charge, the local chiral current is not
conserved. The divergence of the chiral current remains $2 i m_F {\bar
\psi} \gamma_5 \psi$.  How does one reconcile the conservation of
light-front chiral charge with the non-conservation of the axial
current? From the chiral current divergence equation, the light-front
time derivative of the chiral charge is proportional to  $ \int dx^-
d^2 x_\perp {\bar \psi} \gamma_5 \psi$. When the fields in this
integral are expanded in terms of momentum eigenstates, the diagonal
terms $b^\dagger b$ and $d^\dagger d$ --- where $b$ and $d$ are the
quark and antiquark annihilation operators --- vanish, namely the
matrix elements multiplying them vanish. Moreover, the off-diagonal
terms $b^\dagger d^\dagger$, and $bd$ vanish if the zero modes are
absent. Thus it is the absence of zero modes which makes it possible
for the light-front chiral charge to be conserved irrespective of the
non-conservation of the axial current for free massive fermions. The
light-front time derivative of the chiral charge can avoid vanishing
only if ``zero mode quarks and antiquarks'' (quarks and antiquarks
with exactly zero longitudinal momentum) are permitted to exist. But
the cutoffs we use prevent this possibility, and hence chiral symmetry
is exact for all free quarks {\it in the cutoff theory}.

In normal reference frames, the absence of a mass term implies
conservation of the axial vector current and hence a conserved axial
charge. Given a conserved axial current, there are two possibilities:
a) the axial charge annihilates the vacuum, in which case as a
consequence of Coleman's theorem ($``$invariance of the vacuum is the
invariance of the world'') the symmetry is reflected in the spectrum
of the Hamiltonian --- that is, we expect degenerate parity doublets
in the spectrum; or b) the axial charge does not annihilate the
vacuum, in which case one talks about the spontaneous breaking of
chiral symmetry, as a consequence of which massless Nambu-Goldstone
bosons should exist. In the real world, the pion --- the Goldstone
boson --- is very light, and the second possibility is thought to be
realized. The nonzero mass of the pion is thought to arise from the
{\it explicit} symmetry breaking terms in the Lagrangian, namely, the
small current quark masses.

On the light-front, in the absence of interactions, massive quarks in
the cutoff theory do not violate the chiral symmetry of the
light-front Hamiltonian thanks to the cutoffs. Now consider
interactions in the gauge theory.  There remains one explicit chiral
breaking term in the canonical Hamiltonian (see Section IV),
\begin{eqnarray}
	 g m_F \int dx^- d^2 x_\perp \psi_+^\dagger \sigma_\perp
		 \cdot \left(A_\perp  {1 \over \partial^+}\psi_+
		 - {1 \over \partial^+}(A_\perp \psi_+) \right).
\end{eqnarray}
This term, which involves gluon emission and absorption, is linear in
the quark mass and couples the transverse gluon field $A_\perp$ to the
Dirac matrix $\gamma_\perp$ (actually, the Pauli matrix $\sigma_\perp$
in the two-component notation above), which causes a helicity flip. In
the free quark Hamiltonian, $\gamma_\perp$ does not appear, and the
quark mass appears only squared. Now the bare cutoff Hamiltonian has
canonical terms and counterterms.  The chiral charge still annihilates
the vacuum state, which is just a kinematical property of the cutoff
theory, despite the fact that the chiral charge no longer commutes
with even the canonical cutoff Hamiltonian. Thus the vacuum
annihilation property of the chiral charge in the cutoff theory has
nothing to do with the symmetry of the Hamiltonian and does not appear
to have any dynamical consequences. The concept of spontaneous
symmetry breaking seems to have lost its relevance in the cutoff
theory.

But the pion should emerge as an almost massless particle. How does
this become possible on the light-front without zero modes? We may
take a hint from the sigma model with the zero modes removed, as
discussed in Appendix A. In that model terms which explicitly violate
the symmetry must be added to the Hamiltonian to yield a conserved
chiral current, and at the same time the pion must be held massless.
In QCD, the elimination of these modes directly results in effective
interactions.  The effective interactions (the counterterms in the
cutoff theory) can explicitly violate chiral symmetry yet still {\it
persist} in the limit of zero quark mass. Since spontaneous breaking
of chiral symmetry in normal frames is a vacuum effect, we look for
interaction terms that are sensitive to zero mode quarks and
antiquarks. There is a term in the canonical Hamiltonian of the form
\begin{eqnarray}
	 g^2 \int dx^- d^2 x_\perp \psi_+^\dagger
		 \sigma_\perp\cdot A_\perp {1 \over i \partial^+}
		 \Big ( \sigma_\perp\cdot A_\perp \psi_+ \Big ),
\end{eqnarray}
which contains an instantaneous fermion.  Some of the interactions
which correspond to this term are shown diagrammatically in Fig. 2.
These interactions by themselves do not violate chiral symmetry.
However, since they are sensitive to fermion zero modes, the
counterterms for these interactions need not be restricted by chiral
symmetry considerations. The only requirement we can impose is that
they obey light-front power counting criteria.

We show later that there are explicit chiral symmetry breaking terms
which satisfy the power counting restriction.  We add such terms to
our Hamiltonian. As a result of renormalization, in addition to the
emergence of non-canonical explicit symmetry violating terms, the
canonical symmetry violating term of Eq. (2.3) is also renormalized.
Because of the effects of explicit chiral symmetry breaking terms on
this renormalization, the coefficient $m_F$ of this symmetry violating
term in the cutoff Hamiltonian need not be zero, even when the full
relativistic theory has only spontaneous breaking of chiral symmetry.
In fact, in the limit in which chiral symmetry is broken only
spontaneously, the same constituent mass scale may appear in both the
kinetic energy and the symmetry breaking interactions.  This allows
the quark constituent mass to set the scale for most hadron masses,
and yet enables chiral symmetry breaking interactions to be
sufficiently strong to make the pion massless. Thus light-front power
counting criteria and renormalization allow us to introduce effective
interactions which explicitly break chiral symmetry and yet may still
ensure a massless pion in the relativistic limit $ g \rightarrow g_s$.

%%%%%%%%%%%%%%%%%%%%%%%%%%%%%%%%%%%%%%%%%%
\subsection{The Artificial Potential}
%%%%%%%%%%%%%%%%%%%%%%%%%%%%%%%%%%%%%%%%%%

The task of solving the light-front Hamiltonian at the relativistic
value of $g$ is far too difficult to attempt at the present time.
Instead the goal of this paper is to define a plausible sequence of
simpler computations that can build a knowledge base which enables
studies of the full light-front Hamiltonian to be fruitful at some
future date.

A crucial step in simplifying the computation is the introduction
of an artificial potential in the Hamiltonian.

A primary rule for the artificial potential is that it should vanish
at the relativistic limit.  For example, the artificial potential
might have an overall factor $(1-g^2/g_s^2)$ to ensure its vanishing
at $g = g_s$.  This rule leaves total flexibility in the choice of the
artificial potential since no relativistic physics is affected by it.

The first role we propose for the artificial potential is that it
ensure that the bound state structure of the theory at very small $g$
is similar to the actual structure seen in nature.  However, we also
want the weak-coupling behavior of the theory to be similar to QED in
weak coupling. This will ensure that methods of computation already
developed for QED will be applicable.  To ensure these connections we
propose to structure the artificial potential so that bound state
energies all scale as $g^4$ for small $g$, just as QED bound state
energies scale as $e^4$.  We then demand a reasonable match to
experiment when the scale factor $g^4$ is set to $g_s^4$, even when
higher order corrections (of order $g^6$, $g^8$, etc.) are neglected.

The second role of the artificial potential is to remove unfortunate
consequences of the nonzero gluon mass from the gluon exchange potential.
Due to the nonzero mass, the gluon exchange potential falls off too
rapidly in the transverse direction, while being too strong in the
longitudinal direction.  In the longitudinal direction there is an
instantaneous linear potential of order $g^2$ which normally would be
completely cancelled by the one-gluon exchange; but due to the nonzero
mass, the cancellation is incomplete.  Without the artificial potential
there are even instabilities that prevent the existence of stable bound
states at weak coupling.  More details on these instabilities are given in
Section X.B.

The third role we assign to the artificial potential is to give the
weak-coupling theory a structure close to the CQM so that past
experience with the quark model can be used to determine the precise
form of this potential and to fit it to experimental data.  To ensure
this we will define an initial calculation which involves QCD
complications only in a very minimal form.

The final role of the artificial potential is to incorporate a linear
potential in both the longitudinal and transverse directions to ensure
quark confinement for any $g$.  This is important for phenomenology.
It is also needed for studying the roles of a linear potential and
where it might originate, especially in the relativistic limit where
the artificial potential vanishes.

Only broad principles will be laid down here for the artificial
potential. Its construction in detail will require a collaboration
between specialists in the CQM and in QCD perturbation theory.

The basic need is to incorporate the qualitative phenomenology of QCD
bound states into the artificial potential.  This qualitative
phenomenology comes from three sources: kinetic energy, Coulomb-like
potentials, and linear potentials.  We propose that all three terms
should be present in the weak-coupling Hamiltonian and all should have
the same overall scaling behavior with $g$ in bound state
computations, namely $g^4$. The kinetic and Coulomb-like terms can be
constructed directly from the canonical Hamiltonian combined with a
one-gluon exchange term, the latter obtained for the case of zero
gluon mass.  The Coulomb term, if represented in position space, has
the usual form $g^2/r$, except that the definition of $r$ we propose is
\begin{equation}
	 r = \sqrt{\frac{(p^+ \delta x^-)^2}{m_c^2} + \delta x_{\bot}^2}
\end{equation}
and there actually are two terms which are added:
\begin{equation}
	{\cal V}_C = g^2{1\over 2}
         \Big[\frac{m_c}{p^+r} + \frac{m'_c}{p'^+r'}\Big],
\end{equation}
where $\delta x^-$ is the light-front longitudinal separation of two
constituents, $\delta x_{\bot}$ the transverse separation, $m_c$ the
constituent mass, and $p^+$ the constituent longitudinal momentum.
The prime refers to the second of the two constituents.  The $p^+/m_c$
in the definition of $r$ ensures that the dimensions match. The
positive or negative $SU(3)$ charges must also be inserted. See
Appendix C for details.

A few comments regarding the form of the Coulomb potential are in
order here. Our nonrelativistic limit is $g \rightarrow 0 $ and not
the $m_c \rightarrow \infty $ limit studied earlier in Ref. \cite{fs}.
In this limit $p^+$ is held fixed while $\delta x^-$ scales as ${ 1
\over g^2}$. In light-front coordinates, the factor $p^+$ is necessary
for dimensions as already stated and in the nonrelativistic limit
$p^+$ is proportional to the center of mass momentum of the bound
state independent of the relative coordinate $\delta x^-$.
Relativistically, $r$ will need a careful definition to avoid possible
disastrous behavior when $p_1^+ << p_3^+$ or vice versa; this problem
has not been studied.  Also in the relativistic case, $p^+$ does not
commute with $\delta x^-$, so $p^+ \delta x^-$ must be symmetrized to
preserve Hermiticity.

Finally, we need a linear potential term --- terms proportional to
$r$ and $r'$.  In Coulomb bound states, both are of order $1/g^2$.
Hence to achieve an energy of order $g^4$, the linear potential
must have a coefficient of order $g^6$.  Thus the linear potential
term, in position space, would be proportional to $g^6 r$.  To be
precise, and get dimensions straight, its form is
\begin{equation}
	{\cal V}_L = g^6 \beta \Big[\frac{m_c^3 r}{p^+} +
                   \frac{m_c^{\prime 3} r'}{p'^+}\Big],
\end{equation}
with $\beta$ a numerical constant.

The linear potential needs to exist between all possible pairs of
constituents: $qq, q \bar{q}, \bar{q}\bar{q}, qg, \bar{q}g$, and $gg$,
where $q, \bar{q}$ and $g$ stand for quark, antiquark and gluon
respectively.  The linear potential must always be positive
(confining) rather than negative (destabilizing).  We show in Section
X.B. that it cannot involve products of $SU(3)$ charges as the Coulomb
term does.  The Coulomb term could be given a Yukawa structure rather
than the pure $g^2/r$ term. All of the potential has to be Fourier
transformed to momentum space and then restricted to the allowed range
of both longitudinal and transverse momentum after all cutoffs have
been imposed.

The artificial potential must also contain counterterms that remove
the unwanted components of the one-gluon-exchange and
instantaneous-gluon terms.  Thus in addition to the order $g^6$ linear
potential, linear in both the longitudinal and transverse directions,
there is a subtraction to remove the order $g^2$ linear potential in
the longitudinal direction.  The potential removed is the potential
that remains after the incomplete cancellation of the instantaneous
potential in the canonical Hamiltonian by one-gluon-exchange terms.

To ensure that the artificial potential vanishes at $g_s$ without
destroying its weak-coupling features, the subtraction term has to be
treated with care.  We suggest the subtracted linear potential be
multiplied by $(1-g^6/g_s^6)$ so that the subtraction begins to be
negated only in order $g^8$, which is smaller for small $g$ than the
artificial $g^6$ linear potential that needs to be dominant. All other
terms in the artificial potential can be multiplied by $(1-g^2/
g_s^2)$ instead.

To ensure that the Coulomb term shows Coulomb behavior, at least
roughly, at typical bound state sizes, it is important that the mass
used in any Yukawa-type modification of the Coulomb term scale as
$g^2$ rather than being a constant mass.

Finally, the linear potential should be invariant to the full
$SU(6)_W$ flavor-spin symmetry of light-front quarks and antiquarks
discussed by Lipkin and Meshkov\cite{WSPIN}, leaving all $SU(6)_W$
violations to come from quark masses in the kinetic energy, Coulomb
terms, and the more complex renormalization subtractions.  This means
that the quark mass scale appearing in the linear potential will be
the same for up and down quarks.

The artificial potential would be added to the cutoff canonical
Hamiltonian.  Then a complete set of counterterms would have to be
added to ensure that the theory has a limit as the cutoffs are removed.

The artificial potential is designed to allow confined few-body states
to emerge from a field theory. The constraints that dictate its design
are severe, and the type of potential we have discussed is the
simplest we have found that meets these constraints and does not
require large cancellations that thwart a constituent picture.
However, this potential has serious flaws. Probably its worst flaw is
that it confines everything including color singlets to a single
region of space. Any confining interaction that is purely attractive
between all particles does not produce scattering states. QCD must
manage to produce strong attractive forces between color charges as
they separate, without producing such strong forces between color
singlets. In lattice gauge theory this is arranged through gauge
invariance, which forces links to exist between color charges but not
between color singlets. In a field theory calculation without manifest
gauge invariance, there is no such obvious mechanism to turn strong
forces on and off by hand. Higher order calculations will hopefully
provide clues to how QCD produces the important phenomenological
effects of our artificial potential, without producing the unwanted
side-effects.

%%%%%%%%%%%%%%%%%%%%%%%%%%%%%%%%%%%%%%%%%%%%%%%%%%%%%%%%%%%%
%%%%  pc1.tex   (input to QCD paper) 			%%%%
%%%%%%%%%%%%%%%%%%%%%%%%%%%%%%%%%%%%%%%%%%%%%%%%%%%%%%%%%%%%
%%%%    Last modified 1/24/94				%%%%
%%%%%%%%%%%%%%%%%%%%%%%%%%%%%%%%%%%%%%%%%%%%%%%%%%%%%%%%%%%%

%%%%%%%%%%%%%%%%%%%%%%%%%%%%%%%%%%%%%%%%%%%%%%%%%%%%%%%%%%%%
\section{Light-front power counting: Canonical Structure}
%%%%%%%%%%%%%%%%%%%%%%%%%%%%%%%%%%%%%%%%%%%%%%%%%%%%%%%%%%%%

To construct the bare cutoff LFQCD Hamiltonian with counterterms we
will try to follow the procedure adopted in standard canonical
covariant theory as much as possible. In the canonical theory, when we
begin the analysis the bare cutoff Lagrangian is unknown since the
counterterms are unknown prior to the analysis.  One can perform
perturbative calculations and determine the counterterms order by
order in perturbation theory. If one starts with the canonical terms,
which include all the possible terms in accordance with covariance and
power counting, and if one maintains manifest covariance and gauge
invariance at all stages of the calculation, one finds that the
counterterms are also of the same form as the canonical terms. Thus in
the standard procedure the only unknown parameters are just constants,
namely, masses and coupling constants.

Because we are interested in the cutoff light-front Hamiltonian, we
cannot use covariance as a guide. Instead we have to consider power
counting based on the limited kinematical symmetries of the
light-front.   As discussed in Section III.C, one consequence of the
use of light-front coordinates is that counterterms may contain entire
functions of momenta, which is of course a radical difference from
renormalization in an equal-time formulation. Before discussing
renormalization, however, we establish in this section the power
counting rules for determining the possible structure of operators in
the canonical Hamiltonian.

%%%%%%%%%%%%%%%%%%%%%%%%%%%%%%%%%%%%%%%%%%%%%%%%%%%%%%%%%%%%%%%%
\subsection{Light-front power counting}
%%%%%%%%%%%%%%%%%%%%%%%%%%%%%%%%%%%%%%%%%%%%%%%%%%%%%%%%%%%%%%%%

Light-front power counting is in terms of the longitudinal coordinate
$x^-$ and the transverse coordinate $x_\perp$. Why are the two
coordinates treated differently?  The main objective of the power
counting analysis is to deduce the most general structure of
divergences that arise from increasing powers of the interaction
Hamiltonian in perturbation theory. But power counting based on the
kinematical symmetries of the light front is different from power
counting based on the kinematical symmetries in equal-time
coordinates. This is immediately transparent from the light-front
dispersion relation for free particles,  $k^- = {{\textstyle\strut
k_{\perp}^2 + m^2} \over \textstyle\strut k^+}$.  Because the energy
factors into separate $k^+$ and $k_\perp$ dependencies, the
subtractions are not constants.   For example, when $k_\perp$ gets
very large the energy diverges no matter what $k^+$ is. Thus, in
general, we get a divergent constant {\it multiplied by a function of}
$k^+$. In position space this translates into divergences at small
$x_\perp$ being nonlocal in $x^-$ and spread out over the light front.
A similar result follows for the case when $k^+$ gets very small.
This situation is to be contrasted with the equal-time case.  Recall
the relationship between energy and momentum in equal-time theory, $E=
\sqrt{\vec k^2 + m^2}$. In this equal-time form, if $k_\perp\to\infty$
while $k_z$ is fixed, the $k_z$ dependence becomes negligible and
arbitrary functions of $k_z$ cannot arise.

When we analyze the canonical light-front Hamiltonian, we confirm that
indeed it does scale differently under $x^-$- and $x_\perp$-scaling
(strictly speaking, a unique transverse scaling behavior holds only in
the absence of masses). To determine the scaling properties of the
canonical Hamiltonian it is useful to recall the canonical equal-$x^+$
commutation relations obeyed by the field variables in the
two-component formulation \cite{Bj 71,Zh 93b}. For the two-component gluon
field variables, $A_a^i, i=1,2$, we have
\begin{eqnarray}
	\big [ A^i_a(x), A^j_b(y)\big ]_{x^+ = y^+} =
		- {i \over 4} \epsilon(x^- - y^-)
		\delta^2(x_\perp - y_\perp) \delta_{ij}
		\delta_{ab}.  \label{gc}
\end{eqnarray}
For the two-component quark field variables $\xi(x)$ (the
non-vanishing upper components of $\psi_+(x)$) we have
\begin{eqnarray}
	\big\{ \xi_i(x), \xi_j^{\dagger}(y)\big\}_{x^+ = y^+} =
		\delta^3(x - y) \delta_{ij} .  \label{qc}
\end{eqnarray}
Consider the scaling of the longitudinal coordinate
\begin{eqnarray}
	x^- \rightarrow x'^- = s x^-.
\end{eqnarray}
{}From the canonical commutation relations (\ref{gc}) and (\ref{qc}),
under this scale transformation $ A^i(x^-)  \, \rightarrow \,
U_l^\dagger(s) A^i(x^-) U_l(s) \,  = \, A^i(sx^-) \,$ and
$\xi(x^-) \rightarrow U_l^\dagger(s) \xi(x^-) U_l(s) = s^{{1 \over 2}}
\xi(sx^-) $. $U_l(s)$ is the unitary longitudinal scaling operator.  Next
consider the scaling of the transverse coordinates
\begin{eqnarray}
	x_\perp \rightarrow x'_\perp = t x_\perp .
\end{eqnarray}
{}From (\ref{gc}) and (\ref{qc}), $ A^i(x_\perp) \, \rightarrow \,
U_t^\dagger(t) A^i(x_\perp) U_t(t)\, = t \, A^i(t x_\perp)
$ and $ \xi(x_\perp) \, \rightarrow \,
U_t^\dagger(t)\xi(x_\perp) U_t(t)\, = \, t \, \xi(t x_\perp) $. Here
$U_t(t)$ is the unitary transverse scaling operator.
Hence, the power assignments for the field variables are
\begin{eqnarray}
	A_\perp \qquad && : \qquad { 1 \over x_\perp} \nonumber \\
	    \xi \qquad && : \qquad {1 \over \sqrt{x^-}} {1 \over x_\perp}.
\end{eqnarray}
We also need the power assignments for the derivatives
\begin{eqnarray}
	\partial_\perp \qquad &&: \qquad { 1 \over x_\perp} \nonumber \\
		\partial^+ \qquad && : \qquad { 1 \over x^-}
\end{eqnarray}
and the inverse longitudinal derivative
\begin{eqnarray} 	{ 1 \over \partial^+} \qquad : \qquad x^- . \end{eqnarray}

It can be verified that under longitudinal scaling, the canonical
Hamiltonian density is invariant, ${\cal H} \, \rightarrow \,
{\cal H}' \, = \, {\cal H}$ (see Section IV), and the Hamiltonian
\begin{eqnarray}
	H \rightarrow H' = \int dx^{\prime -} d^2x_\perp {\cal H}'  = s H.
\end{eqnarray}
So the Hamiltonian $H = P^-$ scales just like $x^-$, irrespective
of whether there are masses present or not. Similarly, for the
interaction Hamiltonian density in the absence of the helicity
flip interaction, under transverse scaling ${\cal H} \, \rightarrow
\, {\cal H}' \, = { 1 \over t^4} {\cal H}$ and thus
\begin{eqnarray}
	H \rightarrow H' = { 1 \over t^2} H.
\end{eqnarray}
So in the absence of masses, the Hamiltonian scales inversely as
$x^2_{\perp}$.

The Hamiltonian we will consider also includes mass terms.
Under scale transformations masses scale as constants.
The Hamiltonian density for the mass term in the free part,
for example, scales as ${\cal H} \, \rightarrow \,
{\cal H}' = {1 \over t^2} {\cal H}$. Thus the Hamiltonian
does not have a unique scaling behavior when masses are
present. For dimensional analysis we assign
\begin{eqnarray}
	m \qquad : \qquad { 1 \over x_\perp} .
\end{eqnarray}
In the presence of masses, we assign
\begin{eqnarray}
	{\cal H} \qquad && : \qquad {1 \over x_\perp^4}  \nonumber \\
		H \qquad && :  \qquad {x^- \over x_\perp^2} .
\end{eqnarray}
These assignments have meaning only in dimensional analysis.

%%%%%%%%%%%%%%%%%%%%%%%%%%%%%%%%%%%%%%%%%%%%%%%%%%%%
\subsection{Structure of the canonical Hamiltonian
		from power counting}
%%%%%%%%%%%%%%%%%%%%%%%%%%%%%%%%%%%%%%%%%%%%%%%%%%%%

We assume that the canonical LFQCD Hamiltonian density is a
polynomial in the six components $m,A_\perp, \, \xi, \, \partial^+,
\, \partial_\perp, \,$ and ${1 \over \partial^+}$. Then the
most general structure we can build for the canonical Hamiltonian
density, which has dimension ${1 \over (x_\perp)^4}$, is
\begin{eqnarray}
	(\xi \xi^\dagger)^p (A_\perp, \partial_\perp,m)^{4 - 2p}
		(\partial^+)^{-p}.
\end{eqnarray}
Here the expression $(A_\perp, \partial_\perp,m)^{4 - 2p}$ stands for
monomials of order $4-2p$ in any combination of the three variables
$A_\perp, \partial_\perp,m$. The resulting structure is
\begin{eqnarray}
	p=0 && : \qquad A_\perp^4, A_\perp^3 \partial_\perp, A_\perp^2
		\partial_\perp^2, m^2 A_\perp^2 \nonumber \\
	p=1 && : \qquad {1 \over \partial^+}
		(\xi \xi^\dagger) \Big(A_\perp^2 , A_\perp \partial_\perp,
		\partial_\perp^2, m A_\perp, m \partial_\perp, m^2 \Big)
		\nonumber \\
	p=2 && : \qquad \Big({1 \over \partial^+}\Big)^2
		(\xi \xi^\dagger)^2.
\end{eqnarray}
We need not specify the action of $\partial_\perp$ since we assume
that the inverse of $\partial_\perp$ does not occur in the canonical
Hamiltonian. But we need to specify the action of the integration
operator ${1 \over \partial^+}$, which occurs only because of the
constraint equations that eliminate the dependent variables $\psi_-$
and $A^-$. From $\psi_-$ elimination we get the structure
\begin{eqnarray}
	{1 \over \partial^+} \big\{ (\xi, \xi^\dagger)
		(\partial_\perp, A_\perp,m)\big\},
\end{eqnarray}
and from $A^-$ elimination we get the structures
\begin{eqnarray}
	{1 \over \partial^+}\big\{\partial_\perp A_\perp\big\} ,
		({1 \over \partial^+})^2 \big \{ \xi \xi^\dagger,
		A_\perp \partial^+ A_\perp \big \}.
\end{eqnarray}
We will require that inverse powers of $\partial^+$ only appear in
these specific combinations in the canonical Hamiltonian.  Subject to
this constraint, we can include both positive and negative powers of
$\partial^+$.

Specifying the precise way in which ${1 \over \partial^+}$
acts, we can enumerate those terms which obey our canonical
rules:
\begin{eqnarray}
	p=0 && : \qquad (A_\perp)^4, (A_\perp)^3 \partial_\perp,
		(A_\perp)^2 (\partial_\perp)^2, m^2 (A_\perp)^2 ,
		\nonumber \\
	    && \qquad \qquad (A_\perp \partial^+ A_\perp){ 1 \over \partial^+}
                                         (\partial_\perp A_\perp),
                  (A_\perp \partial^+ A_\perp) ({ 1 \over
		\partial^+})^2 (A_\perp \partial^+ A_\perp) \nonumber \\
	p=1 && : \qquad (\xi \xi^\dagger) ({1 \over \partial^+})^2
		(A_\perp \partial^+ A_\perp) , (\xi \xi^\dagger)
		{1 \over \partial^+} ( \partial_\perp A_\perp) ,
		\nonumber \\
	    && \qquad \qquad (m, \partial_\perp, A_\perp) \xi^\dagger
		{ 1 \over \partial^+} \big \{ (m, \partial_\perp,
		A_\perp) \xi\big\} \nonumber \\
	p=2 && : \qquad (\xi \xi^\dagger) ({1 \over \partial^+})^2
		(\xi \xi^\dagger).
\end{eqnarray}
Comparison with the Hamiltonian derived by the standard canonical
procedure (see Section IV) reveals that the free terms $m^2 A_\perp^2
$ and $ m \xi^\dagger {\partial}_\perp { 1 \over \partial^+} \xi $ are
absent. The absence of the first term is because of the presumed gauge
invariance. A term like the latter does appear in the free part when
$\psi_-$ is eliminated, but it is cancelled by a similar term. When
the smoke clears, no $\gamma_\perp$ matrices appear in the free part
of the canonical Hamiltonian. Note that our list excludes terms that
are equivalent through an integration by parts. This integration by
parts is justified because we assume cutoffs are present (see the
discussion in Section V).

%%%%%%%%%%%%%%%%%%%%%%%%%%%%%%%%%%%%%%%%%%%%%%%%%%%%%%%%%%%
%%%  canham.tex   (input to QCD paper)       		%%%
%%%%%%%%%%%%%%%%%%%%%%%%%%%%%%%%%%%%%%%%%%%%%%%%%%%%%%%%%%%
%%%	Last modified 1/24/94				%%%
%%%%%%%%%%%%%%%%%%%%%%%%%%%%%%%%%%%%%%%%%%%%%%%%%%%%%%%%%%%

%%%%%%%%%%%%%%%%%%%%%%%%%%%%%%%%%%%%%%%%%%%%%%%%%%%%%%%%%%%
\section{Light-front QCD Hamiltonian: canonical terms}
%%%%%%%%%%%%%%%%%%%%%%%%%%%%%%%%%%%%%%%%%%%%%%%%%%%%%%%%%%%

In the previous section we motivated the structure of allowed terms in
the canonical Hamiltonian based on power counting. In this section we
use the standard canonical procedure to determine the explicit form of
the Hamiltonian, including color factors. This gives a Hamiltonian
with terms which obey the canonical rules established in Section III.
These canonical rules allow us to extend our Hamiltonian to include a
gluon mass term; and as we demonstrate in Section VIII, our cutoffs
will demand such a mass term to appear through renormalization even if
we did not include it in the starting Hamiltonian.  We choose to
extend the definition of the canonical cutoff LFQCD Hamiltonian to
include a massive gluon from the outset. Of course, this is just the
starting point, for renormalization will eventually force us to add to
our Hamiltonian counterterms which do not obey the canonical rules.

%%%%%%%%%%%%%%%%%%%%%%%%%%%%%%%%%%%%%%
\subsection{Canonical Hamiltonian}
%%%%%%%%%%%%%%%%%%%%%%%%%%%%%%%%%%%%%%

In this Section we follow Ref. \cite{Zh 93a}, to which the reader is
referred for further details.
The QCD Lagrangian is
\begin{equation}
	{\cal L} = - \frac{1}{2} {\rm Tr} (F^{\mu \nu}F_{\mu \nu})
		+ \bar{\psi} ( i \gamma_{\mu} D^{\mu} - m_F ) \psi ,
\end{equation}
where $F^{\mu \nu} = \partial^{\mu} A^{\nu} - \partial^{\nu} A^{\mu} -
ig [A^{\mu} , A^{\nu} ]$ are the gluon field strength tensors and $A^{\mu}
= \sum_a A_a^{\mu} T^a$ are the $3 \times 3$ gluon field matrices, with
$T^a$ the Gell-Mann $SU(3)$ matrices: $[T^a, T^b] = if^{abc}T^c$ and
${\rm Tr} (T^a T^b) = \frac{1}{2}\delta_{ab}$.  The field variable
$\psi$ describes quarks with three colors and $N_f$ flavors, and
$D^{\mu} = \partial^{\mu} - i g A^{\mu} $ are the covariant derivatives,
while $m_F$ is a $N_f \times N_f$ diagonal quark mass matrix.

In light-front coordinates with the
light-front gauge $A_a^+=A_a^0 + A_a^3 =0$,
the Lagrangian can be {\em rewritten} as
\begin{eqnarray}
 {\cal L} = \frac{1}{2}F_a^{+i}(\partial^- A_a^i) +
		i \psi_+^{\dagger} (\partial^- \psi_+)
            - {\cal H} - \left\{ A_a^- {\cal C}_a + \frac{1}{2}
		(\psi_-^{\dagger} {\cal C} + {\cal C}^{\dagger} \psi_- )
		\right\} ,
\end{eqnarray}
where
\begin{eqnarray}
	& & {\cal H} = \frac{1}{2} \left\{ E_a^{-2} + B_a^{-2} \right\}
             + \left\{ \psi_+^{\dagger} \{ \alpha_{\bot} \cdot
		(i\partial_{\bot} + g A_{\bot}) + \beta m_F \} \psi_-
		\right\} \nonumber \\
	& & ~~~~~~~~~~ + \left\{ \frac{1}{2} \partial^+(E_a^- A_a^-)
		- \partial^i (E_a^i A_a^-) \right\}
\end{eqnarray}
is the Hamiltonian density and
\begin{eqnarray}
	& & {\cal C}_a = \frac{1}{2} \partial^+ E_a^- - (\partial^i
		E_a^i + gf^{abc} A_b^i E_c^i) + g \psi_+^{\dagger} T^a \psi_+ \\
	& & {\cal C} = i \partial^+ \psi_-
		- (i \alpha_{\bot} \cdot \partial_{\bot} + g
		\alpha_{\bot} \cdot A_{\bot} + \beta m_F) \psi_+ .
\end{eqnarray}
Here, $E_a^- = -\frac{1}{2} \partial^+ A_a^-$, $E_a^i =
-\frac{1}{2} \partial^+ A_a^i$ and $B_a^- = F_a^{12}$
are components of the light-front color electric field and the
longitudinal component of the light-front color magnetic field,
and $\psi_+$ and $\psi_-$ are the light-front up and down components of
the quark field: $\psi = \psi_+ + \psi_-$, $\psi_{\pm} =
\Lambda_{\pm} \psi = \frac{1}{2} \gamma^0 \gamma^{\pm} \psi$, where
$\Lambda_+ + \Lambda_- = I $,  ${\Lambda_{\pm}}^2 = \Lambda_{\pm}$
and $\Lambda_+ \Lambda_- = 0$.

In (abelian or non-abelian) gauge theory, only two components
--- the transverse components --- of the vector gauge potentials are
physically independent degrees of freedom.  From the equations of
motion, it becomes clear that the independent dynamical
degrees of freedom in LFQCD are the transverse gauge fields
$ A_a^i$ and the up-component quark field $\psi_+$.  The
Lagrangian equations of motion lead to ${\cal C}_a = 0$
and ${\cal C} = 0$, which imply that the longitudinal gauge
fields $ A_a^- $ and the down-component quark field $\psi_-$
are Lagrange multipliers.  Furthermore, in light-front coordinates the
four-component fermion field can be reduced to a two-component
field. The two-component quark field can be explicitly formulated in a
light-front representation of the $\gamma$-matrices defined by
\cite{twoc}
\begin{eqnarray}
	& & \gamma^+= \left[ \begin{array}{cc} 0 & 0 \\ 2i & 0
		\end{array} \right] ~~,~~ \gamma^-= \left[
		\begin{array}{cc} 0 & -2i \\ 0 & 0 \end{array} \right],
		\nonumber \\
	& & ~~~ \nonumber \\
	& & \gamma^i= \left[ \begin{array}{cc} -i \sigma^i & 0 \\ 0 & i
		\sigma^i \end{array} \right] ~~,~~
	\gamma^5= \left[ \begin{array}{cc} \sigma^3 & 0 \\ 0 & -\sigma^3
		\end{array} \right] .
\end{eqnarray}
Then the projection operators $\Lambda_{\pm}$ become
\begin{equation}
	\Lambda_+= \left[ \begin{array}{cc} 1 & 0 \\ 0 & 0 \end{array} \right],
\quad	\Lambda_-= \left[ \begin{array}{cc} 0 & 0 \\ 0 & 1 \end{array} \right]
\end{equation}
and
\begin{equation}
	\psi_+ = \left[ \begin{array}{l} \xi \\ 0 \end{array} \right]~~,~~
	\psi_- = \left[ \begin{array}{c} 0 \\ \left(\frac{1}{i\partial^+}
		\right)[\sigma^i(i \partial^i + gA^i) + im_F] \xi
			\end{array} \right] .
\end{equation}
Hereafter, we shall simply let $\xi$ represent the (two-component) light-front
quark field.  The canonical LFQCD Hamiltonian becomes:
\begin{eqnarray}
	& & H = \int dx^- d^2 x_{\bot} \left\{ \frac{1}{2} (\partial^i
		A_a^j)^2 + gf^{abc}A_a^i A_b^j \partial^i A_c^j
		\right. \nonumber \\
 	& & ~~~~~~~~~~  + \frac{g^2}{4}
		f^{abc} f^{ade} A_b^i A_c^j A_d^i A_e^j  \nonumber \\
	& & ~~~~~~~~~~ + \Bigg[ \frac{ }{ } \xi^{\dagger}
		\{ \sigma_{\bot} \cdot ( i \partial_{\bot} + g A_{\bot})
		- im_F \}  \nonumber \\
	& & ~~~~~~~~~~~~~ \times \left( \frac{1}{i\partial^+} \right)
		 \{ \sigma_{\bot}
		\cdot ( i \partial_{\bot} + g A_{\bot}) + im_F \}
		\xi^{ } \Bigg]  \nonumber \\
	& & ~~~~~~~~~~ + g \partial^i A_a^i \left( \frac{1}{\partial^+}
		\right) (f^{abc} A_b^j \partial^+ A_c^j + 2 \xi^{\dagger}
		T^a \xi ) \nonumber \\
	& & ~~~~~~~~~~ + \frac{g^2}{2} \left( \frac{1}{\partial^+}
		\right) (f^{abc} A_b^i \partial^+ A_c^i + 2 \xi^{\dagger}
		T^a \xi ) \nonumber \\
	& & ~~~~~~~~~~~~~ \times \left. \left( \frac{1}{\partial^+}\right)
		(f^{ade} A_d^j \partial^+ A_e^j + 2 \xi^{\dagger} T^a
		\xi ) \right\} .
\end{eqnarray}
If zero modes are retained there are additional surface terms \cite{Zh
93a}.

%%%%%%%%%%%%%%%%%%%%%%%%%%%%%%%%%%%%%%%%%%%%%%%%%%%%%%%%%%%%%%%%%%%%%%%%%%
\subsection{Counterterms to canonical terms}
%%%%%%%%%%%%%%%%%%%%%%%%%%%%%%%%%%%%%%%%%%%%%%%%%%%%%%%%%%%%%%%%%%%%%%%%%%

Light-front infrared singularities arise when one eliminates the
unphysical degrees of freedom by solving the constraint equations,
which are the source of the operator $1/\partial^+$ in (4.9). This
operator produces tree level, light-front infrared divergences in the
instantaneous four-fermion, two-fermion--two-gluon, and four-gluon
interactions. These divergences require counterterms, which we now
construct.

%%%%%%%%%%%%%%%%%%%%%%%%%%%%%%%%%%%%%%%%%%%%%%%%%%%%
\subsubsection{Instantaneous gluon exchange}
%%%%%%%%%%%%%%%%%%%%%%%%%%%%%%%%%%%%%%%%%%%%%%%%%%%%

First, we consider the instantaneous four-fermion term
\begin{eqnarray}
          H_{qqqq} = - 2g^2 \int dx^- d^2 x^\perp
                      \left\{ (\xi^{\dagger} T^a \xi )
                                               \left( \frac{1}
                {\partial^+} \right)^2 (\xi^{\dagger} T^a \xi) \right\} .
\end{eqnarray}
In terms of the quark color current $j_q^{+a} =  2 \xi^{\dagger}
T^a \xi $ and its partial Fourier transform
\begin{eqnarray}
	j^{+a}_q(x^-,x_\perp) = {1 \over 2 (2 \pi)}
		\int_{- \infty}^{\infty} dp^+ {\tilde j}_q^{+a}
		(p^+, x_\perp) e^{{i \over 2} p^+ x^-},
\end{eqnarray}
we have
\begin{equation}
 H_{qqqq} =  {g^2 \over 8 \pi} \int d^2 x^\perp \int_{- \infty}^{\infty}
	{dp^+ \over (p^+)^2} {\tilde j^{+a}}_q(p^+,x_\perp)
		{\tilde j^{+a}}_q(-p^+,x_\perp).
\end{equation}
Thus we see explicitly that $H_{qqqq}$ has potential divergences.

In the cutoff Hamiltonian $|p^+|$ is restricted to be above $\epsilon$.
That is,
\begin{eqnarray}
	\int_{- \infty}^{+ \infty} dp^+ \rightarrow \int_{- \infty}^{-
		\epsilon} dp^+ + \int_{\epsilon}^{\infty} dp^+.
\end{eqnarray}
To find the divergent part we can do a Taylor expansion of the integrand:
\begin{eqnarray}
	\int {dp^+ \over (p^+)^2} F(p^+,x^\perp) = \int {dp^+ \over (p^+)^2}
		\Big[ F(0,x^\perp)+ {\partial F \over \partial p^+}
		\Big|_{p^+=0} p^+ + ...\Big].
\end{eqnarray}
The linearly divergent term is
\begin{eqnarray}
	{g ^2 \over 8 \pi} {2 \over \epsilon} \int d^2 x_\perp
		{\tilde j}_q^{+a} (0,x_\perp) {\tilde
j}_q^{+a}(0,x_\perp).
\end{eqnarray}

The logarithmically divergent term vanishes because of the symmetric cutoff.
Let us suppose we had used separate cutoffs $\epsilon_+$ and
$\epsilon_-$ for small positive and negative $p^+$, respectively.
Then the potentially logarithmically divergent term is
\begin{eqnarray}
	{ig^2 \over 8 \pi} \ln\left({\epsilon_+\over \epsilon_-}\right)
		\int d^2 x_\perp \int dx^- \int dy^- j^{+a}_q(x^-,
		x_\perp)(x^- - y^-) j^{+a}_q(y^-, x_\perp),
\end{eqnarray}
which is finite if we choose $\epsilon_+$ = $\mu \epsilon_-$, with
$\mu$ some number.  This particular operator vanishes because of
symmetry under $x^- \rightarrow y^-$ exchange, but it forces us to
reconsider our original neglect of zero modes in the solution of the
constraint equations.

We are finding a potentially divergent
sensitivity of the Hamiltonian to modes with small longitudinal
momentum, and we regard this as a signal that finite zero mode
effects may arise.  The potential divergence arises
from the constraint equations and reflects
contributions from $A^-$ close to zero longitudinal momentum.  It is
extremely naive to suppose that the canonical constraint equation for
$A^-$ reproduces the nonperturbative
effects of a zero mode in any simple manner.  We discuss the divergences
that arise from the exchange of small longitudinal momentum gluons with
physical polarization
below, but even here we can think of the counterterms produced by the
constraint equation
as arising from the exchange of small longitudinal momentum
gluons with unphysical polarization.  Once one realizes that an exchange
is involved and that the divergence is independent of transverse
coordinates, it becomes clear that even the assumption that these
counterterms will be local in the transverse direction is naive.
We therefore
allow terms in the Hamiltonian of the form
\begin{eqnarray}
	{ig^2 \over 8 \pi} \int d^2 x_\perp dx^- \int dy^-
		d^2y_\perp j^{+a}_q(x^-,x_\perp)(x^- - y^-)
		{\cal O}_G(x_\perp-y_\perp) j^{+a}_q(y^-, y_\perp).
\end{eqnarray}
${\cal O}_G$ is a function of the transverse variables
and is restricted by dimensional analysis, kinematical boost
invariance, translational invariance,
and invariance under rotations about the longitudinal
axis. ${\cal O}_G$ must be odd under $x_\perp\to y_\perp$ to keep
(4.17) from vanishing, and this is not possible here because gluon
exchange gives only an even number of polarization vectors
$\epsilon^i_\lambda$ with which to contract the transverse indices.
Thus there is indeed no finite counterterm associated with logarithmic
divergences in $H_{qqqq}$, and our discussion serves only to illustrate
one way candidate vacuum interactions can be identified.

As discussed previously, the requirement of boost invariance
means that there can be no finite part associated with the
counterterm for the linear divergence (4.15), whose form
breaks longitudinal boost invariance. Thus the counterterm
for the canonical instantaneous four-fermion interaction is
\begin{eqnarray}
	H^{CT}_{qqqq} = - {g^2 \over 8 \pi} {2 \over \epsilon}
		\int d^2 x_\perp \int dx^- j^{+a}_q(x^-,x_\perp)
		\int dy^- j^{+a}_q(y^-, x_\perp).
\end{eqnarray}
In a similar way, the Hamiltonian counterterms for the instantaneous
two-fermion--two-gluon and the instantaneous four-gluon interactions
are found to be
\begin{eqnarray}
	H^{CT}_{qqgg2} = - {g^2 \over 4 \pi} {2 \over \epsilon}
		\int d^2 x_\perp \int dx^- j^{+a}_q(x^-,x_\perp)
		\int dy^- j^{+a}_{g}(y^-, x_\perp).
\end{eqnarray}
and
\begin{eqnarray} 	H^{CT}_{gggg} = - {g^2 \over 8 \pi} {2 \over \epsilon}
		\int d^2 x_\perp \int dx^- j^{+a}_g(x^-,x_\perp)
		\int dy^- j^{+a}_{g}(y^-, x_\perp),
\end{eqnarray}
where the gluon color current is
$j^{+a}_g = f^{abc} A^{i}_{b} \partial^{+} A^{i}_{c}$.
These three counterterms precisely remove the tree level
light-front infrared divergences.

%%%%%%%%%%%%%%%%%%%%%%%%%%%%%%%%%%%%%%%%%%%%%%%%%%%
\subsubsection{Instantaneous fermion exchange}
%%%%%%%%%%%%%%%%%%%%%%%%%%%%%%%%%%%%%%%%%%%%%%%%%%%

There are several terms in the canonical Hamiltonian which
contain a single inverse derivative ${1 \over \partial^+}$.
With one exception, all these terms can be rewritten so that
${1 \over \partial^+}$ acts on a single field operator. Such
terms do not give rise to divergences since the zero modes are
removed from single operators. The exception is the
two-quark--two-gluon interaction involving an instantaneous
fermion exchange $ H_{qqgg1}$. Here ${1 \over \partial^+}$
acts on a product of field operators and hence can give
rise to a logarithmic divergence (see also Section V).

Denoting $\sigma_\perp\cdot A_\perp(x^-, x_\perp)\,
\xi(x^-,x_\perp)$ by $f(x^-,x_\perp)$, we have
\begin{eqnarray}
	H_{qqgg1} = g^2 \int dx^- d^2 x_\perp f^\dagger(x^-,
		x _\perp) { 1 \over i \partial^+} f(x^-,x_\perp).
\end{eqnarray}
Introducing the partial Fourier transform as before
\begin{eqnarray}
	{\tilde f}(p^+, x_\perp) = \int_{- \infty}^{\infty}
		dx^- e^{{i \over 2} p^+ x^-} f(x^-, x_\perp),
\end{eqnarray}
we find
\begin{eqnarray}
	H_{qqgg1} = {g^2\over 4 \pi} \int d^2 x_\perp
		\int_{- \infty}^{\infty} {dp^+ \over p^+}
		{\tilde f}^\dagger(p^+,x_\perp) {\tilde f}(-p^+, x_\perp).
\end{eqnarray}
A potential logarithmic divergence disappears if we use a
symmetric cutoff.

Once again, though, we must be careful with this logarithmic
divergence. While there is no divergent dependence on the symmetric
longitudinal infrared cutoff $\epsilon$ since the divergence
due to small $p^+$ is cancelled by the divergence due to small
$-p^+$, we still have to worry about non-vanishing contributions
from states with exactly $p^+=0$ --- that is, from exchange of
zero-mode fermions.
Once an infrared cutoff is employed, we have eliminated the
possibility of computing this contribution;
and so we should include in the Hamiltonian terms which can counter
the effects of this exclusion.  Such counterterms have the form
\begin{eqnarray}
	H_{qqgg1}^{CT} = g^2 \int dx^- d^2 x_\perp
		\int dy^- d^2 y_\perp f^\dagger(x^- ,x _\perp)
		{\cal O}_F(x_\perp-y_\perp) f(y^-,y_\perp),
\end{eqnarray}
where dimensional analysis reveals that ${\cal O}_F$ scales as 1.  Since
this term comes from zero-mode
fermion exchange, ${\cal O}_F$ is a function
of transverse variables and the Pauli matrices $\sigma_\perp$.
Since this term arises from vacuum effects, it is not restricted
to have a polynomial dependence on transverse variables and the
renormalized constituent mass. Thus a term like ${\cal O}_F
\sim m^{-1}_F\sigma_\perp \cdot \partial_\perp$
is allowed and should be included in the Hamiltonian.  Such a term
explicitly violates chiral symmetry and need not be small in the
relativistic limit.

This is then the first point at which an arbitrary function
enters the Hamiltonian.
At the simplest level, the function ${\cal O}_F(x_\perp,y_\perp)$
must be determined phenomenologically by fitting bound state
properties.  It is an open question whether theoretical
techniques such as coupling coherence will determine
it {\it a priori}, whether the
requirement of relativistic invariance will determine
it completely {\it a posteriori},
or whether phenomenology will still be necessary
when renormalization is done at higher orders.

%%%%%%%%%%%%%%%%%%%%%%%%%%%%%%%%%%%%%%%%%%%%%%%%%%%%%%%%%%%%%%%%
\subsection{Canonical Hamiltonian: free plus interaction terms}
%%%%%%%%%%%%%%%%%%%%%%%%%%%%%%%%%%%%%%%%%%%%%%%%%%%%%%%%%%%%%%%%
Now we add explicitly the gluon mass term to the LFQCD canonical
Hamiltonian together with the counterterms for instantaneous interactions.
The Hamiltonian is written as a free term plus
interactions
\begin{equation}
        H = \int dx^- d^2 x_{\bot} ({\cal H}_0 + {\cal H}_{int} ),
\end{equation}
with
\begin{eqnarray}
        & & {\cal H}_0 = \frac{1}{2}(\partial^i A_a^j)(\partial^i A_a^j)
		+ \frac{1}{2}m_G^2 A_a^j A_a^j
           + \xi^{\dagger} \left( \frac{-\partial_\perp^2 + m_F^2}{i
		\partial^+} \right) \xi \nonumber \\
        & &
{\cal H}_{int} =  {\cal V}_A + {\cal H}_{qqg} + {\cal H}_{ggg} +
                {\cal H}_{qqgg} + {\cal H}_{qqqq} + {\cal H}_{gggg}.
\end{eqnarray}
Here ${\cal V}_A$ is the artificial potential, and
\begin{eqnarray}
        & & {\cal H}_{qqg} = g \xi^{\dagger} \left\{ - 2 \left(
                \frac{1}{\partial^+} \right) ( \partial_{\bot} \cdot
                A_{\bot})  \right. \nonumber \\
        & & ~~~~~~~~~~~~~~~~~ + ~ \sigma_\perp \cdot A_{\bot} \left(
                \frac{1}{\partial^+} \right) (\sigma_\perp
                \cdot \partial_{\bot}
                + m_F)  \nonumber \\
        & & ~~~~~~~~~~~~~~~~~ + \left. \left( \frac{1}{\partial^+} \right)
                (\sigma_\perp \cdot \partial_{\bot} - m_F)
                 \sigma_\perp \cdot A_{\bot}
                \right\} \xi \\
        & & {\cal H}_{ggg} = g f^{abc} \left\{ \partial^i A_a^j A_b^i A_c^j
                + (\partial^i A_a^i) \left( \frac{1}{\partial^+} \right)
                (A_b^j \partial^+ A_c^j) \right\} \nonumber \\
        & & ~~~ \\
        & & {\cal H}_{qqgg} = g^2 \left\{ \xi^{\dagger} \sigma_\perp \cdot
                A_{\bot} \left( \frac{1}{i \partial^+}
                  \right) \sigma_\perp \cdot
                A_{\bot} \xi \right\} \nonumber \\
        & & ~~~~~~~~~~~~~~  -~2 g^2 \left\{ (f^{abc} A_b^i \partial^+ A_c^i)
           	\left( \frac{1}{\partial^+} \right)^2 (\xi^{\dagger} T^a
		\xi) \right\} + H^{CT}_{qqgg1} + H^{CT}_{qqgg2} \nonumber \\
        & & ~~~~~~~~ = {\cal H}_{qqgg1} + {\cal H}_{qqgg2} \\
        & & {\cal H}_{qqqq} = - 2g^2 \left\{ (\xi^{\dagger} T^a \xi )
		 \left( \frac{1}{\partial^+} \right)^2 (\xi^{\dagger}
		T^a \xi) \right\} + H^{CT}_{qqqq} \\
        & & {\cal H}_{gggg} = \left. \frac{g^2}{4} f^{abc}
                                   f^{ade} \right\{ A_b^i
                A_c^j A_d^i A_e^j \nonumber \\
        & & ~~~~~~~~~~~~~~~~~ \left. - 2 (A_b^i \partial^+ A_c^i) \left(
		\frac{1}{\partial^+} \right)^2 (A_d^j \partial^+ A_e^j)
		\right\}  + H^{CT}_{gggg} \nonumber \\
        & & ~~~~~~~~ = {\cal H}_{gggg1} + {\cal H}_{gggg2} .
\end{eqnarray}
In a careful analysis all of these couplings must be allowed to
renormalize separately and the quark masses in ${\cal H}_0$ and ${\cal
H}_{qqg}$ are not the same.  For simplicity we use one coupling and
one mass in our examples.
It can be seen that all canonical terms are predicted
from the light-front power counting (3.16).
The artificial potential has the form
described in Section II.E.  The Coulomb part will involve
the quark and gluon color currents $j^{+a}_q$ and $j^{+a}_g$
defined above, and the linear part will involve the color
singlet currents $j^+_q = 2\xi^\dagger\xi$ and $j^+_g =
A^i_a \partial^+A^i_a$.

As we have emphasized before, the division of the bare
Hamiltonian into a free part and an interaction part is
arbitrary. We have chosen the free part to be that of
massive quarks and gluons since self mass corrections are
needed in any case.   Note that these terms are
allowed by the power counting criteria.  Our computations
will involve two stages.  The first stage is the computation
of the effective Hamiltonian, which is to be performed
perturbatively.  The artificial potential will be considered
as part of ${\cal H}_I$ for this stage.  The second stage
is the determination of bound states, and as discussed
before we will want to include the (nonrelativistic)
Coulomb and linear parts of
${\cal V}_A$ as part of the unperturbed Hamiltonian for this stage.

%%%%%%%%%%%%%%%%%%%%%%%%%%%%%%%%%%%%%%%%%%%%%%%%%%%%%%%%%%%
%%%   phase.tex   (put in QCD paper)			%%%
%%%%%%%%%%%%%%%%%%%%%%%%%%%%%%%%%%%%%%%%%%%%%%%%%%%%%%%%%%%
%%%  last modified:    1/24/94   			%%%
%%%%%%%%%%%%%%%%%%%%%%%%%%%%%%%%%%%%%%%%%%%%%%%%%%%%%%%%%%%

%%%%%%%%%%%%%%%%%%%%%%%%%%%%%%%%%%%%%%%%%%%%%%%%%%%%%%%%%%%
\section{Power counting: phase space cell analysis
	  and the structure of divergences}
%%%%%%%%%%%%%%%%%%%%%%%%%%%%%%%%%%%%%%%%%%%%%%%%%%%%%%%%%%%
\subsection{Phase space cell division and field operators}
%%%%%%%%%%%%%%%%%%%%%%%%%%%%%%%%%%%%%%%%%%%%%%%%%%%%%%%%%%%

Here we perform the power counting analysis of ultraviolet divergences
and infrared divergences arising from the products of interaction
Hamiltonians. To facilitate this analysis, we first introduce a phase
space cell representation of field variables in terms of wave packet
functions \cite{Wi 65,Wi 71a}.  A realization of these wave packet
functions is given by wavelets. The present considerations are only
qualitative in nature, and hence we do not need their many interesting
mathematical properties\cite{Wlet}; however, they may be needed by
anyone attempting the nonperturbative diagonalization of the
renormalized Hamiltonian.

The motivation for a phase space cell analysis and an example of
phase space cell division is provided in Appendix D in the
context of the more familiar case of equal-time Hamiltonians.
Here we discuss the division of phase space cells in the
light-front case.  For the $``$plus'' momenta we choose
\begin{eqnarray}
	p^+_{max} > p^+_1 > p^+_2 > p^+_3 ... > p^+_{min}
\end{eqnarray}
with $p^+_{i+1} = {1 \over 2} p^+_i $. For the transverse
momenta we choose $ p_{\perp 0} = 0, \; p_{\perp 1} = m, \; $ and
then a sequence $ p_{\perp j} = 2^{j-1} \times m$ until the cutoff
$p_{\perp {max}} $ is reached. Note that we always use
a logarithmic scale for momenta.

For a momentum space cell with centroids $p^+_i$ and $p_{\perp j}$
which is an annulus in the transverse momentum, position space is
divided into a linear grid of cells all with width $\delta x^- = {1
\over p^+_i}$ and $\delta x_{\perp} = { 1 \over p_{\perp j}}$. These
cells are labelled with position indices $l^-$ and $l_\perp =
(l^x,l^y)$. A phase space cell is denoted by the indices $ijl =
(i,j,l^-,l^x,l^y)$.  A wave packet function that belongs to the
complete, orthonormal basis set is denoted in position space by
$\phi_{ijl}(x)$; and its Fourier transform is denoted by  ${\tilde
\phi_{ijl}}(p)$. The $x$-space widths of these functions are $\delta
x^-_i$ and $\delta x_{\perp j}$ and they are assumed to vanish outside
of the cell. Normalization implies that $\phi_{ijl}$ is of order ${1
\over \sqrt{\delta x^-_i}}{1 \over \delta x_{\perp j}}$, while
${\tilde \phi_{ijl}}(p)$ is of order ${ \sqrt{\delta x^-_i}}\,{ \delta
x_{\perp j}}$.

Next we consider field operators. For the quark field operator,
we have
\begin{eqnarray}
	\xi(x) = \sum_s \chi^{\;}_s \int {dk^+ d^2
		k_\perp \over 2 (2 \pi)^3} \Big [ b_{k,s}
		e^{-ik\cdot x} + d^{\dagger}_{k,s}
		e^{ik\cdot x} \Big],
\end{eqnarray}
with
\begin{eqnarray}
	\big\{ b_{k,s}, b^{\dagger}_{k',s'}\big\} =
		\big\{ d_{k,s}, d^{\dagger}_{k',s'}\big\} =
		2 (2 \pi)^3 \delta(k^+ - k'^+) \delta^2 (k_\perp - k'_\perp)
		\delta_{s s'} . \end{eqnarray}
Thus
\begin{eqnarray}
	b \, , \, d \; \sim \; \sqrt{x^-} x_{\perp} .
\end{eqnarray}
We wish to carry out the analysis in such a way that the creation and
destruction operators are of order unity. To achieve this goal, we
introduce phase space cell operators
\begin{eqnarray}
	b_{k,s} \chi^{\;}_{s} = \sum_{l i j}{\tilde \phi}_{ijl}(k)
		b_{ijls} \end{eqnarray}
and
\begin{eqnarray}
	d^{\dagger}_{k,s} \chi^{\;}_{s} = \sum_{ i j l}
		{\tilde \phi}^*_{ijl}(k) d^{\dagger}_{ijls},
\end{eqnarray}
where
\begin{eqnarray}
	b_{ijls} \, , \, d_{ijls} \; \sim \; 1 \; \; {\rm and}
	\; \; \phi_{ijl}(k) \; \sim \; \sqrt{x^-_i} \,x_{\perp j} \; . \end{eqnarray}
Thus
\begin{eqnarray}
	\xi_{s}(x) = \sum_{ijl}\Big [ b_{ijls}
		\phi_{ijl}(x) + d^\dagger_{ijls}
		\phi^{\dagger}_{ijl} (x) \; \Big ]. \end{eqnarray}
It follows that the order of magnitude of $ijl^{th}$ mode's contribution
to the operator $\xi$ is ${1 \over \sqrt{\delta x^-_i}}{1 \over
\delta x_{\perp j}}$.

Next consider the transverse gluon field operator. We write
\begin{eqnarray}
	A_\perp(x) = \sum_{\lambda} \int {dk^+ d^2 k_\perp
		\over 2 \sqrt{k^+} (2 \pi)^3} \;
	\Big [ \epsilon_{\perp \lambda} a_{k,\lambda} e^{-ik\cdot x} +
 		\epsilon^*_{\perp \lambda} a^{\dagger}_{k,\lambda}
		e^{ik\cdot x} \; \Big ]. \end{eqnarray}
with
\begin{eqnarray}
	\big [ a_{k,\lambda}, a^{\dagger}_{k',\lambda'}\big ]
		= 2 (2 \pi)^3 \delta (k^+-k'^+) \delta^2(k_\perp
		- k'_\perp) \delta_{\lambda \lambda'}. \end{eqnarray}
The operators $a_k$ and $a^{\dagger}_k$ are again unsuitable
for our purposes. We define
\begin{eqnarray}
	a_{k,\lambda} = \sum_{ijl} {\tilde \phi}_{ijl}(k)
		a_{ijl \lambda} \end{eqnarray}
where ${\tilde \phi}_{ijl}(k) \sim \sqrt{\delta x^-_{i}} \,
\delta x_{\perp j} $.
Now,
\begin{eqnarray}
	A_\perp(x) = \sum_{ijl \lambda}\int {dk^+ d^2 k_\perp
		\over 2 \sqrt{k^+} (2 \pi)^3} \; \Big [
		\epsilon_{\perp\lambda} {\tilde \phi}_{ijl}(k)
		a_{ijl\lambda} e^{-ik\cdot x} +
      		\epsilon^*_{\perp\lambda} {\tilde \phi}^*_{ijl}(k)
		a^{\dagger}_{ijl \lambda} e^{ik\cdot x} \; \Big ]. \end{eqnarray}
The order of magnitude of ${ijl}^{th}$ mode's contribution
to $A_\perp(x)$ is ${1 \over \delta x_{\perp j}}$.

Because of the cutoffs we impose, the wave packet functions
$\phi_{ijl}(x)$ do not have zero modes; that is,  $ \int dx
\phi_{ijl}(x) =0 $.

We must also specify the order of magnitude of $\partial_\perp$,
$\partial^+$, and $(\partial^+)^{-1}$ when acting on a specified
subset of wavelets in an operator product.  For qualitative purposes
each gradient needs to be applied only to the most rapidly changing
wavelet in the subset, so that $\partial_\perp$ is of order $(\delta
x_\perp)^{-1}$, where $\delta x_\perp$ is the smallest width in the
subset.  A similar result holds for $\partial^+$. The effect of the
inverse operator $(\partial^+)^{-1}$ --- which involves an integration
--- is also determined by the smallest width $\delta x^-$ in the
product.  Qualitatively, it amounts to multiplication of the subset by
the width $\delta x^-$, for the other, broader widths act as constants
over this width. However, there is a special case we must note.

To see this, consider the action of $(\partial^+)^{-1}$ on a product
of field operators. Now the wave packet functions associated with
destruction operators have positive $k^+$ only, and the wave packet
functions associated with creation operators have negative $k^+$ only.
If the product contains only creation operators or only destruction
operators, then the operation of $(\partial^+)^{-1}$ qualitatively
amounts to multiplication by the smaller width among the product in
accord with the rule above. But if the product contains at least one
creation and one destruction operator, then wavelets with Fourier
modes of positive $k^+$ and wavelets with Fourier modes of negative
$k^+$ will both occur in the product. When they have the same mean
value of $k^+$, the product acquires a zero mode and then
$(\partial^+)^{-1}$ no longer produces a localized function but rather
one which covers the whole space. Thus the action of a single $({1
\over \partial^+})$ can lead to a logarithmic infrared divergence and
the action of $({1 \over \partial^+})^2$ can lead to both linear and
logarithmic infrared divergences already in the canonical terms. We
have discussed these divergences and the counterterms which remove
them in Section IV.B.

%%%%%%%%%%%%%%%%%%%%%%%%%%%%%%%%%%%%%%%%%%%%%%%%%%%%%%%%
\subsection{Ultraviolet divergences and counterterms}
%%%%%%%%%%%%%%%%%%%%%%%%%%%%%%%%%%%%%%%%%%%%%%%%%%%%%%%%

Consider ultraviolet divergences arising from the products of
interaction Hamiltonians. Each term in the interaction Hamiltonian
involves an integral over the product of three or four field
operators. Each field operator is expanded in terms of wave packet
functions. First, we consider the ultraviolet divergences arising from
a product of two interaction Hamiltonians. We shall consider the
special case of just two wave packet sectors, one being fixed and the
other being arbitrarily large. This corresponds to a single divergent
loop integral in the familiar language. The fixed wave packet width
corresponds to the fixed external momenta in the loop diagram, and the
varying wave packet width corresponds to the diverging loop momenta.

We denote the widths of the fixed and varying wave packet sectors
by ($\delta x^-$, $\delta x_\perp$) and ($\delta y^-$, $\delta y_\perp$),
respectively. The energy of the varying wave packet sector scales as
${\delta y^- \over \delta y_\perp^2}$. The energy diverges as
$\delta y_\perp \rightarrow 0$, corresponding to the traditional
type of short-distance ultraviolet divergence, which we consider
in this section. The divergences as $\delta y^- \rightarrow \infty$
are light-front infrared divergences, which we consider in the
next section.  Consider ultraviolet divergences which occur
when $\delta y_\perp \ll \delta x_\perp$. We set $\delta x^- =
\delta y^-$.
%(We will come back to the general case later).
The interaction Hamiltonian contains only cubic and quartic
terms in the field operators $\xi^\dagger$, $\xi$, and $A^i$.
We now observe that, as a consequence of momentum conservation
at the vertices, to produce an ultraviolet divergence {\it at
least two of the three or four operators have to refer to a
$y$-sector}.  We also note that a transverse gradient will
yield ${1 \over \delta y_\perp}$ unless it applies to a
product of $x$-sectors only.

Now consider the terms in the interaction Hamiltonian.  If all
operators refer to $y$-sectors, each term scales as ${\delta y^- \over
\delta y_\perp^2}$ except for the mass term (the helicity-flip
interaction), which scales as ${\delta y^- \over \delta y_\perp}$.
But if some operators refer to $x$-sectors, there will be additional
factors of ${\delta y_\perp \over \delta x_\perp}$. For each field
operator ($ \xi^\dagger, \, \xi, \, A_\perp$) belonging to an
$x$-sector and for each derivative $\partial_\perp$ acting on only
$x$-sectors, there is a factor of ${\delta y_\perp \over \delta
x_\perp}$.   This gives the contribution from a single sector with
width $\delta y^-, \delta y_\perp$ as $\delta y_\perp$ gets small. But
many wavelets with different $\delta y_\perp$ can contribute, namely,
all wavelets with different centers that overlap a given $x$-wavelet
(see Fig. 13). Since the transverse space has dimension two, the
number of independent wavelets of linear size $\delta y_\perp$ that
overlap with the wavelet of size $\delta x_\perp$ is of order $(\delta
x_\perp / \delta y_\perp)^2$.

Next consider the second-order perturbation theory formula
\begin{eqnarray}  H_I { 1 \over E.D} H_I, \end{eqnarray} where $E.D$ denotes
the
energy denominator. For qualitative purposes,  we can replace
the energy denominator by the energy of the $y$-sector,
namely, ${\delta y^- \over \delta y_\perp^2}$. Thus the
contribution in second order of a given term to the effective
Hamiltonian resulting from wave packet functions of a given
width $\delta y_\perp$ scales as
\begin{eqnarray} \Big ({\delta y^- \over \delta y_\perp^2}\Big) \times
\Big ({\delta y_\perp^2 \over \delta y^-}\Big) \times
\Big({\delta y^- \over \delta y_\perp^2}\Big ) \times
\Big({\delta x_\perp \over \delta y_\perp}\Big)^2 \times
\Big({\delta y_\perp \over \delta x_\perp}\Big)^n \end{eqnarray}
where $n$ counts the number of $\xi^\dagger$, $\xi$, $A_\perp$, and
$\partial_\perp$ factors that refer to $x$-sectors. If $n$ is
greater than 4 there is no divergence, while for $n$ less
than or equal to four a divergence arises. When $n$
is equal to 4, the contribution from the sector of a
given size $\delta y_\perp$ is a constant. Now we have
to sum over all scale sizes $\delta y_\perp$. The scale
sizes $\delta y_\perp$ allowed are inverse powers of 2 all
the way down to a minimum size ${1 \over \Lambda}$. A sum
over all scale sizes down to ${ 1 \over \Lambda}$ yields
a term of order ($ ln \Lambda$ /$ ln 2$), which gives a
logarithmic divergence as $\Lambda \rightarrow \infty$.

Thus the power counting rule for ultraviolet counterterms is the same
(in $\delta x_\perp$) as for the canonical Hamiltonian; namely, that
there can be at most four operators scaling as $\delta x_\perp^{-1}$.
Even though we did the analysis for two $H_I$'s, it holds for any
number of $H_I$'s, where the first $H_I$ creates one or more $y$-type
constituents, intermediate $H_I$'s maintain at least one $y$-type
constituent, and the last $H_I$ destroys all the remaining $y$-type
constituents. (We are ignoring contributions ``disconnected in $y$''
--- that is, products which contain an $H_I$ with only $x$-sector
constituents.  These diagrams have already acquired counterterms.)
Each energy denominator may be replaced by the energy of the
$y$-sector ${\delta y^- \over \delta y_\perp^2}$ since each
intermediate state contains at least one $y$-type constituent by
assumption. If there are $m$ factors $H_I$, there are $m-1$ factors
${\delta y_\perp^2 \over \delta y^-}$. Each $H_I$ scales like ${\delta
y^- \over \delta y_\perp^2}$ times powers of $( \delta y_\perp /
\delta x_\perp)$. An overall factor $(\delta x_\perp / \delta
y_\perp)^2$, counts the number of $y$-type wave packet functions that
overlap with a single $x$-type wave packet function. Thus the final
rule for the scaling versus $\delta x_\perp$ and $\delta y_\perp$ is
identical for $m$ $H_I$'s as for 2 $H_I$'s --- namely, divergences
occur for products in the effective Hamiltonian of $x$-operators up to
fourth order in $\xi^\dagger,\, \xi, \, A_\perp, \, {\rm and} \,
\partial_\perp$.

%%%%%%%%%%%%%%%%%%%%%%%%%%%%%%%%%%%%%%%%%%%%%%%%%%%%%%%%%
\subsection{Infrared divergence and counterterms}
%%%%%%%%%%%%%%%%%%%%%%%%%%%%%%%%%%%%%%%%%%%%%%%%%%%%%%%%%

Light-front infrared divergences are large-energy divergences as can
be seen from the light-front dispersion relation $ p^- = {p_\perp^2 +
m^2 \over p^+}$, which blows up as $p^+ \rightarrow 0 $. We have
already discussed the counterterms for light-front infrared
divergences that arise in the instantaneous interactions in the
Hamiltonian. In this section we discuss the infrared divergences which
arise in products of the interaction Hamiltonian.

As in the analysis of ultraviolet divergences we will separate out the
divergences caused by a specific $y$-type wave packet function from
the operator structure associated with an $x$-type wavelet. Let us
discuss the differences from the analysis of ultraviolet divergences.
First of all, the $y$-wavelet energy scale ${\delta y^- \over
y_\perp^2}$ is blowing up because $\delta y^-$ is approaching
$\infty$. Thus it will be assumed that $\delta y^- \gg \delta x^-$,
while to avoid confusion with ultraviolet problems we assume $\delta
y_\perp$ is of order $\delta x_\perp$. In contrast to the ultraviolet
case, light-front infrared divergences are caused by a single $y$-type
wavelet with a width $\delta y^-$.  The reason for this is that only
one $y$-type wavelet overlaps with a given $x$-type wavelet, since
$\delta y^-$ is much wider than $\delta x^-$. In this case momentum
conservation at the vertices implies that to produce a light-front
infrared divergence {\it at least two operators in $H_I$ have to be
$x$-type.} Each term in $H_I$ scales as ${\delta y^- \over \delta
y_\perp^2}$ times a factor: a) $({\delta y^- \over \delta x^-})^{1
\over 2} $ for each $\xi$ or $\xi^\dagger$ referring to an $x$-type,
b) $({\delta y^- \over \delta x^-})$ for every $\partial^+$ acting on
an $x$-type operator and a reciprocal factor $({\delta x^- \over
\delta y^-})$ for every $(\partial^+)^{-1}$ operator when applied to a
product including at least one $x$-type operator, and c) $({\delta x^-
\over \delta y^-})$ arising from the range of integration in each
$H_I$ involving an $x$-type operator.  As in the case of ultraviolet
divergences, in the effective Hamiltonian a term with no factor
$({\delta x^- \over \delta y^-})$ diverges logarithmically due to a
sum over different widths $\delta y^-$. However, exceptions to the
above rules occur for terms in the canonical Hamiltonian involving
$({1 \over \partial^+})$ or $({1 \over \partial^+})^2$ operators,
which we will note below.

Let us now catalogue the factors associated with the scaling behavior of
various terms in the interaction Hamiltonian. If we ignore $\partial^+$
or ${1 \over \partial^+}$ factors, then the rules are simple: an $x$-type
$\xi$ produces a factor $({\partial y^- \over \partial x^-})^{1/2}$,
an $x$-type $A_\perp$ produces no factor, and the presence of any $x$-type
operator in
$H_I$ produces $({\delta x^- \over \delta y^-})$ as result of the change of
integration width. Thus, ignoring $\partial^+$ and ${1 \over \partial^+}$
factors, we can compute the impact of any product of $x$-type operators on
the scaling behavior of $H_I$ :
\begin{eqnarray} \xi \xi  \qquad && : \qquad {\rm no \; factor} \nonumber \\
A_\perp \xi \qquad && : \qquad (\delta x^- / \delta y^-)^{1 \over 2} \nonumber
\\
A_\perp \xi \xi \qquad && : \qquad {\rm no \; factor} \nonumber \\
A_\perp A_\perp \qquad && : \qquad (\delta x^- / \delta y^-) \nonumber \\
A_\perp A_\perp \xi \qquad && : \qquad (\delta x^- / \delta y^-)^{1 \over 2}
\nonumber \\
A_\perp A_\perp \xi \xi \qquad && : \qquad {\rm no \; factor } \nonumber \\
\xi \xi \xi \qquad && : \qquad (\delta y^- / \delta x^-)^{1 \over 2} \nonumber
\\
\xi \xi \xi \xi  \qquad && : \qquad (\delta y^- / \delta x^-). \end{eqnarray}

Next we classify again the possible $x$-type operators in $H_I$ by the
degree of divergence in $H_I$ that results, when the effects of
$\partial^+$ and ${1\over\partial^+}$ factors are included. Here it is
assumed that $ \partial^+$ or $(1 / \partial^+)$ is applied to a
product that contains at least one $x$-type operator. It is useful to
recall the structure of terms in the canonical Hamiltonian (see
Section IV), which we consider one by one in the following.

When all the operators in the Hamiltonian refer to the $y$-sector, the
Hamiltonian scales as $\delta y^-$.  We have to determine the scale factor
that arises as a result of replacing two or more $y$-type operators by
$x$-type operators. Let us first consider some specific examples. In the
following we denote the operators that belong to the $x$-type by a
subscript $(x)$ and those belonging to the $y$-type by a subscript $(y)$
respectively.
Now consider the term in the Hamiltonian $\int dz^- d^2 z_\perp
\xi^\dagger A_\perp { 1 \over \partial^+} \xi$.   Here we write the
integration variable as $z$ in order to distinguish $dz$ from the
widths $\delta x$ and $\delta y$.
\begin{enumerate}
\item  For the case $\int dz^- d^2 z_\perp (A_\perp )_{(y)}
\xi^\dagger_{(x)}{ 1 \over \partial^+} \xi_{(x)} $,
from our rules we get
$ \delta y^-  \; \times \;
  \; ({\delta y^- \over \delta x^-}) \;  {\rm (for \; two \;} \xi '{\rm s})
 \;  \times \;
({\delta x^- \over \delta y^-})\;  ({\rm for} \;
 { 1 \over \partial^+}) \;  \times \;
({\delta x^- \over \delta y^-})$  (arising from the range of integration).
Thus the scaling
behavior is like $\delta x^-$ which we write as $\delta y^- \;  \times \;
({\delta x^- \over \delta y^-})$. Hence the scale factor is
$({\delta x^- \over \delta y^-})$.

\item For the case $\int dz^- d^2 z_\perp (A_\perp \xi^\dagger)_{(x)}
{ 1 \over \partial^+} \xi_{(y)}$
we get from our rules $ \delta y^- \;   \times \;
 ({\delta y^- \over \delta x^-})^{1 \over 2} \; ({\rm for}\; \xi^\dagger) \;
 \times \;
\; ({\delta x^- \over \delta y^-}) \;$  ( arising from the range of
integration) .
Thus the scaling behavior is like $(\delta x^- \, \delta y^-)^{1 \over 2}
$ which we write as
$ \delta y^- \; \times \; ({\delta x^- \over \delta y^-})^{ 1 \over 2}$ .
Hence the scale factor is $({\delta x^- \over \delta
y^-})^{1 \over 2} $.
\end{enumerate}

Now there may be exceptions to these rules when $(\partial^+)^{-1}$
or $(\partial^+)^{-2}$ acts on a product of operators, for then it would
pick out a zero mode as described above.  However, we have eliminated
this possibility by imposing a longitudinal cutoff $\epsilon$ on such
terms and then adding a counterterm to remove the $\epsilon$-dependence.
Because of this, integration by parts is allowed and can be used to
determine the order of magnitude of
$(\partial^+)^{-1}$ or $(\partial^+)^{-2}$. The exception to the usual
rule occurs, then, when
there are not $x$-sector wavelets on each side of
the inverse gradient,
for integration by parts allows us to re-write the product
so that the inverse operator acts entirely on $y$-type wavelets.  Then
it picks out a $y$-sector width and so does not produce the factor
$(\delta x^- /\delta y^-)$ or $(\delta x^- /\delta y^-)^2$ expected
from the above rules.  Thus, for example,
consider the term $ \int dz^- d^2 z_\perp
(\xi^\dagger A_\perp)_{(y)} { 1 \over \partial^+} (\xi
A_\perp)_{(x)} $.  With the zero mode eliminated,
we are free to
interchange the operation of ${
1 \over \partial^+}$ to get
$\int dz^- d^2 z_\perp {1 \over \partial^+} \big
((\xi^\dagger A_\perp)_{(y)} \big )
(A_\perp \xi)_{(x)}$.  Thus
the scaling is
$ \delta y^- \; \times \;({\delta y^- \over \delta x^-})^{ 1 \over 2}$ (from
$\xi$) $ \times $
$({\delta x^- \over \delta y^-})$ (arising from the range of integration).
Hence the scaling behavior is
$(\delta x^- \, \delta y^-)^{1 \over 2}$, which we write as
$ \delta y^- \; \times \; ({\delta x^- \over \delta y^-})^{ 1 \over 2}$; and
so the scale factor is $({\delta x^- \over \delta
y^-})^{1 \over 2} $.

In the following we determine the scale factor for various terms in
the interaction Hamiltonian when two or more operators in them are
replaced by $x$-type operators, the remaining operators being of
$y$-type. Each ${ 1 \over \partial^+} $ may apply to $y$ operators
too, e.g, ${1 \over \partial^+} \xi$ might mean ${ 1 \over
\partial^+}(A_\perp \xi)$ with $A_\perp$ of $y$-type. The scale factor
is unchanged by any such embedded $y$ operators.

$\underline{Products \; of \;   \xi \; and  \; \xi: }$

\begin{eqnarray} \xi \xi \qquad && : \qquad {\rm no \; factor} \nonumber \\
(\partial^+)^{-2} (\xi \xi) \qquad && : \qquad {\rm no \;
factor \; \; (exception \; applies)}
\nonumber \\
\xi (\partial^+)^{-1} \xi \qquad && : \qquad (\delta x^- / \delta y^-)
\nonumber \\
\xi (\partial^+)^{-2} \xi \qquad && : \qquad (\delta x^- / \delta y^-)^{2}.
\end{eqnarray}

$ \underline{Products \; of \; A_\perp \; and \; \xi : }$

\begin{eqnarray} A_\perp \xi \qquad && : \qquad (\delta x^- / \delta
y^-)^{1\over 2} \nonumber \\
(\partial^+)^{-1} (A_\perp \xi) \qquad && : \qquad (\delta x^- / \delta y^-)^{1
\over 2} \;
({\rm exception \; applies}) \nonumber \\
\xi {(\partial^+)}^{-2}(\partial^+ A_\perp) \qquad && : \qquad (\delta x^- /
\delta y^-)^{3 \over 2} \nonumber \\
(\partial^+ A_\perp) (\partial^+)^{-2} \xi \qquad && : \qquad (\delta x^- /
\delta y^-)^{3 \over 2} \nonumber \\
\xi {(\partial^+)}^{-2} A_\perp \qquad && : \qquad (\delta x^- / \delta y^-)^{5
\over 2} \nonumber \\
A_\perp {(\partial^+)}^{-2} \xi \qquad && : \qquad (\delta x^- / \delta
 y^-)^{5 \over 2} \nonumber \\
\xi {(\partial^+)}^{-1} A_\perp \qquad && : \qquad (\delta x^- / \delta y^-)^{3
\over 2} . \end{eqnarray}

$\underline{ Products\; of \; \xi, \xi, \; and \; A_\perp \; :}$

\begin{eqnarray} \xi^2 (\partial^+)^{-2} A_\perp \qquad && : \qquad
(\delta x^- / \delta y^-)^{2} \nonumber \\
\xi^2 (\partial^+)^{-2} \partial^+ A_\perp \qquad && : \qquad
(\delta x^- / \delta y^-) \nonumber \\
 A_\perp (\partial^+)^{-2} \xi^2  \qquad && : \qquad
(\delta x^- / \delta y^-)^2 \nonumber \\
(\partial^+ A_\perp) (\partial^+)^{-2} \xi^2  \qquad && : \qquad
(\delta x^- / \delta y^-) \nonumber \\
\xi^2 (\partial^+)^{-1}  A_\perp   \qquad && : \qquad
(\delta x^- / \delta y^-) \nonumber \\
\xi (\partial^+)^{-1}  A_\perp \xi  \qquad && : \qquad
(\delta x^- / \delta y^-) \nonumber \\
\xi A_\perp(\partial^+)^{-1}   \xi  \qquad && : \qquad
(\delta x^- / \delta y^-) . \end{eqnarray}

$\underline{ Products \; of \; A_\perp \; and \;  A_\perp\; :}$

\begin{eqnarray} A_\perp A_\perp \qquad && : \qquad (\delta x^- / \delta y^-)
\nonumber \\
A_\perp (\partial^+)^{-2} A_\perp \qquad && : \qquad (\delta x^- / \delta
y^-)^3  \nonumber \\
A_\perp (\partial^+)^{-2} \partial^+ A_\perp \qquad && : \qquad
 (\delta x^- / \delta y^-)^2 \nonumber \\
(\partial^+ A_\perp) (\partial^+)^{-2} \partial^+ A_\perp \qquad && : \qquad
 (\delta x^- / \delta y^-) \nonumber \\
(\partial^+)^{-2} (A_\perp \partial^+  A_\perp) \qquad && : \qquad
 {\rm no \; factor \; (exception \; applies)}  \nonumber \\
A_\perp \partial^+ A_\perp \qquad && : \qquad {\rm no \; factor} \nonumber \\
A_\perp {1 \over \partial^+} A_\perp \qquad && : \qquad
(\delta x^- / \delta y^-)^2 . \end{eqnarray}

$\underline{ Products \; of \; \xi, \; A_\perp, \; and \; A_\perp\; :}$

\begin{eqnarray} \xi (\partial^+)^{-2} A_\perp \partial^+ A_\perp \qquad && :
\qquad
(\delta x^- / \delta y^-)^{3 \over 2} \nonumber \\
A_\perp \xi (\partial^+)^{-1} A_\perp  \qquad && : \qquad
(\delta x^- / \delta y^-)^{3 \over 2} \nonumber \\
A_\perp  (\partial^+)^{-1} \xi A_\perp  \qquad && : \qquad
(\delta x^- / \delta y^-)^{3 \over 2} . \end{eqnarray}

$\underline {Products \; of \; \xi, \; \xi, \; and \; \xi \; :}$

\begin{eqnarray} \xi (\partial^+)^{-2} \xi^2 \qquad && : \qquad (\delta x^- /
\delta
y^-)^{3 \over 2} \nonumber \\
\xi^2 (\partial^+)^{-2} \xi \qquad && : \qquad (\delta x^- / \delta y^-)^{3
\over 2} . \end{eqnarray}

The above list splits into four types: a) Products which contain no
${\delta x^- \over \delta y^-}$
factors. They are all components of the color charge density, either the quark
or the gluon component. These products can occur in any number of $H_I$'s
without reducing the divergence of the overall product. If these products are
the only source of $x$-type operators, the overall divergence behaves as
$\delta y^-$, that is, linearly. b) Products which contain a factor of
$(\delta x^- / \delta y^-)^{1 \over 2}$. These products involve a single
$\xi$. Since only an even number of $\xi$'s can appear in the effective
Hamiltonian, two such products are required from two separate $H_I$'s; and the
result is to generate a factor $(\delta x^- / \delta y^-)$, which implies a
logarithmic divergence. If more than two such products occur there is no longer
any divergence. c) Products which generate a factor $(\delta x^- / \delta
y^-)$. If any such product occurs once, a linear divergence is converted to a
logarithmic divergence; and if they occur more than once, there is no
divergence. Finally, d) products that scale as a higher than linear power of
$(\delta x^- / \delta y^-)$. Such products cancel any divergence
--- even if present linearly.

As we have already explained, counterterms for linear divergences cannot have
finite parts. Thus we are interested in logarithmic divergences. Even in this
class, the power-counting analysis may be an
overestimate since, for example, as seen in Section III.B,
a logarithmic divergence may disappear when a
symmetric cutoff is used.

%%%%%%%%%%%%%%%%%%%%%%%%%%%%%%%%%%%%%%%%%%%%%%%%%%%%%%%%%
\subsection{Examples}
%%%%%%%%%%%%%%%%%%%%%%%%%%%%%%%%%%%%%%%%%%%%%%%%%%%%%%%%%
Now we provide examples of the determination of the divergence
structure in perturbation theory using
the qualitative phase space cell analysis described above.
We consider the second-order shift in
the energy of a gluon coming from a two-gluon
intermediate state.

As candidates for $H_I$, we take
\begin{eqnarray}
	H_{I(1) } = g f^{abc} \int dz^- d^2 z_\perp \partial^i
		A^j_a A^i_b A^j_c,
\end{eqnarray}
\begin{eqnarray}
	H_{I(2)} = -g f^{abc} \int dz^- d^2 z_\perp \Big (
		{1 \over \partial^+} \partial^i A^i_a \Big)
		A^j_b \partial^+ A^j_c.
\end{eqnarray}

We separate the field operators into low-momentum parts
which contain wavelets of width $\delta x_\perp,
\delta x^-$ and high-momentum parts which contain wavelets
of width $\delta y_\perp, \delta y^-$ .  In the discussion below we
drop factors of $g$ and color factors.

%%%%%%%%%%%%%%%%%%%%%%%%%%%%%%%%%%%%%%%%%%%%%%%%%%%%%%%%%%%%%
\subsubsection{Ultraviolet divergence}
%%%%%%%%%%%%%%%%%%%%%%%%%%%%%%%%%%%%%%%%%%%%%%%%%%%%%%%%%%%%%
To avoid confusion with infrared divergences, we set $\delta
x^- = \delta y^-$. For a candidate $H_I$, we choose $H_{I(1)}$.
Remember that to produce an ultraviolet divergence at least two
operators have to belong to the high sector.

a) Consider $H_{I(1)} \approx \int dz^- d^2 z_\perp (\partial^i
A^j  A^i)_{(y)} A^j_{(x)} $
and the energy shift $\Delta E \approx  H_{I(1)} { 1 \over E.D} H_{I(1)}$.
Referring back to the scaling formula (5.14), $n=2$. Hence the scaling
behavior of the energy shift is
${1 \over (\delta y_\perp)^2} \; \times \;  (\delta y_\perp)^2  \; \times \;
{1 \over (\delta y_\perp)^2}\; \times \;
({\delta x_\perp \over \delta y_\perp})^2 \; \times \;
({\delta y_\perp \over \delta x_\perp})^2$,
that is, $ \Delta E \approx {1 \over (\delta y_\perp)^2} $, a quadratic
ultraviolet divergence.

b) Consider $H_{I(1)} \approx \int dz^- d^2 z_\perp \partial^i A^j_{(x)} (A^i
A^j)_{(y)} $ and the energy shift $\Delta E \approx  H_{I(1)} { 1 \over E.D}
H_{I(1)}$. In this case $n=4$ and hence the scaling behavior of the energy
shift is ${1 \over (\delta y_\perp)^2} \; \times \;  (\delta y_\perp)^2
 \; \times \; {1 \over (\delta y_\perp)^2}\; \times \;
({\delta x_\perp \over \delta y_\perp})^2 \; \times \;
({\delta y_\perp \over \delta x_\perp})^4$, that is, $ \Delta E
\approx {1 \over (\delta x_\perp)^2} $, a logarithmic
ultraviolet divergence.

%%%%%%%%%%%%%%%%%%%%%%%%%%%%%%%%%%%%%%%%%%%%%%%%%%%%%%%%%%
\subsubsection{Infrared divergence}
%%%%%%%%%%%%%%%%%%%%%%%%%%%%%%%%%%%%%%%%%%%%%%%%%%%%%%%%%%
To avoid confusion we set $\delta x_\perp = \delta y_\perp$. Remember that to
produce an infrared divergence at least two operators have to belong to the
low sector. Thus, for example, we may take
\begin{eqnarray} H_{I(1)} \approx \int dz^- d^2 z_\perp \partial^i A^j_{(y)}
(A^i A^j)_{(x)}, \end{eqnarray}
\begin{eqnarray} H_{I(2)}  \approx \int dz^- d^2 z_\perp \Big ( { 1 \over
\partial^+}
\partial^i A^i_{(y)} \Big) (A^j \partial^+ A^j)_{(x)}. \end{eqnarray}

a) Consider the energy shift $\Delta E \approx H_{I(2)}
{ 1 \over E.D} H_{I(2)}$. According to our rules, $H_{I(2)} $
scales like $\delta y^-$. The energy denominator produces a
factor ${1 \over \delta y^-}$. Another $H_{I(2)}$ produces a
factor $\delta y^-$. Thus $ \Delta E \approx \delta y^-$, which
results in a linear infrared divergence.

b) Consider the energy shift $\Delta E \approx H_{I(2)}
{ 1 \over E.D} H_{I(1)}$. According to our rules, $H_{I(1)}$
scales like $ \delta y^- \; \times ({\delta x^- \over \delta y^-})$.
The energy denominator produces a
factor ${1 \over \delta y^-}$.  $H_{I(2)}$ produces a factor $\delta
y^-$. Thus $ \Delta E \approx \delta x^-$, which results in a logarithmic
infrared divergence.

To complete the analysis one must determine what happens when $\delta
y_\perp \ll \delta x_\perp$ and simultaneously $\delta y^- \gg \delta
x^-$, as well as other double orderings involving drastically
different transverse and longitudinal widths.  We have not completed
analyses for all such cases.

%%%%%%%%%%%%%%%%%%%%%%%%%%%%%%%%%%%%%%%%%%%%%%%%%%%%%%%%%%%%%%%%%
\subsection{The structure of counterterms: summary}
%%%%%%%%%%%%%%%%%%%%%%%%%%%%%%%%%%%%%%%%%%%%%%%%%%%%%%%%%%%%%%%%%

We have found that the power counting rule for ultraviolet
counterterms is the same (in $\delta x_\perp$) as for the canonical
Hamiltonian, that there can be at most four operators scaling as ${1
\over \delta x_\perp}$ . However, the counterterm has a complex
nonlocality in the longitudinal coordinates. The reason for this is
that the longitudinal distance scales for $x$-type and $y$-type
wavelets overlap  even when ultraviolet transverse divergences are
produced. Thus the ultraviolet counterterms are built out of products
up to fourth order in $\xi, \;  \xi^\dagger, \; A_\perp, \; {\rm and}
\; \partial_\perp$ (see Eq. (3.16)) and have completely arbitrary
longitudinal structure. As a result of the constraint from
longitudinal boost invariance only vertex counterterms can involve
{\it a priori} unknown functions of longitudinal variables.

The counterterms for infrared divergences, on the other hand, involve
arbitrary numbers of quark and gluon operators. They have a complex
nonlocality in the transverse variables. This is in contrast to the
divergent counterterm for the canonical instantaneous four-fermion
interaction which is local in the transverse direction. If we identify
the counterterms arising from infrared gluons (small longitudinal
momentum) as the source of transverse confinement, then the unknown
nonlocal transverse behavior would have to include confining effects
at large transverse separation.

{}From a physical point of view the appearance of long-range many-body
interactions involving arbitrary numbers of quarks and gluons is
inevitable if one allows a confining two-body interaction.  It has
long been appreciated in the CQM that if one uses confining two-body
interactions, unphysical long-range van der Waals forces inevitably
arise above 1+1 dimensions \cite{WAAL}.  In the CQM various schemes
allow the potential to distinguish between colored objects in
different hadrons, but we have no such operators at our disposal in a
field theory.  Therefore the only way to cancel the residual
long-range multipole interactions between color singlets is with
long-range many-body interactions.  It would be a failure of our
approach if such operators were not allowed by the same arguments that
lead to a confining two-body interaction.  The possible impact such
operators have on the predictive power of the theory is not known yet.

There is a question of whether the artificial potential will
drastically alter the analysis of infrared divergences.  We do not
study this problem here.

As mentioned in Section II, one might worry that the appearance of
functions of momenta in the counterterms could destroy the predictive
power of the theory and lead to non-renormalizability. However, these
functions are needed to restore Lorentz covariance and gauge
invariance as $g \rightarrow g_s$, and without such counterterms
physical quantities will not even approach finite limits as cutoffs
are removed for any coupling.

We do not expect any new parameters to appear in the properly
renormalized theory; so all new counterterms must actually be
determined by the finite set of canonical free parameters. This
problem has been studied in Ref. \cite{PW93}, where a set of
relationships called ``coupling coherence'' \cite{CC} were used to fix
this relationship. Coupling coherence rests on the observation that
only canonical variables should run independently with the cutoffs in
renormalization group equations.  To lowest orders in perturbation
theory it has been explicitly demonstrated that coupling coherence
fixes the bare Hamiltonian in scalar field theory, and that Lorentz
covariance is restored by the resultant counterterms \cite{P93}.
Whether coupling coherence helps solve the infrared problem has yet to
be demonstrated, but it is at least able to deal with spontaneous
symmetry breaking \cite{PW93}.

%%%%%%%%%%%%%%%%%%%%%%%%%%%%%%%%%%%%%%%%%%%%%%%%%%%%%
%%%%%%%%%%%%%% cutoff.tex (input to QCD paper) %%%%%%
%%%%%%%%%%%%%%%%%%%%%%%%%%%%%%%%%%%%%%%%%%%%%%%%%%%%%
%%%%%%%%%%%%  last modified 1/24/94           %%%%%%
%%%%%%%%%%%%%%%%%%%%%%%%%%%%%%%%%%%%%%%%%%%%%%%%%%%%%

%%%%%%%%%%%%%%%%%%%%%%%%%%%%%%%%%%%%%%%%%%%%%%%%%%%%%
\section{A CUTOFF SCHEME FOR THE HAMILTONIAN }
%%%%%%%%%%%%%%%%%%%%%%%%%%%%%%%%%%%%%%%%%%%%%%%%%%%%%

In the previous sections we presented a qualitative analysis of
divergences based on wavelets. Now we set up a precise momentum space
framework for renormalization of the LFQCD Hamiltonian.  We employ the
``similarity renormalization scheme'' \cite{GW93a,GW93b}, which
allows one to avoid the difficulties often encountered in traditional
perturbation renormalization schemes when using plane wave states to
study the structures of divergences.  Thus, although we use plane wave
states to determine the form of the effective Hamiltonian, we may
still set up the renormalization to follow the qualitative wavelet
analysis of Section V.  Of course, the assumption here is that the
artificial Coulomb and linear potentials will not affect the structure
of the divergences.  As we have seen, the canonical Hamiltonian
already leads to divergences at the tree level due to $k^+$ getting
small. Beyond tree level, products of Hamiltonians also lead to
divergences both due to $k^+$ getting small and $k_\perp$ getting
large.  We need to introduce a regulator before we can investigate
these divergences and construct the corresponding counterterms. This
is done here.

%%%%%%%%%%%%%%%%%%%%%%%%%%%%%%%%%%%%%%%%%%%%%%%%%%%%%%%%%%%%%%
\subsection{Cutoffs on constituents}
%%%%%%%%%%%%%%%%%%%%%%%%%%%%%%%%%%%%%%%%%%%%%%%%%%%%%%%%%%%%%%

Since the final step of the renormalization process involves the
nonperturbative diagonalization of the Hamiltonian, we need to develop
a cutoff procedure that is applicable to the Hamiltonian as a whole
instead of just to perturbative calculations order by order. One way
to accomplish this is to cut off the single particle momenta appearing
in the field variables themselves. Cutoffs on the constituent momenta
$k^+$ and $k_\perp$, which we call ``constituent cutoffs'', obviously
violate the longitudinal and transverse boost symmetries of the light
front. What does this violation mean?  The direct effect of the cutoff
is to eliminate states that would be present without the cutoff. For a
given center of mass momentum, a key parameter is the lowest mass
state that is not eliminated by the cutoff.  This lowest mass
depends on the center of mass momentum chosen, and is degraded if
$P_\perp$ increases or $P^+$ decreases.

If constituent cutoffs destroy light-front boost invariance except for
a limited range of center of mass momenta, then, why do we want to use
such cutoffs?  One could instead employ --- as has been suggested
elsewhere, see e.g.,  \cite{P93,fs,Le 83,DLY 70} --- cutoffs which
preserve both transverse and longitudinal boost invariance.  We shall
call such cutoffs ``Jacobi cutoffs'' for they act on the Jacobi or
internal momenta of a constituent, which are defined relative to the
center of mass momenta.  Unfortunately, Jacobi cutoffs inevitably
refer to extensive quantities of a state and thereby introduce
nonlocalities in the Hamiltonian.  If the cutoff on constituent $i$
depends on the total momentum of the state, then a dependence on all
spectators $j$ will be introduced in any matrix element. Thus any
counterterm may change when another spectator is included, which
greatly complicates the renormalization process\cite{P93}. So there is
a stiff price to pay for insisting on a regulator which preserves
explicit boost symmetry.

We prefer to employ a constituent cutoff scheme because of the
conceptual simplicity of its implementation.  This is especially
important for LFQCD, where the final diagonalization of the
Hamiltonian will be done on a computer.  In addition, we will be
encountering nonlocal effective potentials which result from the
similarity renormalization scheme,  and it will be important to avoid
nonlocalities from the regulator itself. Choosing constituent cutoffs
ensures there is a limited possibility of confusing between different
sources of nonlocalities. Moreover, as we show below, one can choose
constituent cutoffs which would allow boost invariance to be
maintained to a good approximation within a large domain of center of
mass momenta. In the next section, we discuss the details of the
cutoff scheme and map out the center of mass domain (the values of
$P^+$ and $P^\perp$) in which when a cutoff constituent appears in a
state, the internal mass of that state is guaranteed to be above an
effective cutoff $\Lambda^2$.

%%%%%%%%%%%%%%%%%%%%%%%%%%%%%%%%%%%%%%%%%%%%%%%%%%%%%%%%%%%%%%
\subsection{Details of the cutoff scheme}
%%%%%%%%%%%%%%%%%%%%%%%%%%%%%%%%%%%%%%%%%%%%%%%%%%%%%%%%%%%%%%

We start with the formula for the total mass of an eigenstate of
$H_0$ (where $H_0$ is the free part of the Hamiltonian) of $n$
constituents with momenta ($k_i^+, k_{i\perp}$) and masses $m_i$:
\begin{eqnarray} M^2 = P^+ \sum_i {m_i^2 + k_{i\perp}^2 \over k_i^+} -
P_\perp^2, \end{eqnarray}
where $P^+ = \sum_i k_i^+$ and $P_\perp = \sum_i k_{i\perp}$.
If we introduce the internal or Jacobi momenta
$ x_i = {k_i^+ \over P^+}$ and $q_{i\perp} = k_{i\perp} - x_i P_\perp$, then
\begin{eqnarray} M^2 = \sum_i {m_i^2 +q_{i\perp}^2 \over x _i}. \end{eqnarray}
Thus the square of the total mass of the system is simply a sum
of internal mass
terms, which is a manifestation of the kinematical boost symmetries.

We need to prove that the partial sums of internal mass terms can always be
lowered by replacing two constituents by a single constituent with the same
total $q_\perp$ and $x$ and the lower of the constituent masses;
that is, \begin{eqnarray}
{m_1^2 + q_{1\perp}^2 \over x_1} +
{m_2^2 + q_{2\perp}^2 \over x_2}  \ge
 {(m_1+m_2)^2 + (q_{1\perp} + q_{2\perp})^2 \over x_1 +  x_2} . \end{eqnarray}
Let $ \xi_1 = {x_1 \over x_1 + x_2}$,
$\xi_2 = {x_2 \over x_1 + x_2} = 1-\xi_1$,
$q_{1\perp} = r_{\perp} + \xi_1 q_\perp$,
$q_{2\perp} = -r_{\perp} + \xi_2 q^\perp$, where $q_\perp = q_{1\perp} +
q_{2\perp}$.
Then
\begin{eqnarray}
{m_1^2 + q_{1\perp}^2 \over x_1} +
{m_2^2 + q_{2\perp}^2 \over x_2}  && =
 {1 \over x_1 + x_2}
\left[
{m_1^2 + r_{\perp}^2 \over \xi_1} +
{m_2^2 + r_{\perp}^2 \over \xi_2}   + q_\perp^2 \right]
\nonumber \\
&& \ge
{1 \over x_1 + x_2}\left[
{m_1^2 \xi_2 + m_2^2 \xi_1 \over \xi_1 \xi_2}
+ q_\perp^2 \right].  \end{eqnarray}
But the minimum value of ${m_1^2 \over \xi_1} + {m_2^2 \over \xi_2}$
is $(m_1+m_2)^2$.
Hence
\begin{eqnarray}
{m_1^2 + q_{1\perp}^2 \over x_1} +
{m_2^2 + q_{2\perp}^2 \over x_2}
\ge {(m_1+m_2)^2 + (q_{1\perp} + q_{2\perp})^2 \over x_1 +  x_2}
\ge {m^2 + (q_{1\perp} + q_{2\perp})^2 \over x_1 +  x_2},\end{eqnarray}
where $m$ is the lower of the constituent masses.  It follows then
that any multi-constituent state with a given total
$x$ and $q_\perp$ must have an invariant mass which is greater
than or equal to that of the state containing just two constituents
of lowest mass $m$ and having the same total $q^\perp$ and $x$.

We set constituent momentum cutoffs such that for states of center of
mass momentum ($P^+_0$, $P_\perp =0$) the cutoff momenta first appear
in states when the $(mass)^2$ is $ 2 \Lambda^2$ and then evaluate the
range of center of mass momenta ($P^+$, $P_\perp$) for which cutoff
momentum constituents first appear at a $(mass)^2$ of $\Lambda^2$ or
higher. Thus all states with $(mass)^2 < 2\Lambda^2$ and total
momentum $(P^+_0,0)$ are kept; and we must then restrict the range of
$(P^+,P_\perp)$ so as to ensure that no state with this total momentum
and $(mass)^2 < \Lambda^2$ has a constituent beyond the cutoff
boundary. This gives us a large domain to test whether covariance can
be restored to good approximation with appropriate counterterms.

By our theorem, if a constituent of
momentum ($k^+_1$, $ k_{1\perp}$) and mass $m_1$ appears in a state of center
of
mass momentum ($P^+_0, 0$), then the state of minimum total mass containing
this constituent contains a second constituent with momentum
($k^+_2$, $k_{2\perp}$), where $k_2^+ = P_0^+ - k_1^+$ and $k_{1\perp} = -
k_{2\perp}$, and mass $m$, the minimum of the constituent masses. The mass of
this state is
\begin{eqnarray} (mass)^2 = {m_1^2+k_{1\perp}^2 \over x_1} +
{m^2+k_{1\perp}^2 \over 1 - x_1}  \end{eqnarray}
with $x_1 = {k_1^+ \over P_0^+}$, $x_2 = {k_2^+ \over P_0^+}$. We therefore
need to characterize the {\it boundary curve} in $x_1,k_{1\perp}$ space on
which
\begin{eqnarray} {m_1^2 + k_{1\perp}^2 \over x_1} +
{m^2 + k_{1\perp}^2 \over 1- x_1}  =
 {m_1^2 \over x_1} + {m^2 \over 1- x_1} +{k_{1\perp}^2 \over x_1 (1-x_1)}
 = 2 \Lambda^2. \end{eqnarray}
Solving this equation for $k_{1\perp}^2$ we have
\begin{eqnarray} k_{1\perp}^2 = 2 \Lambda^2 x_1 (1-x_1) - m_1^2 (1-x_1) - m^2
x_1. \end{eqnarray}
For $\Lambda^2 \gg m_1^2$, we find
$k_{1\perp}^2$ achieves its maximum value very
close to ${\Lambda^2 \over 2}$, when $x_1 = {1 \over 2}$. To ensure no
deterioration in the cutoff for momenta $P^+ > P^+_0$, we define the cutoff
boundary curve to be
\begin{eqnarray} \label{eq:b9} k_{1\perp}^2 =  {\Lambda^2 \over 2} - m^2
\qquad  {\rm for} \quad  x_1 = {k_1^+ \over P_0^+} > {1 \over 2}
\quad  ( {\rm including} \; x_1 > 1), \end{eqnarray}
while for $x_1 < {1 \over 2} $
\begin{eqnarray} \label{eq:b10}
k_{1\perp}^2 = 2 \Lambda^2 x_1 (1-x_1) - m^2, \end{eqnarray}
which applies only in the domain $x_1 > {m^2 \over 2 \Lambda^2}+ {\cal
O}({m^4 \over \Lambda^4})$ for which
$k_{1\perp}^2 \ge 0$.  We will ignore corrections to the bound on $x$
that are ${\cal O}({m^4 \over \Lambda^4})$, but a
careful analysis must retain these.  $x_1 < {m^2 \over 2 \Lambda^2}$
is not allowed.  Note that $x_1$ is defined in terms of $P_0^+$, even
for $P^+ \neq P^+_0$.

The reader may at first feel somewhat uneasy about the statement
``including $x_1 > 1$.''
Consider for simplicity a two-body system of equal-mass particles. First
consider the frame $P^+=P^+_0$ and $P_\perp=0$. The internal mass of the state
is
\begin{eqnarray} M^2 = \left({m^2 +k_{1\perp}^2 \over k_1^+} +
{m^2 +k_{2\perp}^2 \over k_2^+} \right) P_0^+ . \end{eqnarray}
Define $x_1 ={ k_1^+ \over P_0^+}$ and $x_2 = { k_2^+ \over P_0^+}$.
Thus in the frame  $P^+=P_0^+$, $0 < x_1 < 1 $, where we have dropped
terms of order $m^2/\Lambda^2$.
Next consider the frame
$P^+ = {\tilde P}^+$. Let $ c = {{\tilde P}^+ \over P_0^+}$. Thus in this
frame $ 0 < x_1 < c $ and $c$ is greater than 1 if ${\tilde P^+} > P_0^+$.
We also have $x_2 = c - x_1$ and $0<x_2<c$.
If we did not extend the boundary as exhibited in (\ref{eq:b9}),
however, the $x_i$'s
would be limited as $0 < x_i < 1$; and
one could
not have a two constituent state with ${\tilde P}^+ > 2P^+_0$.  Such a
cutoff would likely be too restrictive.

We next determine the boundary curve in $P^+$ and $P_\perp$ space for which
the cutoff mass is $\Lambda^2$, namely, constituents on the cutoff boundary
occur only in states of $(mass)^2 = \Lambda^2$ or higher.
If the center of mass momentum of a state is $(P^+,P_\perp)$ and
we consider one constituent $(k^+_1,k_{1\perp})$ on the cutoff boundary
$(\ref{eq:b9}-\ref{eq:b10})$
with $x_1 = k^+_1/P^+_0$, then from our theorem above
we know that the invariant mass of this state must be at least
\begin{eqnarray}
\label{eq:bm2}
M^2 = \frac{m^2 + (k_{1\perp} - \xi_1 P_\perp)^2}{\xi_1} +
\frac{m^2 + (k_{2\perp} - \xi_2 P_\perp)^2}{\xi_2}
= \frac{m^2 + (k_{1\perp} - \xi_1 P_\perp)^2}{\xi_1(1-\xi_1)},
\end{eqnarray}
where $\xi_1 = k_1^+/P^+$, $\xi_2 = k_2^+/P^+$, and the second
equality follows from the relations $k_2^+ = P^+ - k^+_1$ and
$k_{2\perp} = P_\perp - k_{1\perp}$.  We ask, what must
$(P^+,P_\perp)$ be to ensure that the minimum of this mass
over all allowed values of $k_1^+$, with $k_{1\perp}^2$ on
the boundary, is $\Lambda^2$?

First, note that $M^2$ is a minimum when $k_{1\perp}$ and $P_\perp$
are in the same direction.  Let $\gamma = P^+/P_0^+$, so that
$\xi_1 = x_1/\gamma$.  Using (\ref{eq:bm2}) and
the requirement that $M^2$ be greater than $\Lambda^2$, we find
\begin{eqnarray}
x_1^2 P_\perp^2 - 2x_1\gamma \mid k_{1\perp}\mid \; \mid P_\perp\mid
\ge \Lambda^2 x_1 (\gamma - x_1) - \gamma^2\big(k_{1\perp}^2+m^2\big).
\end{eqnarray} Dropping
the solution which excludes $P_\perp = 0$, we have
\begin{eqnarray}
0 \le \mid P_\perp \mid \le {1\over x_1}\left[ \gamma \mid k_{1\perp}
\mid
- \left( \Lambda^2 x_1 (\gamma - x_1) - \gamma^2 m^2 \right)^{1/2}
\right].
\end{eqnarray}
Substituting for $k_{1\perp}$ from (\ref{eq:b9}-\ref{eq:b10}),
dropping
terms of order $m^2/\Lambda^2$, and minimizing the right-hand side
with respect to $x_1$, we find
\begin{eqnarray}
0 \le \mid P_\perp \mid \le
\Lambda \left[ (1-\gamma)(2\gamma - 1)\right]^{1/2} \qquad \rm{for}
\quad  1/2 \le \gamma \le 3/4.
\end{eqnarray}
and
\begin{eqnarray}
0 \le \mid P_\perp \mid \le  \frac{\Lambda}{2\sqrt{2}}\qquad \rm{for} \quad
\gamma \ge 3/4,
\end{eqnarray}
Using the definition of $\gamma$, this implies that only states with
$P^+ \ge P^+_0/2$ are allowed.

%%%%%%%%%%%%%%%%%%%%%%%%%%%%%%%%%%%%%%%%%%%%%%%%%%%%%%%%%%%%%%
\subsection{Cutoffs and field operators}
%%%%%%%%%%%%%%%%%%%%%%%%%%%%%%%%%%%%%%%%%%%%%%%%%%%%%%%%%%%%%%

To summarize the results of the previous section,
the cutoff boundary on each
constituent's momenta $(k^+_i,k_{i\perp})$
is specified by
\begin{eqnarray} k_{i\perp}^2 = 2 \Lambda^2 {k_i^+ \over P_0^+}
(1-{k_i^+\over P_0^+}) - m^2 \qquad {\rm for} \quad
 {m^2 \over 2 \Lambda^2}P_0^+ < k_i^+  < {P_0^+ \over 2}
\end{eqnarray}
 and
\begin{eqnarray} k_{i\perp}^2 = {\Lambda^2 \over 2} - m^2 \qquad {\rm
for}\quad
 {P_0^+ \over 2}<
k_i^+  , \end{eqnarray}
where $m$ is the lowest of the constituent masses.
Then, in a considerable range of center of mass domain ($P_\perp=0, P^+= {P^+_0
\over 2}$ to $P_\perp ={\Lambda  \over 2 \sqrt{2}},
P^+ \geq {3P^+_0 \over 4}$), when a
cutoff constituent appears in a state, the internal mass of the state is
guaranteed to be at least $\Lambda^2$.  This domain for physics with
cutoff $(mass)^2\ge \Lambda^2$ is shown in Fig. 4.

At all stages of the calculation we
will also want to provide a buffer zone outside this constituent
momentum cutoff boundary extending $k_{i\perp}^2$ roughly a factor of
2 and $k^+_i$ a factor of 1/2.  We can accomplish this by setting the
outside of the buffer zone at
\begin{eqnarray} k_{i\perp}^2 = 4 \Lambda^2 {k_i^+\over P^+_0}
(1-{k^+_i\over P_0^+}) - m^2 \qquad {\rm for} \quad
 {m^2 \over 4 \Lambda^2}P_0^+ < k^+_i  < {P_0^+ \over 2}
\end{eqnarray}
 and
\begin{eqnarray} k_{i\perp}^2 = \Lambda^2  - m^2 \qquad {\rm  for}\quad
 {P_0^+ \over 2}< k^+_i. \end{eqnarray}
The buffer zone allows the use of a smoothing cutoff in order to
let interactions die gradually.
In (6.17-6.20),
the cutoff $\Lambda$ eliminates both ultraviolet transverse and
infrared longitudinal degrees of freedom.  In order to define such
a constituent cutoff which still limits the invariant mass of
a state, however, we are forced to introduce a longitudinal
momentum scale $P_0^+$.  Dependence on this scale must also
be eliminated as part of the renormalization process.

We now have the boundary on the degrees of freedom kept in
$k_\perp$ and $k^+$.  This is shown in Fig. 5.
These cutoff boundaries are to be employed in the integrals
over momenta in the momentum-space expansion of the
field operators.  The quark field operator is
\begin{eqnarray}
\xi(x) =  \sum_s \chi_s \int_c {dk^+ d^2k_\perp\over 16\pi^3}
\left[ b_s(k)e^{-ik\cdot x} + d^\dagger_s(k)e^{ik\cdot x}\right],
\end{eqnarray}
and the gluon field operator is
\begin{eqnarray}
A^i(x) =  \sum_\lambda \int_c {dk^+ d^2k_\perp\over
16\pi^3\sqrt{k^+}}
\left[ \epsilon_\lambda^i a_\lambda(k)e^{-ik\cdot x} +
\epsilon_\lambda^{i*} a^\dagger_\lambda(k)e^{ik\cdot x}\right],
\end{eqnarray}
where color indices have been suppressed.
If we choose sharp momentum cutoffs, the momentum integrals may be explicitly
written as
\begin{eqnarray}
\int_c dk^+ d^2k_\perp && \equiv \int_{k^+_{min}}^{P^+_0/2} dk^+
\int d^2k_\perp
\theta(2\Lambda^2 x(1-x) - m^2 - k_\perp^2) \nonumber \\
&& \qquad +\int_{P^+_0/2}^\infty dk^+
\int d^2k_\perp
\theta({\Lambda^2\over 2} - m^2 - k_\perp^2),
\end{eqnarray}
where now $k^+_{min} = {m^2\over 2\Lambda^2}P_0^+$.

With sharp momentum cutoffs it is possible to develop non-analyticities in the
structure of counterterms. We illustrate this with the following example.

Consider the type of transverse momentum integration that occurs in the gluon
self-mass. In its simplest form this integral is
\begin{eqnarray} I &&= \int d^2 k_{1 \perp} d^2 k_{2 \perp} \,
\delta^2 (P_\perp - k_{1 \perp}
- k_{2 \perp}) \; \theta \big(\Lambda^2 - k_{1 \perp}^2\big) \;
\theta \big(\Lambda^2 - k_{2 \perp}^2\big), \\
&& = \int d^2 k_{1 \perp}
\; \theta \big(\Lambda^2 - k_{1 \perp}^2\big) \;
\theta \big(\Lambda^2 - (P_\perp - k_{2 \perp})^2\big), \end{eqnarray}
with $\Lambda$ a very large parameter.
Introducing the variable $ \kappa_\perp $ by $ k_{1 \perp} = \kappa_\perp + {1
\over 2} P_\perp, \; k_{2 \perp} = - \kappa_\perp + { 1 \over 2} P_\perp$,
\begin{eqnarray} I = { 1 \over 2} \int_{0}^{2 \pi} d \phi \; \kappa_{max}^2 ,
\end{eqnarray}
with (for $P_\perp << \Lambda$)
\begin{eqnarray} \kappa_{max} \simeq Minimum \big[ \Lambda - { 1 \over 2} \mid
P_\perp \mid cos
\phi, \Lambda + { 1 \over 2 } \mid P_\perp \mid cos \phi \big ]. \end{eqnarray}
Then,
\begin{eqnarray} I_{divergent} = \pi \Lambda^2 - 2 \Lambda \mid P_\perp \mid.
\end{eqnarray}
The sharp cutoff has given rise to a non-analytic divergence.
To avoid such non-analyticities in the structure of counterterms,
analytic cutoff functions are
preferred. For example, the momentum integrals can be cutoff as
\begin{eqnarray}
\int_c dk^+ d^2k_\perp \equiv \int_0^{\infty} dk^+
\int d^2k_\perp
{\cal C}\Big({k_\perp^2 + m^2 \over
k_\perp^2 + m^2 + 4\Lambda^2 x/(1 + 4x)}\Big),
\end{eqnarray}
where $x = k^+/P^+_0$. ${\cal C}(y)$ is a  cutoff
function which equals 1 for $y =0$ and decreases analytically to 0 for $y$
= $1$.  The integration over longitudinal momentum
is thus cutoff at a minimum $k^+_{\min}= {m^2\over 2\Lambda^2}P_0^+$.
As long as ${\cal C}$ is analytic the self-mass will have to be analytic in
$P_\perp$; the $\mid P_\perp \mid$ divergence will disappear.

%%%%%%%%%%%%%%%%%%%%%%%%%%%%%%%%%%%%%%%%%%%%%%%%%%%%%%%%
%%%%%%%  similarity.tex  (input to QCD paper)      %%%%%
%%%%%%%%%%%%%%%%%%%%%%%%%%%%%%%%%%%%%%%%%%%%%%%%%%%%%%%%
%%%%%%%   last modified:  1/24/93                 %%%%%
%%%%%%%%%%%%%%%%%%%%%%%%%%%%%%%%%%%%%%%%%%%%%%%%%%%%%%%%

%%%%%%%%%%%%%%%%%%%%%%%%%%%%%%%%%%%%%%%%%%%%%%%%%%%%%%%%
\section{ Similarity Renormalization Scheme}
%%%%%%%%%%%%%%%%%%%%%%%%%%%%%%%%%%%%%%%%%%%%%%%%%%%%%%%%

The bare cutoff Hamiltonian will be solved in two stages. The first
stage is a renormalization stage in which effects at relativistic
momenta are computed using perturbation theory.  These effects are
computed to a specified order in $g$ in perturbation theory.  For the
simplest computation the renormalization calculation is limited to
second-order terms in $g$.  To study the next level of complexity, the
similarity renormalization program can be carried out to fourth order
in $g$, but that is not attempted here.

The output of the renormalization stage is an effective Hamiltonian
which is dominated in weak coupling by nonrelativistic relative
momenta.  The bound states of the effective Hamiltonian in weak
coupling are pure $q\bar q$, pure $qqq$, or pure $gg$ states, exactly
as predicted in the CQM.  All major effects of gluon emission and
absorption --- or more complex processes --- are absorbed into the
effective Hamiltonian for these states.  Once the effective
Hamiltonian is constructed, the computation of bound state energies
involves readily executed numerical computations.

A new procedure for the renormalization of Hamiltonians has been
developed by G{\l}azek and Wilson to best meet the needs of
light-front computations \cite{GW93a,GW93b}.  The new procedure is
needed to simplify qualitative calculations leading to the effective
Hamiltonian.  Of especial importance is an ability to handle linear
potentials correctly.  The problem with a linear potential is that
even if it has a small coefficient (even of order $g^6$), the linear
potential becomes of order 1 or even larger simply by going to large
enough distances.  But once a linear potential has become large enough
it can no longer be treated perturbatively and must instead be
incorporated into an unperturbed Hamiltonian, for either qualitative
or quantitative analysis.  Since a crucial question is how linear
potentials do or do not develop, a procedure which handles them
sensibly is of critical importance.

Another requirement for the new renormalization procedure is that no
small energy denominators should arise in any order of perturbation
theory.  That is, energy denominators of the form $1/(E-E')$ should
never arise with $|E-E'|$ being far smaller that either $E$ or $E'$.
Normal perturbative treatments of field theoretic Hamiltonians lead
constantly to such small energy denominators.  But for exploring the
complex phenomena of light-front renormalization, which will require
crude and qualitative analyses, it is desirable to avoid encountering
the possibly large effects resulting from such small denominators.
Such large effects are very difficult to estimate using qualitative
arguments.  Large effects arise in particular when kinematic
constraints force $E-E'$ to have a single sign yet allow it to be
small.  A typical example is collinear emission of gluons at large
transverse momenta.  In this case, the incoming and outgoing
constituents all have large light-front energies due to their large
momenta.  Nevertheless, the final state energy is only slightly above
the initial state energy, yet cannot ever be lower than the initial
state energy.

The third requirement we place on the new renormalization procedure is
that the effective Hamiltonian should have a Jacobi-type cutoff
severely restricting internal momenta while the center of mass momenta
are restricted only by the original bare cutoff. This will allow us to
verify the restoration of longitudinal and transverse boost invariance
in the effective Hamiltonian itself.

The first stage computation also identifies all cutoff dependencies in
the bare Hamiltonian, and therefore can be used to identify all needed
counterterms to eliminate the cutoff dependence in perturbation
theory, including dependencies that violate as well as conserve boost
invariance.

\subsection{The similarity transformation of Hamiltonians}

For now our discussion will be valid for any Hamiltonian with non-negative
eigenvalues. Specialization to the light-front case occurs below in
Section VII.B.
The basic idea of the ``similarity renormalization scheme'' is
to develop a sequence of infinitesimal unitary
transformations that transform the initial bare Hamiltonian
$H_B$ to an effective Hamiltonian $H_\sigma$, where $\sigma$
is an arbitrarily chosen energy scale.  When the 0
transformations are multiplied together, one arrives at a
transformation $S_\sigma$:
\begin{equation}
\label{eq:sim1}
H_\sigma = S_\sigma H_B S_\sigma^\dagger.
\end{equation}
%where unitarity implies that $S^{-1} = S^\dagger$ and
%$H_\sigma$ has the same spectrum as $H_B$.
The basic goal for the transformation $S_\sigma$ is that
$H_\sigma$ should be in band-diagonal form relative to the
scale $\sigma$.  What this means, qualitatively, is that matrix
elements of $H_\sigma$ involving energy jumps much larger
than $\sigma$ (other than jumps between two large but
nearby energies) will all be zero, while matrix elements
involving smaller jumps or two nearby energies remain in
$H_\sigma$.

We require that for $\sigma\to\infty$,
$H_\sigma \to H_B$ and
$S_\sigma \to 1$.  The effective Hamiltonian we seek involves
$H_\sigma$ with $\sigma$ of order the quark or gluon mass.
The similarity operator $S_\sigma$ may be expressed in terms
of the anti-Hermitian
generator of infinitesimal changes of scale $T_\sigma$ as
\begin{equation}
S_\sigma = {\cal T}
\exp(\int_\sigma^\infty d\sigma^\prime\, T_{\sigma^\prime}),
\end{equation}
where ${\cal T}$ puts operators in order of increasing $\sigma$.
Clearly, this satisfies the above limit condition.  Using
\begin{equation}
T_\sigma = S_\sigma {d S_\sigma^\dagger\over d\sigma} =
-{d S_\sigma\over d\sigma} S_\sigma^\dagger,
\end{equation}
the differential form of (\ref{eq:sim1}) is
\begin{equation}
\label{eq:sim2}
{d H_\sigma\over d\sigma} = [ H_\sigma, T_\sigma ],
\end{equation}
which is subject to the boundary condition
$\lim_{\sigma\to\infty}H_\sigma = H_B$ from above.
We shall never need to explicitly construct the similarity operator
$S_\sigma$ in what follows.

We need to specify the action of $T_\sigma$.  We write
$H_B = H_0 + H^B_I$,
$H_\sigma = H_0 + H_{I\sigma}$, let $E_i$ and $E_j$ be
eigenvalues of $H_0$, and denote any given matrix element
of $H_\sigma$ as $H_{\sigma ij}$.  The differential
equations of the similarity renormalization scheme \cite{GW93b} that
determine both $H_\sigma$ and $T_\sigma$ are,
\begin{eqnarray}
\label{eq:sim3}
{d H_{\sigma ij}\over d\sigma} &=& f_{\sigma ij} [H_{I\sigma},T_\sigma]_{ij}
+ {d\, \over d\sigma}(\ln f_{\sigma ij}) H_{\sigma ij},\\
\nonumber
T_{\sigma ij} &=& {1\over E_j-E_i}\left\{ (1-f_{\sigma ij})
[H_{I\sigma},T_\sigma]_{ij}
- {d\, \over d\sigma}(\ln f_{\sigma ij}) H_{\sigma ij}\right\},
\end{eqnarray}
where $f_{\sigma ij} = f(x_{\sigma ij})$
with
\begin{equation}
x_{\sigma ij} = {|E_i - E_j|\over E_i + E_j + \sigma}.
\end{equation}
These equations are consistent with (\ref{eq:sim2}).
The function $f(x)$ should be chosen
as follows:
\begin{eqnarray}
\nonumber
0 \le x \le {1\over 3},&&\quad  f(x) = 1\qquad \qquad \qquad
{\it (near\ diagonal\ region)};\\
{1\over 3} \le x \le {2\over 3},&&\quad  f(x) {\rm \ drops\ from\ }
1\ {\rm to}\ 0\ \quad {\it (transition\ region)};\\
\nonumber
{2\over 3} \le x \le 1,&&\quad  f(x) = 0\qquad \qquad \qquad
{\it (far\ off\ diagonal\ region)}.
\end{eqnarray}
$f(x)$ is to be infinitely differentiable throughout
$0\le x\le 1$, including the transition points $1/3$ and $2/3$.

It will be shown that with these definitions $H_{\sigma ij}$
is zero in the far off-diagonal region
and $T_{\sigma ij}$ is zero in the near-diagonal region.
The second claim is immediately obvious from the form
of $T_{\sigma ij}$ in (\ref{eq:sim3}), for $1$$-$$f$ and
$d_\sigma$($\ln f$) vanish when $0 \le x \le 1/3$.
To see that $H_{\sigma ij}$ vanishes when $x\ge 2/3$,
we first note that since $H_0$ is diagonal,
$d_\sigma(\ln f)H_{0ij}$ vanishes identically
($f$ is $1$ and $d_\sigma$$f$ is 0 for $x = 0$);
and from this result we can
indeed declare that the unperturbed part $H_0$ of
$H_\sigma$ does not vary with $\sigma$, with $T_\sigma$ being
of order $H_I$ or higher.  Now
we can re-write the equations (\ref{eq:sim3}) as
\begin{eqnarray}
\label{eq:sim4}
&&\qquad\qquad\quad{d \,\over d\sigma} \left\{
{1\over f_{\sigma ij}}H_{I\sigma ij} \right\}
= [H_{I\sigma},T_\sigma]_{ij},\\
\nonumber
&& T_{\sigma ij} = {1\over E_j-E_i}\left\{ (1-f_{\sigma ij})
[H_{I\sigma},T_\sigma]_{ij}
- ({df_{\sigma ij} \over d\sigma})
{1\over f_{\sigma ij}} H_{I\sigma ij}\right\}.
\end{eqnarray}
We will see below that $H_I/f$ remains finite even when $f=0$.
Solving the first equation with the stated boundary condition
at $\sigma\to\infty$ gives
\begin{equation}
\label{eq:sim5}
H_{I\sigma ij}
= f_{\sigma ij}\left\{H^B_{Iij} - \int_\sigma^\infty d\sigma^\prime
[H_{I\sigma^\prime},T_{\sigma^\prime}]_{ij}\right\}.
\end{equation}
Since $f(x)$ vanishes when $|x|\ge 2/3$, we find that
$H_{\sigma ij}$ does indeed vanish in the far off-diagonal
region.

That $T_{\sigma ij}$ is zero in the near-diagonal region means
that a perturbative solution to $H_{\sigma ij}$ in terms of
$H^B_{Iij}$ will never involve vanishing energy denominators,
which are a potential source of large errors in other perturbative
renormalization schemes.  That $H_{\sigma ij}$ is zero in
the far off-diagonal region will help identify divergent terms
and determine the form of counterterms necessary to remove
these divergences.  In LFQCD, we are considering
cutoffs which severely break the symmetries of the canonical
theory.  The renormalization process is therefore sure
to be quite complicated, and the advantage of the similarity
renormalization procedure is that it allows a
clean identification and separation of divergences.

Finally, we have to discuss renormalization conditions. The bare LFQCD
Hamiltonian has a limited structure --- canonical terms, an artificial
potential, and renormalization terms.  The renormalization
counterterms are to be determined by fixing outputs: both physical
masses and renormalized coupling constants defined in terms of
measurable parameters, and a subset of the constraints of Lorentz
covariance.  These outputs are to be obtained perturbatively by
solving the Hamiltonian bound state problem.  But for now we mainly
want the functional form of the counterterms, which has to precede
determination of their strengths.  The main requirement for
counterterms in $H_B$ is that they remove divergences in the integral
over the commutator $[H_{I\sigma^\prime},T_{\sigma^\prime}]$.

We can summarize the equations for $H_{I\sigma}$ and $T_\sigma$
as
\begin{equation}
H_{I\sigma} = H^B_{I\sigma} + {\underbrace{[H_{I\sigma^\prime},
T_{\sigma^\prime}]}}_R,
\end{equation}
where $H^B_{I\sigma ij} = f_{\sigma ij} H^B_{Iij}$ and the
linear operation $R$ is
\begin{equation}
{\underbrace{X_{\sigma^\prime ij}}}_R
= - f_{\sigma ij}\int_\sigma^\infty d\sigma^\prime
X_{\sigma^\prime ij}.
\end{equation}
Using the equation for $H_{I\sigma}$ allows us to write
\begin{equation}
T_\sigma = H^B_{I\sigma T} + {\underbrace{[H_{I\sigma^\prime},
T_{\sigma^\prime}]}}_T,
\end{equation}
where
\begin{equation}
H^B_{I\sigma Tij} = -{1\over E_j-E_i}
\left({d\,\over d\sigma}f_{\sigma ij}\right) H^B_{Iij}
\end{equation}
and
\begin{equation}
{\underbrace{\ X_{\sigma^\prime ij}\ }}_T
= -{1\over E_j - E_i}\left({d\,\over d\sigma}f_{\sigma ij}\right)
\int_\sigma^\infty d\sigma^\prime X_{\sigma^\prime ij}
+ {1\over E_j - E_i}\left(1-f_{\sigma ij}\right)
X_{\sigma ij}.
\end{equation}
This equation structure saves us from writing the solution
for $T_\sigma$ separately because it is obtained by substitution
from the equation for $H_{I\sigma}$; namely,
$H^B_{I\sigma}\to H^B_{I\sigma T}$, and all higher
order terms have the substitution ${\underbrace{...}}_{R\to T}$.
So we can start writing the iterated solution for
$H_{I\sigma}$:
\begin{eqnarray}
\label{eq:sim6}
H_{I\sigma} &=& H^B_{I\sigma} + {\underbrace{[H^B_{I\sigma^\prime},
H^B_{I\sigma^\prime T}]}}_R \\
\nonumber &&
+ {\underbrace{[{\underbrace{[H^B_{I\sigma^{\prime\prime}},
H^B_{I\sigma^{\prime\prime} T}]}}_{R^\prime} ,
H^B_{I\sigma^\prime T}]}}_R, \\
\nonumber &&
+ {\underbrace{[H^B_{I\sigma^\prime},
{\underbrace{[H^B_{I\sigma^{\prime\prime}},
H^B_{I\sigma^{\prime\prime} T}]}}_{T^\prime}]}}_R
 + \dots
\end{eqnarray}
The counterterms in $H^B_I$ must be chosen to cancel the
divergences
which occur in integrals over intermediate states
at higher order in $H^B_I$, and such counterterms must
also then be included in higher-order iterations.
If a limit to this process exists, the Hamiltonian is
said to be renormalizable.

\subsection{Discussion of the scheme}

Now we may discuss the similarity renormalization scheme in more
detail.   The ultimate aim of this scheme is to transform the
Hamiltonian $H_B$ into a manageable, band-diagonal form $H_\sigma$.
$H_B$ is the bare cutoff Hamiltonian, forced to be finite by the
imposition  of some cutoff $\Lambda$.  The differential transformation
framework produces a set of Hamiltonians $H_{\sigma^\prime}$, where
$\sigma^\prime$ ranges from ${\cal O}(\Lambda)$ down to some scale
$\sigma$ and thereby dresses the Hamiltonian.  One step in this
process is the determination of the form of the counterterms which
must be included in $H_B$ so that each of the matrix elements of the
transformed Hamiltonian $H_{\sigma ij}$ has no large dependence on
$\Lambda$.  This is the renormalization process: as we send
$\Lambda\to\infty$, we get a renormalized, scale-dependent effective
Hamiltonian $H^R_\sigma$.  This does not finish the renormalization of
the Hamiltonian, however, for the finite parts of the counterterms in
$H_B$ will produce in $H_\sigma^R$ unknown constants and functions of
momenta which must be adjusted to reproduce physical observables and
to restore the symmetries which were broken by the cutoff $\Lambda$.
These quantities are to be fixed, then, by solving $H_\sigma^R$; and
one should be able to do this with a combination of few-body
Hamiltonian methods and weak-coupling diagrams.

We show this renormalization scheme pictorially
in Fig. 6.  The grey
area is the region where $\sigma$ is large enough so that
$H_\sigma$ is equivalent to the bare Hamiltonian $H_B$.
The necessary condition is that $|x_{\sigma ij}| \le 1/3$
for all $i$ and $j$, which is true for
\begin{eqnarray}
\label{eq:bsig}
\sigma \ge {\Lambda\over 2} - {5\mu\over 2},
\end{eqnarray}
where $\Lambda$ is the maximum and $\mu$ is the minimum
energy of the free states of the cutoff Hamiltonian
$H_B$.  By lowering the scale $\sigma$ below this region
for a given $\Lambda$, we start transforming $H_B$ --- eventually
producing a band-diagonal Hamiltonian at some sufficiently
low scale.  Thus we form a new ``Triangle of Renormalization,''
as shown in Fig. 6.  First, we eliminate degrees of
freedom by fixing the cutoff at some value $\Lambda_1$, and
we denote the corresponding bare Hamiltonian as $H^1_B$.
Now we perform the unitary similarity transformation to
bring the Hamiltonian to a band-diagonal form characterized
by some scale $\sigma_0$, which we denote as $H_0^1$ in the figure.
Next, we increase the cutoff to some value $\Lambda_2$ to
get $H_B^2$ and then transform via $\sigma$ to get
$H^2_0$, defined at the same similarity scale $\sigma_0$
as was $H^1_0$.  As we increase $\Lambda_N$ in this way,
we end up with a sequence of band-diagonal Hamiltonians
$H^N_0$.  The renormalized Hamiltonian at scale $\sigma_0$
is the limit of this sequence, $H^R_0 = \lim_{N\to\infty}H^N_0$.

Of course, we could use the similarity transformation to
bring the effective Hamiltonian to any scale $\sigma_n$ and obtain a set
of sequences $\{H^N_n\}$.  As we change $\sigma$, we change
the characteristic scale; but the physics is invariant with
respect to this change for large enough $\Lambda_N$.
The utility of the similarity renormalization scheme is that
the transformation is invertible, as indicated by the
double-arrowed lines in Fig. 6.  Thus, if we find a
Hamiltonian that is finite and $\Lambda$-independent for
any one scale $\sigma_n$, the differential similarity
framework guarantees that we can obtain a Hamiltonian that is
finite and cutoff independent for all $\sigma$.  Note that
we require that each matrix element of $H_\sigma$ be
cutoff independent for external momenta which are small in
comparison to the cutoff,
which is more restrictive than just
requiring this of the eigenvalues of $H_\sigma$.

We now discuss the choice of counterterms.  We start with the bare
cutoff Hamiltonian $H_B$, which consists of the canonical
Hamiltonian with cutoff $\Lambda$ plus a set of counterterms
which have an explicit $\Lambda$-dependence arranged so that
physical observables obtained from $H_B$ are $\Lambda$-independent.
These counterterms are at the outset undetermined.  As we
perform the similarity transformation, we get an effective Hamiltonian
$H_\sigma$, where the scale $\sigma$ replaces $\Lambda$.  This is
done through a precise differential framework.  Qualitatively,
we are separating high-energy degrees of freedom (of order $\Lambda$)
from low-energy degrees of freedom (of order $\sigma$).  In the
language of the phase space cell analysis of Section V,
we have $\Lambda \sim$ transverse (longitudinal) widths
$(\delta y_\perp)^{-1} (\delta y^-)$
respectively and
$\sigma \sim$ transverse (longitudinal) widths
$(\delta x_\perp)^{-1} (\delta x^-) $ respectively.
Thus by varying $\Lambda$, we
determine which divergences have to be eliminated from $H_\sigma$.
The counterterms which eliminate these divergences must be added
to $H_B$, and so can have no dependence on $\sigma$.  The structure
of each counterterm $H^{CT}_{Bij}$
is limited by the structure of the divergence
it must cancel --- namely, by the external legs associated with
states $i$ and $j$.
Along with
the diverging part of each counterterm, we may in principle associate
an unknown finite piece, whose precise
form is to be determined
by fitting data and restoring symmetries.  These terms
must be restricted so that a piece which removes an ultraviolet
divergence, for example, will have a precise behavior with respect
to external transverse momenta, but may
have a finite piece associated with it which
contains an unknown function of the external longitudinal momenta ---
provided that function does not break boost invariance.  A similar
rule applies to counterterms for infrared divergences.  However,
remember that the individual constituents associated with
external momenta do not correspond to physical
states, and so we cannot include counterterms for apparent
infrared divergences
which would be eliminated upon integration with a test function.

Finally, we need to fill in some details in order to match the general
presentation of the similarity scheme given here to the particular
choices of light-front coordinates and the cutoffs on constituent
momenta described in Section VI.  The bare LFQCD Hamiltonian has
a cutoff $\Lambda$
%which is essentially a cutoff on invariant mass and
which has the dimensions of transverse momentum.  The cutoff
$\Lambda$ of Section VI is designed to simultaneously eliminate
constituents with large light-front energies due to either
large transverse or small longitudinal momenta.  We want to define
the light-front similarity scale to have the same dimensions as
$\Lambda$ and to be a true invariant mass scale.  The scale
$\sigma$ is defined through (7.6), which we now re-define for the
light-front as
\begin{eqnarray}
\label{eq:xsiglf}
x_{\sigma ij} = \frac{|P^-_j - P^-_i|}{P^-_i + P^-_j +P^-_\sigma}.
\end{eqnarray}
Here, $P^-_i$ and $P_j^-$ are the (off-mass-shell) light-front
free energies of the states $i$ and $j$ (namely,
sums over the states' constituents' light-front free energies);
and
\begin{eqnarray}
P^-_\sigma \equiv \frac{-2 P_\perp^2 + \sigma^2}{P^+}.
\end{eqnarray}
Since ($P^+,P_\perp$) is conserved in any process --- that is,
$H_{ij} \propto \delta(P^+_i-P^+_j)\delta^2(P_{i\perp}-P_{j\perp})$ --- we
see that $x_{\sigma ij}$ as given in (7.18) is independent of the
total longitudinal momentum $P^+$ and transverse momentum $P_\perp$.
Defined in this way, the
light-front similarity scale $\sigma^\prime$ interpolates from
a mass scale ${\cal O}(\Lambda)$ down to some final scale $\sigma$,
which sets the mass scale of the bound states of the effective
Hamiltonian.   Moreover, from (6.1) and (6.2) we see
that (\ref{eq:xsiglf}) reduces to
\begin{eqnarray}
x_{\sigma ij} = \frac{|M_j^2 - M_i^2|}{M_i^2 + M_j^2 + \sigma^2},
\end{eqnarray}
with $M_i$ and $M_j$ the total masses of the states $i$ and $j$.
So the similarity transformation replaces the cutoff $\Lambda$, which
explicitly breaks boost invariance, with a boost invariant
mass scale $\sigma$.

%%%%%%%%%%%%%%%%%%%%%%%%%%%%%%%%%%%%%%%%%%%%%%%%%%%%%%%%%%%%%%%%%%%%%%
%%%%%%%%%%%% file ``gluon.tex''  input to QCD paper %%%%%%%%%%%%%%%%%%
%%%%%%%%%%%%%%%%%%%%%%%%%%%%%%%%%%%%%%%%%%%%%%%%%%%%%%%%%%%%%%%%%%%%%%
%%%%%%%%%%%%       Last modified: 1-24-94           %%%%%%%%%%%%%%%%%%
%%%%%%%%%%%%%%%%%%%%%%%%%%%%%%%%%%%%%%%%%%%%%%%%%%%%%%%%%%%%%%%%%%%%%%

%%%%%%%%%%%%%%%%%%%%%%%%%%%%%%%%%%%%%%%%%%%%%%%%%%%%%%%%%%%%%%%%%%%%%%
\section{Example calculation: Second-order gluon mass correction}
%%%%%%%%%%%%%%%%%%%%%%%%%%%%%%%%%%%%%%%%%%%%%%%%%%%%%%%%%%%%%%%%%%%%%%

In this section we provide an example of the application of the new
regulator scheme and the similarity transformation perturbation theory
to second order in the coupling constant. We calculate the correction
to the gluon mass coming from intermediate two-gluon states, which is
depicted in Fig. 7. This correction contains severe divergences.

First, we need to find the explicit form for $H^{(2)}$ in (\ref{eq:sim6}).
After substitution from the previous equations and use of the Hermiticity
of $H^B_I$, we find
\begin{equation}
\label{eq:sim7}
	H_{I\sigma ij}^{(2)} = {\underbrace{[H^B_{I\sigma^\prime},
		H^B_{I\sigma^\prime T}]_{ij}}}_R = \sum_k H^B_{Iik}
		H^B_{Ikj}\left\{ \frac{g_{\sigma jik}}{P^-_i-P^-_k}
		+ \frac{g_{\sigma ijk}}{P^-_j-P^-_k} \right\},
\end{equation}
where the sum over intermediate states $k$ includes an
integration over momenta.  The function $g$ is
\begin{equation}
	g_{\sigma ijk} = f_{\sigma ij} \int_\sigma^\infty d\sigma^\prime
		f_{\sigma' ik} \frac{d\;}{d\sigma^\prime} f_{\sigma' jk},
\end{equation}
and it serves to keep $i$ and $j$ in the near-diagonal and transition
regions ($x_{\sigma ij}< 2/3$) and $j$ and $k$ in
the transition and far off-diagonal regions ($x_{\sigma jk}> 1/3$).
The identity $f_{\sigma ij}=f_{\sigma ji}$ and integration
by parts give
\begin{equation}
	g_{\sigma ijk} + g_{\sigma jik}
		= f_{\sigma ij}(1 - f_{\sigma ik} f_{\sigma jk}),
\end{equation}
so that we can write
\begin{eqnarray}
\label{eq:sim8}
	\frac{g_{\sigma jik}}{P^-_i-P^-_k} +
		\frac{g_{\sigma ijk}}{P^-_j-P^-_k} &=& {1\over 2}f_{\sigma ij}
		(1-f_{\sigma ik}f_{\sigma jk})\left\{{1\over P^-_i-P^-_k}
		+ {1\over P^-_j-P^-_k}\right\} \\
	\nonumber && \qquad
		- \frac{P^-_i-P^-_j}{2(P^-_i-P^-_k)(P^-_j-P^-_k)}
		(g_{\sigma ijk}-g_{\sigma jik}).
\end{eqnarray}
The first term is the same as that obtained in the Bloch
renormalization scheme \cite{P93}, but now modified so that the energy
denominators never get small.
The second term in (\ref{eq:sim8}) vanishes as $P^-_i \to P^-_j$.
Divergences occur when $P^-_k$ is much greater than $P^-_i$ and
$P^-_j$, in which case the first term in (\ref{eq:sim8}) will
give the dominant contribution to (\ref{eq:sim7}).

For the gluon mass correction, the initial and final states
have the same momenta, so that
\begin{equation}
	H_{I\sigma ij}^{(2)} = \sum_k H_{Iik}^BH_{Ikj}^B
		\frac{1 - f_{\sigma i k}^2}{P^-_i - P^-_k} .
\end{equation}
We denote the single gluon state as $| i \rangle = | P,a,\lambda\rangle$,
where $P=(P^+,P_{\bot})$, $a$, and $\lambda$ are its momentum, color
index, and helicity index. Using
the diagrammatic rules for the canonical vertices{\cite{Zh 93b}}, we have:
\begin{eqnarray}
	H_{I \sigma ij}^{(2)} && = \langle P,b,\lambda'
		| H_{I \sigma}^{(2)} | P,a,\lambda
		\rangle \nonumber \\
	&& = - {1 \over 2}{1 \over P^+} g^2
		(f^{acd}f^{bdc}) \int_c {dk_1^+ d^2 k_{1\bot} \over
		16\pi^3}{dk_2^+ d^2 k_{2\bot} \over 16\pi^3}
		{ \theta (k_1^+) \over k_1^+}{\theta (k_2^+)
		\over k_2^+} \nonumber \\
&& \qquad \qquad \qquad { 16 \pi^3 \delta^3(P-k_1-k_2) \over
		P_i^- - P_k^-} \big [ 1 - f^2 (x_{\sigma ik}) \big]
		\nonumber \\
	&& \; \;  \epsilon_{\lambda'}^{*i_4} \Bigg\{ \delta_{i_3 i_2}
		\Big[ (k_2^{i_4} - k_1^{i_4}) - { P^{i_4} \over P^+} (k_2^+
		- k_1^+) \Big] + \delta_{i_2 i_4} \Big[ -(k_2^{i_3}
		+ P^{i_3}) + { k_1^{i_3} \over k_1^+} (k_2^+ + P^+) \Big]
		\nonumber \\
	&& ~~~~~~~~~~~~~~~~~~~~~~~~~~~~~~~~~~ +
		\delta_{i_4 i_3} \Big[ (P^{i_2} + k_1^{i_2}) - { k_2^{i_2}
		\over k_2^+} (P^+ + k_1^+) \Big] \Bigg\} \nonumber \\
	&& \; \; \times \Bigg\{ \delta_{i_2 i_3} \Big[ (k_2^{i_1}
		- k_1^{i_1}) - { P^{i_1} \over P^+} (k_2^+ - k_1^+) \Big]
		+ \delta_{i_1 i_3} \Big[ (k_1^{i_2} + P^{i_2}) - { k_2^{i_2}
		\over k_2^+} (k_1^+ + P^+) \Big] \nonumber \\
	&&~~~~~~~~~~~~~~~~~~~~~~~~~~~~~~~~~~~ + \delta_{i_1 i_2} \Big[ -
		(P^{i_3} + k_2^{i_3}) + { k_1^{i_3} \over k_1^+} (P^+ + k_2^+)
		\Big] \Bigg\} \epsilon_\lambda^{i_1} \nonumber \\
	&& =  {g^2 C_A \over 2P^+ 16 \pi^3} \delta^{ab}
 \int_c dk_1^+ d^2 k_{1\bot}
		dk_2^+ d^2 k_{2\bot} \delta^3(P-k_1-k_2)
{\cal M}_{2 \lambda \lambda'}
		(P,k_1,k_2) \big( 1 - f^2_{\sigma ik} \big)
\end{eqnarray}
where the factor ${1 \over 2} $ is the symmetry factor,
$P_i^- = {m_G^2 + (P_\perp)^2 \over P^+}$,
$ P_k^- = k_1^- + k_2^- = { m_G^2 + k_{1\perp}^2 \over k_1^+} +
{m_G^2 + k_{2\perp}^2 \over k_2^+} $,
$f_{\sigma ik} = f(x_{\sigma ik})$ with $f$ given by (7.7)
and $x_{\sigma ik}$ given by (7.17-19),
$C_A \delta^{ab} = f^{acd} f^{bcd} = N \delta^{ab}$,
and ${\cal M}_{2\lambda \lambda'}$ denotes the two vertices with
the energy denominator.

To find the mass correction, we set $P_\perp=0$ and define
\begin{eqnarray}
	H_{I\sigma ii}^{(2)}|_{P_\perp=0} \equiv {\delta m_G^2\over P^+}
	    \delta_{\lambda \lambda^\prime}\delta^{ab},
\end{eqnarray}
where the mass
shift is $\delta m_G^2 = \delta m_{G\Lambda}^2 + \delta m_{G\sigma}^2$.

Integration over $k_2$ and
the cutoff scheme of Section VI lead to
\begin{eqnarray}
	\delta m_G^2 \delta_{\lambda \lambda'}
 = - {g^2 N\over 16\pi^3}
	\int dk_1^+ d^2 k_{1\bot} \theta_\epsilon^\Lambda(k_1)
		\theta_\epsilon^\Lambda(P - k_1)
		{\cal M}_{2\lambda \lambda'}(P,k_1,P-k_1)
		\big( 1 - f^2_{\sigma ik} \big),
\end{eqnarray}
where
\begin{eqnarray} \nonumber
{\cal M}_{2\lambda \lambda'}(P,k_1,P-k_1) &
=& \frac{1}{k_1^+}\frac{1}{P^+-k_1^+}
\Big[ {k_{1\bot}^2 + m_G^2\over k_1^+}
+ {k_{1\bot}^2 + m_G^2\over P^+ - k_1^+}
- { m_G^2\over P^+}\Big]^{-1}\\ && \quad
\times 4\Big( k_{1 \perp}^2 \Big[  \big(\frac{P^+}{k_1^+}\big)^2
+ \big(\frac{P^+}{P^+ - k_1^+}\big)^2 \Big] \delta_{\lambda \lambda'}+
2 k_{1 \perp}. \epsilon^*_{\lambda'} k_{1 \perp} .
\epsilon_{ \lambda} \Big)
\end{eqnarray}
and the cutoff function $\theta_\epsilon^\Lambda(k)$
is defined via (6.23) as
\begin{eqnarray}
	\theta_\epsilon^\Lambda(k) &&\equiv \theta (k^+ - \epsilon)
		\Big[\theta \Big({P_0^+ \over 2} - k^+ \Big) \theta \Big(
		2 \Lambda^2 {k^+ \over P_0^+} \Big( 1 -  {k^+ \over P_0^+}
		\Big) - m^2 - k_\bot^2 \Big) \nonumber \\
	&& ~~~~~~~~~~~~~~~~~~~~~~~~ + \theta\Big(k^+ - {P_0^+ \over 2}\Big)
		\theta \Big( {\Lambda^2 \over 2} - m^2 - k_\bot^2 \Big)
		\Big],
\end{eqnarray}
with $\epsilon \simeq { m^2 P_0^+ \over 2 \Lambda^2}$, $\Lambda$ the
bare cutoff, and $m$ the smaller of $m_G$ and $m_F$.
We use sharp cutoffs here only for the purpose of illustration.
We stress that smooth cutoffs are preferred so as to avoid
non-analyticities.

We write
\begin{eqnarray}
	\delta m_G^2 \delta_{\lambda \lambda'}
= - { g^2 N \over 4\pi^3 }
	       I_{\lambda \lambda^\prime},
\end{eqnarray}
with
\begin{eqnarray}	I_{\lambda \lambda'} &=&
	\int dx d^2 k_{1\bot}
	\theta_\epsilon^\Lambda\big(xP^+,k_{1\bot} \big)
	\theta_\epsilon^\Lambda\big((1-x)P^+,-k_{1 \bot}\big)
\nonumber \\
 	&&\qquad \Big( 2 k_{1 \bot}.\epsilon^*_{ \lambda'}
k_{1 \bot}. \epsilon_{ \lambda}
  + \big( \frac{1}{x^2} + \frac{1}{(1-x)^2}\big) k_{1 \perp}^2
\delta_{\lambda \lambda'}
\Big) \nonumber \\
&& \qquad 		{1 \over k^2_{1 \bot} + m_G^2[1-x+x^2]}
		\Big[ 1 - f^2 (x_{\sigma ik}) \Big],
\end{eqnarray}
with
\begin{eqnarray}
   x = {k_1^+ \over P^+} \; \; {\rm and } \; \;
	x_{\sigma ik} = \frac{k_{1 \bot}^2 + m_G^2(1-x+x^2)}
	   {k_{1 \bot}^2 + m_G^2(1+x-x^2) + \sigma^2}.
\end{eqnarray}
So we see that the external (total) momentum $P^+$ enters only through the
cutoffs.
The term independent of $f_{\sigma ik}$ in (8.12) gives rise to divergent
contributions as the cutoff $\Lambda$ becomes large, as well as
finite contributions, and the second term involving $f_{\sigma ik}^2$
is a finite, $\sigma$-dependent contribution ($f_{\sigma ik}$ cuts off
the integration before it reaches the bare cutoff $\Lambda$).

Evaluating $I$ for the case
$P^+ > P^+_0$, for example, the divergent part of the mass shift is
\begin{eqnarray}
	\delta m_{G\Lambda}^2 = -  N {g^2 \over 4 \pi^2 } \big (
	F_1 + F_2 + F_3 + F_4   \big). \end{eqnarray}
$F_1$ is the quadratic ultraviolet divergent part,
\begin{eqnarray}
	F_1 =   \Lambda^2 \Big[ {1 \over 2} - { P_0^+ \over 6P^+}
		- {4P^+ \over P_0^+} - {4P^+ \over P_0^+}
		\Big({ 2 P^+ \over P_0^+} -1\Big ) \ln { 2 P^+ -
		P_0^+ \over 2 P^+} \Big] . \end{eqnarray}
$F_2$ is the logarithmic ultraviolet divergent part,
\begin{eqnarray}
	F_2 = - m_G^2 \ln \Lambda^2  \Bigg [ { 5 \over 6}
		- 2 \ln{P^+ \over P_0^+} \Bigg ] . \end{eqnarray}
$F_3$ is the mixed ultraviolet and infrared divergent part,
\begin{eqnarray}
	F_3 = 4 \Lambda^2 {P^+ \over P_0^+} \ln {P_0^+ \over 2 \epsilon}
		-2 m_G^2  \Big[ {P^+ \over \epsilon}  -  \ln {P_0^+ \over
		\epsilon} \Big] \ln \Lambda^2 . \end{eqnarray}
$F_4$ is the pure infrared divergent part,
\begin{eqnarray}
	F_4 && = 2 m_G^2 {P^+  \over \epsilon} \ln {P_0^+ \over \epsilon}
  		- 2\Big( m^2 + m_G^2 + m_G^2 \ln {2 \over m_G^2} \Big)
		{P^+ \over \epsilon} \nonumber \\
	&&~~~~~~ - m_G^2 \ln^2 {P_0^+\over \epsilon} - 2 m_G^2
		\Big(1 - {P^+\over P_0^+} - \ln {2 \over
		m_G^2 } \Big) \ln {P_0^+ \over \epsilon}.
\end{eqnarray}
Note that the mass shift is
negative, as it must be in second-order perturbation theory.

The coefficients of the ultraviolet quadratic and logarithmic
divergences are multiplied by functions of longitudinal momenta which
are in accord with the power counting rules but which are not boost
invariant. We have to subtract them entirely, and the finite parts of
these counterterms are just arbitrary constants since an arbitrary
function of longitudinal momenta violates longitudinal boost
invariance. The divergences arising from small longitudinal momenta
also violate longitudinal boost invariance because of the cutoff
scheme, and so these divergences are also to be subtracted away,
leaving room for only an arbitrary constant.

The evaluation of the second term in (8.12) depends explicitly on the
choice of the function $f$. For qualitative purposes, we may choose
a step function, namely,
\begin{equation} f(x) = \theta({1 \over 2} -x )
\end{equation}
which results in the effective cutoffs for the Jacobi momenta
\begin{equation}
  \kappa_{\perp max}^2 = (\sigma^2+ 3 m_G^2)x(1-x)
	- m_G^2 ~ ,~~~ 	\frac{m_G^2}{(\sigma^2+ 3 m_G^2)} \leq x
	\leq 1 - \frac{m_G^2}{\sigma^2 + 3 m_G^2}.
\end{equation}
These yield a result bounded by $\sigma$ and finite as $\Lambda
\rightarrow \infty$.

Thus, with the cutoff theory, even if we set the bare gluon mass to
zero, the gluon mass correction $\delta m_{G}^2$ does not vanish.
After subtracting the infinite mass correction, there remains a finite
contribution which depends on the mass scale parameter $\sigma$.  This
dependence is altered by spectators because $x_{\sigma i j}$ changes
when spectator energies change.

%%%%%%%%%%%%%%%%%%%%%%%%%%%%%%%%%%%%%%%%%%%%%%%%%%%
%%%%%%% g2c.tex  (input to QCD paper)         %%%%%
%%%%%%%%%%%%%%%%%%%%%%%%%%%%%%%%%%%%%%%%%%%%%%%%%%%
%%%%%%%  Last modified 1/24/94   %%%%%%%%%%%%%%%%%
%%%%%%%%%%%%%%%%%%%%%%%%%%%%%%%%%%%%%%%%%%%%%%%%%%%

%%%%%%%%%%%%%%%%%%%%%%%%%%%%%%%%%%%%%%%%%%%%%%%%%%%
\section{Infrared counterterms in lowest-order
perturbation theory}
%%%%%%%%%%%%%%%%%%%%%%%%%%%%%%%%%%%%%%%%%%%%%%%%%%%

In Section IV we discussed the infrared counterterms in the canonical
Hamiltonian itself. Products of the interaction Hamiltonian lead to
light-front infrared (small $k^+$) divergences in addition to the
ultraviolet (large $k_\perp$) divergences.  Note that we do not have
the conventional infrared divergence (small $k_\perp$) since gluons
are massive. As we have pointed out there is a crucial difference
between these infrared divergences and ultraviolet divergences.

In constructing counterterms for the infrared divergences in the
products of the interaction Hamiltonian, we find strong cutoff
dependence already in tree level amplitudes --- for example, in the
${\cal O}(g^2)$ quark--antiquark scattering amplitude. Here, what one
has in fact are infrared {\it singularities}. When we diagonalize the
Hamiltonian, these amplitudes are integrated with wave packet
functions, and {\it nonintegrable} singularities in the amplitudes
give rise to energy divergences. Thus to find the infrared counterterm
it is not enough just to isolate the infrared singular amplitudes; we
have to integrate the amplitude over the external leg with a wave
packet function. For QED, in contrast, where the external legs
correspond to physical particles, all tree level singularities have to
cancel.  No such cancellations will be assumed here.

In this section we outline the construction of infrared (small $k^+$)
counterterms up to order $g^2$. The infrared divergences arise from
either an exchanged infrared gluon  or an exchanged infrared fermion
in the intermediate state. The counterterms arising from infrared
gluons are of interest when looking for the origin of confinement, and
the counterterms arising from infrared fermions are of interest when
looking for the origin of spontaneous chiral symmetry breaking.

%%%%%%%%%%%%%%%%%%%%%%%%%%%%%%%%%%%%%%%%%%%%%%%%%%%%%%%%
\subsection{Counterterms from infrared gluons}
%%%%%%%%%%%%%%%%%%%%%%%%%%%%%%%%%%%%%%%%%%%%%%%%%%%%%%%%
\subsubsection{Counterterms in the
		quark--antiquark sector}
%%%%%%%%%%%%%%%%%%%%%%%%%%%%%%%%%%%%%%%%%%%%%%%%%%%%%%%%

The interaction Hamiltonian $H_{qqg}$ gives rise to an effective
Hamiltonian in second order whose matrix element for the
$q\bar q$ states
$|i \rangle = |p_1, s_1, \alpha_1; p_2, -s_2, \alpha_2 \rangle$
and $|j \rangle = |p_3, s_3, \alpha_3; p_4, -s_4, \alpha_4 \rangle$
(see Fig. 8) is given by
\begin{eqnarray}
	H_{I\sigma ij}^{(2)} && = -g^2 \; T_{\alpha_3 \alpha_1}^a
		T_{\alpha_4 \alpha_2}^a {\cal M}_{2ij} \; \Big[ {\cal F}_2 \;
		f_{\sigma ij} \; (1 - f_{\sigma ik} f_{\sigma j k}) +
		{\cal O} \big((P^-_k)^{-3}\big)\Big],
\end{eqnarray}
where $p_n,s_n,\alpha_n$ are the momentum, helicity and color
index of the $n$th quark,
\begin{eqnarray} {\cal M}_{2ij} && = \chi^\dagger_{s_3} \Big [ 2 {p_1^{i_1} -
	p_3^{i_1} \over p_1^+ - p_3^+} - i m_F \Big(
	{1 \over p_1^+} - {1 \over p_3^+} \Big) \sigma^{i_1}
	- \Big( \sigma^{i_1} {\sigma_\perp \cdot p_{1 \perp}
	\over p_1^+} + {\sigma_\perp \cdot p_{3 \perp} \over p_3^+}
	\sigma^{i_1} \Big) \Big] \chi_{s_1} \nonumber \\
  && ~~~~~~~~ \times\chi^\dagger_{-s_2} \Big[ 2 {p_4^{i_1} - p_2^{i_1}
	\over p_4^+ - p_2^+} - i m_F \Big({1 \over p_4^+} -
	{1 \over p_2^+} \Big) \sigma^{i_1} - \Big(\sigma^{i_1}
	{\sigma_\perp \cdot p_{4 \perp} \over p_4^+} + {\sigma_\perp
	\cdot p_{2 \perp} \over p_2^+} \sigma^{i_1} \Big)\Big]
	\chi_{-s_4}, \end{eqnarray}
\begin{eqnarray}
	{\cal F}_2 && = {\theta_\epsilon^\Lambda(p_1 - p_3) \over p_1^+
		- p_3^+} {1 \over 2} \Big[{ 1 \over p_1^- - p_3^- - q^-}
		+ { 1 \over p_4^- - p_2^- - q^- } \Big] \nonumber \\
	&& ~~~~~~~  + {\theta_\epsilon^\Lambda(p_3 - p_1) \over p_3^+
		- p_1^+} { 1 \over 2} \Big[{ 1 \over p_2^- - p_4^- +
		q^- } + { 1 \over p_3^- - p_1^- + q^- } \Big] ,
\end{eqnarray}
and
\begin{eqnarray}
	f_{\sigma ij} = f (x_{\sigma i j}) ~~~\; {\rm with} ~~~ \;
		x_{\sigma i j} = {\mid P^+ P^-_j - P^+ P^-_i \mid
		\over P^+ P^-_j + P^+ P^-_i -2 P_\perp^2 + \sigma^2}.
\end{eqnarray}
The function $\theta_\epsilon^\Lambda(k)$, which arises from the
cutoffs on the intermediate gluon, is defined in Section VII:
\begin{eqnarray}
	\theta_\epsilon^\Lambda(k) &&\equiv \theta (k^+ - \epsilon)
		\Big[\theta \Big({P_0^+ \over 2} - k^+ \Big) \theta \Big(
		2 \Lambda^2 {k^+ \over P_0^+} \Big( 1 -  {k^+ \over P_0^+}
		\Big) - m^2 - k_\bot^2 \Big) \nonumber \\
	&& ~~~~~~~~~~~~~~~~~~~~~~~~ + \theta\Big(k^+ - {P_0^+ \over 2}\Big)
		\theta \Big( {\Lambda^2 \over 2} - m^2 - k_\bot^2 \Big)
		\Big],
\end{eqnarray}
with $\epsilon = { m^2 P_0^+ \over 2 \Lambda^2}$, $\Lambda$ the
bare cutoff, and $m$ the smaller of $m_G$ and $m_F$.
Here we have also defined
\begin{eqnarray}
	&& p^- = {m_F^2 + p_{ \perp}^2 \over p^+} ~,
	 ~~~~ q^- = {m_G^2 + (p_{1 \perp} -p_{3 \perp})^2 \over p_1^+
		- p_3^+}, \end{eqnarray}
$P_k^-$ denotes the intermediate state energy:
\begin{eqnarray}
	&& P_k^- = p_3^- + q^- + p_2^- ~~~ {\rm for}~~ p_1^+ > p_3^+ ~~~
	{\rm and}~~~ P_k^- = p_1^- - q^- + p_4^- ~~~ {\rm for}~~ p_1^+ < p_3^+.
\end{eqnarray}
The expression for ${\cal F}_2$ can be further simplified,
\begin{eqnarray}
	{\cal F}_2 && = {\theta_\epsilon^\Lambda (p_1 - p_3) +
		\theta_\epsilon^\Lambda (p_3 - p_1) \over p_1^+
		- p_3^+} {1 \over 2} \Big[{ 1 \over p_1^- - p_3^-
		- q^- } + { 1 \over p_4^- - p_2^- - q^- } \Big] .
		\label{g2qq1}
\end{eqnarray}

To identify the infrared counterterms, we look for singular terms.
We have, explicitly,
\begin{eqnarray} P^-_i = p_1^- + p_2^-, \; P^-_j = p_3^- + p_4^- .
\end{eqnarray}
The intermediate state energy is dominated by that of the infrared gluon,
namely,
\begin{eqnarray} 	P^-_k \approx  q^- \; {\rm for} \; p_1^+ > p_3^+ ~~~, ~~~
	P^-_k \approx  -q^- \; {\rm for} \; p_3^+ > p_1^+.  \end{eqnarray}
As the gluon longitudinal momentum fraction reaches $\epsilon$,
both $x_{\sigma ik}$
and $x_{\sigma j k}$ become greater than ${2 \over 3}$; and as a
consequence $ 1 - f_{\sigma ik} f_{\sigma jk} \rightarrow 1$.
The leading singular term is
\begin{eqnarray}
&&	4 g^2 T_{\alpha_3 \alpha_1}^a T_{\alpha_4 \alpha_2}^a
		\delta_{s_1 s_3} \delta_{s_2 s_4} \Big( 1 - {m_G^2
		\over (p_{1 \perp}- p_{3 \perp})^2 + m_G^2} \Big)
		{f_{\sigma ij} \over (p_1^+ - p_3^+)^2} \nonumber \\
&& \qquad \qquad 	\times	\Big[ \theta^\Lambda_\epsilon(p_1-p_3)
		+ \theta^\Lambda_\epsilon(p_3-p_1)\Big].
\end{eqnarray}
To find the infrared
divergence we integrate over $p_3$ with a test function
$\phi(p^+_3,p_{3\bot})$, which we may assume is nonzero
only in the range where $f_{\sigma ij} = 1$.  This evaluation is
complicated by the cutoff function.
Using the definition
of $\theta^\Lambda_\epsilon$ in (9.5), we find
a divergent piece
\begin{eqnarray}
I \equiv \int_\epsilon^{P^+_0/2} {dk^+\over (k^+)^2} \int d^2k_\perp
\phi(p+k) \theta\Big({m^2\over\epsilon}k^+
\big( 1-{k^+\over P^+_0} \big) - m^2 - k_\bot^2\Big)
{k_\bot^2\over k_\bot^2 + m_G^2},
\end{eqnarray}
where $k = p_3 - p_1$; and a similar piece for $p^+_1 - p^+_3 > \epsilon,$
which is the same as (9.12) except that
$\phi(p+k) \to \phi(p-k)$.   To find the
small-$\epsilon$ dependence in I, we write $k^+=\epsilon x$, switch
the order of integration, and expand in $\epsilon$.
The divergence has the form
\begin{eqnarray}
		{2m^2 \over\epsilon} \int d^2k_\bot
		{k_\bot^2\over(k_\bot^2+m^2)(k_\bot^2+m_G^2)}
		\phi(p_1^+,p_{1\bot} + k\bot).
\end{eqnarray}
Hence we need to include a counterterm for this linear
divergence in $H_B$:
\begin{eqnarray}
	4 g^2 T_{\alpha_3 \alpha_1}^a T_{\alpha_4 \alpha_2}^a
		\delta_{s_1 s_3} \delta_{s_2 s_4}
	{2m^2 \over \epsilon}
	{(p_{1\bot}-p_{3\bot})^2
	\over \Big((p_{1\bot}-p_{3\bot})^2+m^2\Big)
	\Big((p_{1\bot}-p_{3\bot})^2+m_G^2\Big)}
	\delta(p_1^+ - p_3^+).
\end{eqnarray}
Without our cutoffs and for zero gluon mass we would find
exactly the same counterterm
but with the opposite sign as
the counterterm for the instantaneous four-fermion interaction
in the canonical Hamiltonian, and hence no ${\cal O}(g^2)$
counterterm would be
necessary for the linear infrared divergence.
But our cutoff scheme, which is dependent on the choice of
massive constituents and thereby eliminates zero modes, does
not allow complete cancellation to occur.

There are also ${1 \over q^+}$ type singularities which lead to
logarithmic infrared divergences.  But these divergences are cancelled
from the two $\theta$-functions in (9.8). Thus there is no counterterm
necessary to remove a logarithmic infrared divergence in $q\bar q$
scattering to second order \cite{G 93}. However, as discussed in
Section IV.B for the canonical Hamiltonian counterterms, the use of a
symmetric cutoff to ensure the cancellation of divergences from small
positive and negative longitudinal momenta excludes any possibility of
finite contributions from exact zero modes. To counter this
elimination of zero modes, we can include a finite term analogous to
(4.17).  Here the transverse structure in ${\cal M}_2$ --- for
example, the product of the first and last terms in (9.2) --- may keep
such terms from vanishing, in contrast to terms corresponding to
instantaneous zero-mode gluon exchange.  The finite counterterms also
involve a product of two fermion color charges as one needs to start
building true confining potentials. True confining potentials confine
only non-zero color charge states as opposed to the artificial
potential which confines all states.

%%%%%%%%%%%%%%%%%%%%%%%%%%%%%%%%%%%%%%%%%%%%%%%%%%%%%%%%%%%
\subsubsection{Counterterms in the quark--gluon sector}
%%%%%%%%%%%%%%%%%%%%%%%%%%%%%%%%%%%%%%%%%%%%%%%%%%%%%%%%%%%

Now we consider the $q g$ states.  The effective Hamiltonian
in second order (see Fig. 9) is given by
\begin{eqnarray}
	H_{I\sigma ij}^{(2)} && = i g^2 f^{abc} \;T_{\alpha_2 \alpha_1}^b
		{\cal M}_{2ij} \; \Big[ {\cal F}_2 \; f_{\sigma ij} \;
		(1 - f_{\sigma ik} f_{\sigma jk}) + {\cal O}\big(
		(P^-_k)^{-3}\big) \Big], \end{eqnarray}
for the states
$|i \rangle = |k_1, \lambda_1, a; p_1, s_1, \alpha_1 \rangle$ and
$|j \rangle = | k_2, \lambda_2, c; p_2, s_2, \alpha_2 \rangle$, where
\begin{eqnarray}
	{\cal M}_{2ij} && =  \epsilon_{\lambda_2}^{* i_2} \Big [
		\delta^{i_1 i_3} \Big( (k_1+q)^{i_2} - {k_2^{i_2}
		\over k_2^+} (q^+ + k_1^+) \Big) + \delta^{i_2 i_3}
		\Big((k_2-q)^{i_1} - {k_1^{i_1} \over k_1^+}
		(k_2^+ - q^+) \Big) \nonumber \\
	&& \qquad \qquad + \delta^{i_1 i_2} \Big(-(k_1+k_2)^{i_3} +
		{q^{i_3} \over q^+} (k_1^+ + k_2^+) \Big) \Big ]
		\epsilon_{\lambda_1}^{i_1} \nonumber \\
	&& \qquad \times \chi^{\dagger}_{s_2} \Big[{2 q^{i_3} \over q^+}
		- i m_F \Big({1 \over p_1^+}- { 1 \over p_2^+} \Big)
		\sigma^{i_3} - \Big(\sigma^{i_3} {\sigma_\perp \cdot
		p_{1 \perp} \over p_1^+} + {\sigma_\perp \cdot p_{2
		\perp} \over p_2^+} \sigma^{i_3}\Big) \Big]\chi_{s_1} , \\
	&& 		\nonumber \\
	{\cal F}_2 && = {\theta_\epsilon^\Lambda (p_1 - P_3) +
		\theta_\epsilon^\Lambda (p_3 - p_1) \over k_1^+
		- k_2^+} {1 \over 2} \Big[
		{ 1 \over k_1^- - k_2^- - q^-} + { 1 \over p_2^-
		- p_1^- - q^-} \Big],
\end{eqnarray}
and
\begin{eqnarray}	k^-_n = {m_G^2 + k_{n\perp}^2 \over k^+_n}~~, ~~~~
	q^- = {m_G^2 +  (k_{1 \perp} - k_{2 \perp})^2 \over k_1^+
	- k_2^+ } . \end{eqnarray}

Again, to identify the infrared counterterms, we look for singular
terms. As in the case of the quark--antiquark sector, the intermediate
state energy is dominated by the infrared gluon and the factor $(1 -
f_{\sigma i k} f_{\sigma jk})$ is replaced by 1.
The most singular term (leaving out the factor $f_{\sigma ij}$) is
\begin{eqnarray}
&&	2 i g^2 f^{abc} T_{\alpha_2 \alpha_1}^b \delta^{i_1 i_2}
	\delta_{\lambda_1 \lambda_2} \delta_{ s_1 s_2} {(k_1^+
	+ k_2^+) \over (k_1^+ - k_2^+)^2} \Big( 1 - {m_G^2
	\over m_G^2 + (k_{1 \perp} - k_{2 \perp})^2} \Big) \nonumber \\
&& \qquad \qquad  \times		\Big[ \theta^\Lambda_\epsilon(p_1-p_3)
		+ \theta^\Lambda_\epsilon(p_3-p_1)\Big].
\end{eqnarray}
To find the divergence we must again integrate over $k_2$ with a test
function, which again yields a linearly divergent term as above. Here
also the cutoffs and nonzero gluon mass spoil any complete
cancellation of this divergence with the counterterm for the two
quark--two gluon instantaneous interaction.

After the subtraction of the linear divergence, a linearly singular
term (proportional to $m_G^2$) survives; but as in the case of the
$q\bar q$ scattering amplitude, there is no divergence after
integrating the amplitude with a test function. There are also other
$1/q^+$ terms in (9.18), but the logarithmic divergences they produce
are cancelled from the two $\theta^\Lambda_\epsilon$-functions.
However, as with the $q\bar q$ sector, even though the logarithmic
divergence for small $q^+$ is cancelled by that for small $-q^+$, we
need to include a counterterm to account for the possibility of finite
effects from exchange of gluons with exactly $q^+=0$.

%%%%%%%%%%%%%%%%%%%%%%%%%%%%%%%%%%%%%%%%%%%%%%%%%%%%%%%%%
\subsection{Counterterms from infrared fermions}
%%%%%%%%%%%%%%%%%%%%%%%%%%%%%%%%%%%%%%%%%%%%%%%%%%%%%%%%%

In this subsection, we consider the contributions to the matrix elements
of the effective Hamiltonian in the quark-gluon sector arising from quark
exchange.  First, we discuss the contribution from an intermediate
one-quark--two-gluon state, as is shown in Fig. 10a.
The matrix element is given by
\begin{eqnarray}
	H_{I\sigma ij}^{(2)} && = g^2 \; (T^a T^b)_{\alpha_2 \alpha_1}
		{\cal M}_{2 ij} \; \Big[{\cal F}_2 \; f_{\sigma i
		j} \; (1 - f_{\sigma ik} f_{\sigma j k}) + {\cal O}
		\big((P^-_k)^{-3})\big) \Big],
\end{eqnarray}
where
\begin{eqnarray}
	{\cal M}_{2ij} && = \chi^\dagger_{s_2} \epsilon_{\lambda_2}^{*\bot}
		\cdot \Big [ 2 {k_{1 \perp}  \over k_1^+ } + i m_F
		\Big( {1 \over p_2^+} - {1 \over p_3^+} \Big) \sigma_\perp
		- \Big(\sigma_\perp {\sigma_\perp \cdot p_{3 \perp} \over
		p_3^+} + {\sigma_\perp \cdot p_{2 \perp} \over p_2^+}
		\sigma_\perp \Big) \Big] \nonumber \\
	&& \qquad \times \Big [ 2 {k_{2 \perp}  \over k_2^+} + i m_F
		\Big({1 \over p_3^+} - {1 \over p_1^+} \Big) \sigma_\perp
		- \Big(\sigma_\perp {\sigma_\perp \cdot p_{1 \perp} \over
		p_1^+} + {\sigma_\perp \cdot p_{3 \perp} \over p_3^+}
		\sigma_\perp \Big) \Big ] \cdot \epsilon^{\perp}_{\lambda_1}
		\chi_{s_1}, \end{eqnarray}
\begin{eqnarray}	{\cal F}_2 && =  \theta_\epsilon^\Lambda (p_1^+ - k_2^+) {1
\over 2}
		\Big[ { 1 \over p_1^-  - p_3^- - k_2^- } + { 1 \over p_2^-
		- p_3^- - k_1^-} \Big],  \end{eqnarray}
with
$ p^+_3 = p^+_1 - k_2^+, \; p_{3 \perp } = p_{1 \perp } - k_{2 \perp }$ .

In the limit $p_3^+ \rightarrow 0$, we have
\begin{eqnarray}
	H_{I \sigma}^{(2)} \approx  -g^2 T^a T^b \chi^\dagger
		(\sigma^{\bot} \cdot \epsilon^{* \bot})(\sigma^{\bot}
 		\cdot \epsilon^{\bot}) \chi {\theta_\epsilon^\Lambda
		(p_1^+ - k_2^+) \over p_1^+ - k_2^+} \end{eqnarray}
which leads to a logarithmic divergence when we integrate the effective
Hamiltonian with wave packet functions which are functions of external
momenta.

Next we consider the contributions to the matrix elements of the
effective Hamiltonian in the quark-gluon sector arising from antiquark
exchange,
that is, from an intermediate two-quark--one-antiquark state, as is
shown in Fig. 10b.
The matrix element is given by
\begin{eqnarray}
	H_{I\sigma ij}^{(2)} && = -g^2 \; (T^a T^b)_{\alpha_2 \alpha_1}
		{\cal M}_{2ij} \; \Big[ {\cal F}_2 \; f_{\sigma ij} \;
	(1 - f_{\sigma ik} f_{\sigma j k}) + {\cal O}\big((P^-_k)^{-3}
		\big) \Big],
\end{eqnarray}
where
\begin{eqnarray}
	{\cal M}_{2ij} && = \chi^\dagger_{s_2} \epsilon_{\lambda_2}^{*
		\bot} \cdot \Big [ 2 {k_{1 \perp}  \over k_1^+ } +
		i m_F \Big({1 \over p_2^+} + {1 \over p_3^+} \Big)
		\sigma_\perp - \Big( \sigma_\perp {\sigma_\perp \cdot
		p_{3 \perp} \over p_3^+} + {\sigma_\perp \cdot p_{2
		\perp} \over p_2^+} \sigma_\perp \Big) \Big ]
		\nonumber \\
	&& \qquad \times \Big[ 2 {k_{2 \perp}  \over k_2^+} - i m_F
		\Big( {1 \over p_3^+} + {1 \over p_1^+} \Big)
		\sigma_\perp - \Big(\sigma_\perp {\sigma_\perp \cdot
		p_{1 \perp} \over p_1^+} + {\sigma_\perp \cdot p_{3
		\perp} \over p_3^+} \sigma_\perp \Big) \Big] \cdot
		\epsilon^{\perp}_{\lambda_1} \chi_{s_1}, \\
	{\cal F}_2 && =  \theta_\epsilon^\Lambda (k_2^+ - p_1^+){1 \over
		2} \Big[ {1 \over k_1^- - p_2^- - p_3^-} + {1 \over k_1^-
		- p_1^- - p_3^-} \Big] \end{eqnarray}
with
$ p^+_3 =  k_2^+ - p^+_1, \; p_{3 \perp } =   k_{2 \perp }- p_{1 \perp }$ .

In the limit $p_3^+ \rightarrow 0$, we have
\begin{eqnarray}
	H_{I\sigma ij}^{(2)} \approx  g^2 T^a T^b \chi^\dagger (\sigma^{\bot}
		\cdot \epsilon^{*\bot} )(\sigma^{\bot} \cdot \epsilon^{\bot} )
		\chi {\theta_\epsilon^\Lambda ( k_2^+ - p_1^+) \over
		k_2^+ -  p_1^+} \end{eqnarray}
which leads to a logarithmic divergence when we integrate the effective
Hamiltonian with wave packet functions which are functions of external
momenta.

Note that if we keep both types of intermediate states, the singularity
acquires a principal value prescription since the $\theta$-functions
sum to one, and thus no counterterm is needed. This is because the
singularity occurs when the intermediate state energy is dominated
by the exchanged quark or antiquark, in which case it does not
matter if the other particles in the intermediate state are quarks
or gluons.  However, it is clear that the finite part of these
diagrams will be affected differently, for the function $1 -
f_{\sigma ik}f_{\sigma jk}$ in (9.23) will now cut off the two
diagrams differently since the intermediate state energies are
different when the exchanged quark and antiquark have finite energy.
Thus we expect that at higher order there will be diagrams which contain
quark or antiquark exchange of finite energy and have divergences
associated with another process.  Counterterms for these diagrams
will have a different dependence for the intermediate $qgg$ and
$\bar qqq$ states and so a logarithmic divergence will not acquire
a principal value prescription as here in
second order.  The counterterms necessary to remove such higher-order
divergences will provide a possible source for explicit chiral symmetry
breaking terms in the renormalized Hamiltonian.

Moreover, as with gluon exchange above, despite the cancellation of
near-zero $k^+$ logarithmic divergences, we must include a finite
term which might restore any physics lost by the elimination of exact
zero modes.  As discussed in Section IV.B, there are finite non-canonical
terms which have the structure of these logarithmic divergences and
which satisfy the requirements of power
counting and boost invariance yet break chiral symmetry.  Since such
terms necessarily contain the effects normally associated with the
vacuum, it is no surprise that they provide a source for chiral
symmetry breaking (see the discussion of the sigma model in
Appendix A).

Finally, a similar result applies for the matrix element describing
quark--antiquark annihilation into two gluons, which involves the
quark exchange shown in Fig. 11a and a similar diagram with an
intermediate antiquark.
We should here include
a finite term so as to
counter the removal of zero-mode fermion exchange, and this may include
a piece which explicitly breaks chiral symmetry.
It follows that
the chiral symmetry violating term in the $q\bar q g$ coupling will
have an arbitrary size due to renormalization from such a
$q\bar q gg$ counterterm, as is depicted in Fig. 11b.
This term then
need not vanish in the limit of only spontaneous chiral symmetry
breaking.

%%%%%%%%%%%%%%%%%%%%%%%%%%%%%%%%%%%%%%%%%%%%%%%%%%%%%%%%%
\subsection{Many-body infrared counterterms}
%%%%%%%%%%%%%%%%%%%%%%%%%%%%%%%%%%%%%%%%%%%%%%%%%%%%%%%%%

While we provide no thorough discussion of higher-order counterterms,
we need to clarify one qualitatively new feature that arises beyond
second order.  In Section IX.A we showed that the matrix elements for
gluon exchange include pieces that diverge like $1/(q^+)^2$, where
$q^+$ is the longitudinal momentum exchange. This leads to divergences
even at tree level. These divergences require counterterms, and we
have argued that the finite part of such counterterms may contain
confining interactions unless boost invariance precludes this.

One example should adequately illustrate what happens at higher orders.
In Fig. 12a we show a fourth order tree diagram in which two gluons are
exchanged between three quarks.  To evaluate the matrix elements
that arise in the effective Hamiltonian we must integrate over small
$q^+$.  Keeping only the most singular parts of the matrix elements, we
encounter longitudinal integrals of the form

$$\int_\epsilon {dq_1^+ \over q_1^+} \int_\epsilon
{dq_2^+ \over q_2^+} \biggl(
{1 \over q_1^+}\biggr)^2 \biggl({1 \over q_2^+}\biggr)^2 q_1^+ q_2^+
\biggl({M_1^2 \over q_1^+}+{M_2^2 \over q_2^+}\biggr)^{-1} \;. \eqno(9.28)$$

\noindent The entire product of matrix elements is relatively
complicated, but this integral is easily analyzed.  There is a
$log^2(\epsilon M_1^2/M_2^2)$ divergence, and a $log^2(\epsilon
M_2^2/M_1^2)$ divergence.  There is actually a nested divergence in
this diagram, which would be subtracted if one adds the appropriate
third order diagram; however, even after this subtraction one is left
with $log^2(\epsilon)$ divergences.

Another simple example is shown in Fig. 12b, in which there is a gluon
exchanged between two quarks while one quark interacts through an
instantaneous potential with a spectator.  Regardless of the form of
the instantaneous interaction, this diagram diverges logarithmically
unless a symmetric cutoff on longitudinal momenta is used.  As argued
above, the presence of this potential logarithmic divergence indicates
that the elimination of zero modes may produce an interaction with the
operator structure in Figure 11b.  This three-body interaction may be
long range in the transverse direction, and it may be as strong as the
two-body confining interactions.  As we pointed out earlier, such
long-range many-body interactions are required to cancel unphysical
van der Waals forces if there is a confining two-body interaction
\cite{WAAL}.

%%%%%%%%%%%%%%%%%%%%%%%%%%%%%%%%%%%%%%%%%%%%%%%%%%%%
%%%%%%%  future.tex  (input to QCD paper)       %%%%
%%%%%%%%%%%%%%%%%%%%%%%%%%%%%%%%%%%%%%%%%%%%%%%%%%%%
%%%%%%%  last modified 1/24/94    		%%%%
%%%%%%%%%%%%%%%%%%%%%%%%%%%%%%%%%%%%%%%%%%%%%%%%%%%%

%%%%%%%%%%%%%%%%%%%%%%%%%%%%%%%%%%%%%%%%%%%%%%%%%%%%
\section{LFQCD Bound State Computations}
%%%%%%%%%%%%%%%%%%%%%%%%%%%%%%%%%%%%%%%%%%%%%%%%%%%%

Consider how bound states are handled in QED\cite{QED}. There one
includes in the unperturbed Hamiltonian an instantaneous potential
whose form is obtained from the general two-fermion four-point
function.  The precise choice of the potential to include in the
unperturbed part is a matter of some art, since one must weigh the
competing demands of making the unperturbed calculation as simple as
possible while at the same time including a large part of the relevant
physics and ensuring that perturbative corrections are small.
Invariably, the bound state calculation reduces to the usual
nonrelativistic Schr\"odinger equation in the appropriate limit, and
the corresponding nonrelativistic states are used in the perturbative
expansion.  The precise details of any particular calculation will
vary, and the most efficient methods can only be arrived at after some
experience.

We want to take the same approach to bound states here, but we know
from phenomenological constituent quark models that the confining
potentials we need for the bound state calculation are far different
from the Coulomb-like potentials
that arise at low order in the usual relativistic
perturbation theory. We expect counterterms for the light-front
infrared divergences to be the ultimate source of the needed
potentials.  For artificially small coupling, however, the artificial
potential is the dominant confinement mechanism.  The similarity
renormalization scheme and massive constituents with cutoffs allow us
to treat this confining potential in a weak coupling framework.

In this section, we outline the construction of the effective LFQCD
Hamiltonian for low-energy hadrons and discuss the increasing levels
of complexity in the LFQCD bound state computations in this
formulation.

%%%%%%%%%%%%%%%%%%%%%%%%%%%%%%%%%%%%%%%%%%%%%%%%%%%%%%%%
\subsection{The effective LFQCD Hamiltonian
		and bound states}
%%%%%%%%%%%%%%%%%%%%%%%%%%%%%%%%%%%%%%%%%%%%%%%%%%%%%%%%

In the similarity renormalization scheme, the effective LFQCD
Hamiltonian that will be used to compute the hadronic bound
states is
\begin{eqnarray}
	H_{\sigma i j} = \;
Lim_{\Lambda \rightarrow \infty} \;
f_{\sigma i j} \Big [ H^B_{ij}
		+ \sum_k { 1 \over 2} H^B_{I ik} H^B_{I kj}
		\Big({g_{\sigma ijk} \over P^-_j - P^-_k} +
		{ g_{\sigma jik} \over P^-_i - P^-_k}\Big)
		+ \cdots \Big ] .
\end{eqnarray}
$H_{\sigma}$ causes transitions only between states staying ``close
to the diagonal'' due to the factor $f_{\sigma ij}$, and the effects
of transitions to and from intermediate states which are ``far
off-diagonal'' have explicitly appeared in the effective Hamiltonian
$H_\sigma$ as a perturbative expansion.

Recall that $H^B_{ij}$ has the cutoff $\Lambda$ which violates both
longitudinal and transverse boost invariance. First we identify the
counterterms that must be included in $H^B$ so that the matrix
elements of the effective Hamiltonian $H_{\sigma ij}$ have  no
divergent dependence on $\Lambda$. Then we send $\Lambda \rightarrow
\infty$ so that cutoff dependence is removed from the effective
Hamiltonian $H_\sigma$.  As in standard perturbative renormalization
theory, $\Lambda$ dependence will be removed order by order in
$g_\sigma$, where $g_\sigma$ is the running coupling constant at the
similarity scale in Hamiltonian matrix elements. This avoids having to
solve the nonperturbative bound state problem to identify and remove
$\Lambda$ dependence.

There is a question whether $H_{\sigma ij}$ satisfies boost invariance
after $\Lambda \rightarrow \infty$. There can be finite terms in
$H_\sigma$ which violate boost invariance yet cannot justifiably be
subtracted. Such terms might result, for example, as a byproduct of
divergent terms of the form $ ln \; {\Lambda^2 \over P_\perp^2} \;
f(P^+)$, where $P_\perp$ and $P^+$ are external momenta. The $ln \;
\Lambda^2$ divergence must be subtracted. But the finite term $ ln \;
P_\perp^2 \; f(P^+)$ cannot be subtracted because no arguments exist
that would justify such a non-analytic $P_\perp$ dependence resulting
from the effects of states above the cutoff. In this case violations
of boost invariance can only disappear at special values of $g$ where
the coefficient of the boost violating terms vanish. Clearly one such
special value has to be $g =g_s$, as part of the restoration of
covariance at $g_s$.

In the following it is assumed that the effective Hamiltonian
$H_\sigma$ has boost invariance simply as a result of taking the $\Lambda
\rightarrow \infty$ limit. $H_\sigma$ is generated as the first step
in solving the bare cutoff Hamiltonian $H_B$. The second step is to
construct bound states from $H_\sigma$.  We may write the bound state
equation as
\begin{eqnarray}
\sum_j H_{\sigma ij} \psi_{N\sigma j}
= {\cal E}_N \psi_{N\sigma i},
\end{eqnarray}
where the sum over $j$ includes a sum over Fock space sectors
and integrals over momenta.  The light-front energy of the
$Nth$ eigenstate is
\begin{eqnarray}
{\cal E}_N = (M_N^2 + P_{N\perp}^2)/P^+_N,
\end{eqnarray}
where $P^+_N$ and $P_{N\perp}$ are the total longitudinal and
transverse momenta of the eigenstate and $M_N$ is its mass.

We will solve this field theoretic bound state problem in the standard
fashion, using bound state perturbation theory.  This requires us to
identify a part of $H_\sigma$, $H_{\sigma 0}$, which is treated
nonperturbatively to produce bound states.  The essential
simplification that makes further calculation possible is that
$H_{\sigma 0}$ does not contain any interactions that change particle
number, so that the methods of few-body quantum mechanics can be used
to solve this initial nonperturbative problem.  All field theoretic
corrections that arise from particle creation and annihilation will
then be treated perturbatively, as in QED.  The potentials that appear
in $H_{\sigma 0}$ must be chosen to make it possible for bound state
perturbation theory to converge; and as stated above, this requires
some art even in QED.

The bare cutoff Hamiltonian is divided as $H_B = H^B_0 + H^B_I$.  For
the determination of the effective Hamiltonian $H_\sigma$, the
unperturbed part of the bare Hamiltonian $H^B_0$ is chosen to be that
of free massive quarks and gluons with the standard relativistic
dispersion relation in light-front kinematics. The interaction part
$H^B_I$ then contains the canonical interaction terms $H_{int}$,
renormalization counterterms $H_B^{CT}$, and the artificial potential
$V_A$.  The artificial potential is itself separated as $V_A = V_0 +
V_A^{CT}$, with Coulomb and linear parts, $V_0 = V_C + V_L$, and
counterterms $V_A^{CT}$ which remove unphysical effects resulting from
the choice of a massive gluon. For the bound state calculation, we
want to choose the starting Hamiltonian to contain as much of the
physics in as simple a form as possible.  So we define
\begin{eqnarray}
H_{\sigma 0 ij} = f_{\sigma ij} \big(H^B_{0ij} + V_{0ij}\big),
\end{eqnarray}
and write the unperturbed bound state solutions as
$\psi^{(0)}_{N\sigma j}$.  These bound states will be pure $q\bar q$,
pure $qqq$, or pure $gg$, as in the CQM.

The starting Hamiltonian is determined from the
the Fourier transforms of
\begin{eqnarray}
	H^B_0 &&= \int dx^-d^2 x_{\bot} \Big\{ {1 \over 2}
		(\partial^i A_a^j)^2 + m_G^2 A_a^{i2} +
		\xi^{\dagger} \Big( {-\partial_{\bot}^2 + m_F^2
		\over i \partial^+} \Big) \xi \Big\}, \\
	V_0 && = \int dx^- d^2 x_{\bot} dy^- d^2 y_{\bot}
		\Big\{ - {1\over 4\pi}
		j^{+a} (x) {\cal V}_C (x,y) j^{+a}(y)
		\nonumber \\
	&& ~~~~~~~~~~~~~~~~~~~~~~~~~~~~~~~~~~ +
		 j^+ (x) {\cal V}_L (x,y) j^+(y) \Big\},
\end{eqnarray}
where ${\cal V}_{C}$ and ${\cal V}_{L}$ are given by Eqs.(2.6) and
(2.7), $j^{+a}= j^{+a}_q + j^{+a}_g$ is the color vector charge
density, and $j^+= j^+_q + j^+_g$ is the color singlet charge density.
This starting Hamiltonian determines the spectrum to order $g_\sigma^4$
and produces the zeroth order bound states that are used to analyze
radiative corrections starting at order $g_\sigma^6$.

The interaction part of the effective Hamiltonian can now be analyzed
perturbatively, and it includes effective
potentials that contain the effects of interactions involving far
off-diagonal states as well as renormalization counterterms, which are
determined order by order. Consider the effective Hamiltonian
$H_\sigma$ generated to second order in the renormalized coupling constant. The
counterterms $H^{CT}_B$ up to this order have been discussed in the
Sections IV.B, VIII, and IX.  They are the counterterms for the
canonical instantaneous interactions, which can be combined as
\begin{eqnarray}
	H^{CT}_{B2} = - {g^2 \over 4 \pi \epsilon} \int d^2 x_{\bot}
		\Big( \int dx^- j^{+a} (x^-,x_{\bot}) \Big)^2,
\end{eqnarray}
and the counterterms from one-gluon exchange given in momentum space
in (9.14) and (9.19), plus a similar term in the gluon-gluon sector.
There are also quark and gluon mass counterterms, some of which we
have given explicitly in Section VIII. The effect of these counterterms
is to completely cancel the leading radiative corrections from
instantaneous gluon exchange, all of which diverge or produce
pathologies.  In addition, as has been
discussed in Sections IV and IX, we need to include terms which
counter the elimination of zero modes, like (4.24):
\begin{eqnarray}
	H^B_{c3} = g^2 \int dx^- d^2 x_{\bot} dy^- d^2 y_{\bot}
		\xi^{\dagger} (\sigma_{\bot} \cdot A_{\bot})
		{\cal O}_F(x,y)(\sigma_{\bot} \cdot A_{\bot}) \xi .
\end{eqnarray}
Such a counterterm can break chiral symmetry, for example, if ${\cal
O}_F \sim m_F^{-1} (\sigma_{\bot} \cdot \partial_{\bot})$.  At this
order, ${\cal O}_F$ must be determined phenomenologically; it remains
to be seen whether coupling coherence and/or the restoration of
Lorentz covariance will completely determine it at higher orders and as
one lets $g \rightarrow g_s$. As
discussed in Section IX, there may be many more such counterterms
resulting from the elimination of zero modes. We have not written them
down explicitly, but this must be done to compute order $g_\sigma^6$
shifts in the hadron spectrum.

What is the dependence of this effective Hamiltonian $H_\sigma$ on the
scale $\sigma$? Suppose we restrict the states to only a $q {\bar q}$
pair. Then only number conserving (potential-like) terms in $H_\sigma$
contribute. Now if we let the scale $\sigma $ approach a very large
number, only the original interactions in $H_B$ survive since the
factor $(1 - f_{\sigma ik } f_{\sigma jk})$ $ \rightarrow 0$. Thus the
canonical four-fermion interaction together with all counterterms
survive, whereas the transversely smeared four-fermion interaction
diminishes in strength. In the limit $\sigma \rightarrow \infty$ we
are recovering $H_B$ which is no surprise. In this limit the effects
of higher Fock sectors (for example $q{\bar q}g$) can be recovered
only by including them explicitly, and so it clearly becomes a poor
approximation to include only the $q\bar q$ sector when $\sigma$ becomes
too large.

By lowering the similarity scale $\sigma$, we reduce the allowed
range of gluon momenta which can contribute, for example, to the
binding of a quark and antiquark in a meson. These effects must appear
elsewhere; and we see that through the similarity transformation they
are added directly to the Hamiltonian via the second and higher order
terms in $H_B$ in (10.1).  Thus by lowering $\sigma$, we put the bare
gluon exchange effects of $H_B$ into a $q\bar q$ potential in the
effective Hamiltonian $H_\sigma$ perturbatively. This clearly changes
the character of the bound state calculation.  It changes from a field
theoretic computation with arbitrary numbers of constituents to a
computation dominated by an effective $q\bar q$ potential. If we
choose the similarity scale $\sigma$ to be just above the hadronic
mass scale, the major effects come from the $q\bar q$ sector.   The
resulting nonrelativistic calculation will not see the scale $\sigma$
in first approximation, since only states close to the diagonal ---
for which $f_{\sigma ij} =1$ in (10.4) --- will contribute.

Consider Coulombic bound states with a Hamiltonian of the form $H =
{p^2 \over m} - {g^2 \over r}$. $ p \sim {1 \over r}$ and the energy
of the bound state scales like $g^4$. Since $r$ scales like ${ 1 \over
g^2}$, if we add to the Hamiltonian a linear potential of the form $
\alpha \, r$ and insist that the energy still scales like $g^4$ we
infer that $\alpha \sim g^6$.

The bound state equation for the unperturbed
effective Hamiltonian $H_{\sigma 0}$ can be written as
\begin{eqnarray}
\Big\{\big[{1\over  p^+} p^2 + {1\over p^{\prime +}} p^{\prime 2}\big]
- {g^2\over 4\pi}
\Big[ {m_F\over p^+ r} +
{m_F\over p^{\prime +} r'}\Big]
+ \big({g^2\over 4\pi}\big)^3{\beta }
\Big[ {m_F^3 r\over p^+} +
{m_F^3 r'\over p^{\prime +}}\Big] \Big\} \Psi = { 4 m_F {\cal E} \over P^+}
\Psi,
\end{eqnarray}
where $P^+$ is the center of mass longitudinal momentum. Using $p^+
\sim { 1 \over 2} P^+ $ and $p^{\prime +}\sim { 1 \over 2} P^+$, the
fully nonrelativistic bound state energy is
\begin{eqnarray}
{\cal E} = \Big\langle {p^2\over m_F} -
{g^2\over 4\pi r} +
\beta \big({g^2\over 4\pi}\big)^3 m_F^2 r
 \Big\rangle.
\end{eqnarray}
Now as expected, the uncertainty principle shows that ${\cal E}$ is minimized
when $r$ is of order ${ 1 \over g^2}$, and $p$ is of order $g^2$ and hence
${\cal E}$ is of order $g^4$. Numerical computations can easily give precise
results.

%%%%%%%%%%%%%%%%%%%%%%%%%%%%%%%%%%%%%%%%%%%%%%%%
\subsection{Potential instability problem}
%%%%%%%%%%%%%%%%%%%%%%%%%%%%%%%%%%%%%%%%%%%%%%%%%%

As we have argued in Section II.E, we add an artificial potential to
our effective Hamiltonian to ensure that the bound state structure is
similar to that of the CQM.  An additional benefit of including an
artificial potential is that it allows the removal of a potential
resulting from the choice of a massive gluon which might otherwise
cause an instability in the bound state calculations.

We have shown in the previous sections that the introduction of
a nonzero gluon mass leads to the incomplete cancellation
of the linear interaction in the longitudinal direction
between the canonical instantaneous interaction and
the interaction arising from one-gluon exchange.  The resulting
interaction between the color charge densities is proportional to
\begin{eqnarray}
	g^2 T^a T^a { m_G^2 \over m_G^2 + (p_{1\bot} - p_{3\bot})^2 } { 1 \over
		(p_1^+ - p_3^+)^2 }.
\end{eqnarray}
In the nonrelativistic limit, the transverse momentum difference
in the denominator can be neglected in comparison to $m_G^2$,
and this part of the effective Hamiltonian reduces to
a pure linear interaction in the longitudinal direction.
This linear interaction may cause an instability in the bound
states when they are expanded to include higher
Fock space sectors. Taking the
expectation value of this term in a color singlet state consisting
of a quark and an antiquark, namely, $| s_2 \rangle = b^{i \dagger}
d_i^\dagger | 0 \rangle$, we obtain the linear potential
contribution to the energy. But the expectation value of
the same interaction in a color singlet state consisting of
a quark, an antiquark and gluon, namely, $| s_3 \rangle =
\big [ T^a \big ]_i^j b^{i \dagger} d_j^\dagger a^{a \dagger}
| 0 \rangle$, is the same as the previous case except that
it is multiplied by the factor $- {1 \over 2} C_A + C_F
= - { 1 \over 6}$ for $SU(3)$. This results in an linear
potential which gives a {\it negative} contribution to the
energy. This negative contribution leads to a possible
instability for many-body bound states.

To remove the above instabilities in hadronic bound states,
we need to subtract this linear interaction, as discussed in Section II.E.
We add this subtraction to the artificial
potential with the form
\begin{eqnarray}
	{g^2 \over 4} \Big( 1 - {g^6 \over g^6_s} \Big) \int d^2 x_{\bot}
		dx^- dy^- ~ j^{+a}(x^-,x_{\bot}) |x^- - y^-| j^{+a}
		(y^-, x_{\bot})
\end{eqnarray}
Here we multiply by a coefficient $(1 - g^6/g^6_s)$ rather than
$(1-g^2/g^2_s)$ to ensure that the order-$g^6$ linear artificial
potential in (10.7) is always dominant for weak coupling.
The introduction of an artificial linear
potential in both the longitudinal and transverse directions
then stabilizes the bound states for small running coupling
constant $g$. It is not known yet whether this instability recurs for $g$ near
$g_s$ and if so how to counter it.

%%%%%%%%%%%%%%%%%%%%%%%%%%%%%%%%%%%%%%%%%%%%%%%%%%%%%%%%%%
\subsection{Sources of Complexity in the QCD computations}
%%%%%%%%%%%%%%%%%%%%%%%%%%%%%%%%%%%%%%%%%%%%%%%%%%%%%%%%%%

A basic goal of this paper is to structure a sequence of computations
in LFQCD with growing levels of complexity.  The major sources of the
complexity will be reviewed here.

All computations envisioned here have two stages.  The first stage is
a computation of an effective Hamiltonian $H_{\sigma}$.  We have
defined $\sigma$ to be a mass cutoff parameter, a mass above which
states diminish in importance for further computations.  The goal of
the first stage is to compute $H_{\sigma}$ with $\sigma$ being just
above the hadronic mass scale.  In the simplest computation,
$H_{\sigma}$ consists of a truncated version of the canonical
Hamiltonian (namely $H_{I\sigma}^B$) combined with further corrections
(from the commutator in (7.10)) computed only to order $g^2$, as in
(10.1). From this, then, bound state energies are to be computed only
to order $g^4$.  This is a straightforward computation with the main
concern to fix free parameters through a comparison with physical
spectra after setting $g^4$ equal to $g_s^4$.  The fit at this stage
is extremely crude.

At this point the results are analogous to those for positronium with
only a Coulomb interaction.  There are additional terms in $H_\sigma$
coming from one gluon exchange that are order $g^2$, but because of
their dependence on $r$ they do not correct the energy until order
$g^6$ or higher.  They produce spin and orbital splittings, but before
these splittings can be fully computed one must determine corrections
to the kinetic energy of order $g^2$, corrections to the Coulomb
interaction of order $g^4$, and one must allow corrections to the
artificial linear potential of order $g^8$ in $H_\sigma$. Just as in
QED, operators must be classified both on the basis of explicit powers
of $g$ multiplying them and on their implicit dependence on $g$
arising from the fact that $r$ scales like $1/g^2$ in the bound
states.  At each stage one must consistently retain all terms in
$H_\sigma$ that produce corrections of the same order in $g$ after
both explicit and implicit dependence on $g$ is determined.  This is
standard bound state perturbation theory, with a linear potential
added and forced to contribute at the same leading order as a Coulomb
interaction, and with the underlying field theory producing additional
interactions.

Actually there is an important subtlety here.  One gluon exchange falls
exponentially fast at large $r$ because of the gluon mass.  To obtain an
order $g^6$ spin splitting the gluon mass in the exchange
interaction must be of order $g^2$ or zero, not of order one.  We allow
the artificial potential to contain spin-dependent interactions
at weak coupling that vanish as $g \rightarrow g_s$, with a range
governed by a mass
that is order $g^2$.  At no point do we allow the introduction of
unphysically light constituents to obtain mass splittings, because
there is no mechanism to prevent copious production of such
light constituents in all hadrons
as $g \rightarrow g_s$, and there is no evidence for the proliferation of
light hadrons such constituents would necessitate even at weak
coupling.

The calculation of all the terms in $H_\sigma$ required to compute
binding energies to order $g^6$ is complicated because of the large
number of perturbative diagrams involved and the complexity of the
cutoffs.  However, there is a very interesting question that arises.
The question is whether longitudinal boost invariance will be
irretrievably violated by the fourth-order results.  If the
fourth-order results include logarithmic divergences due to the
infrared cutoff, divergences which also have non-trivial dependencies
on transverse momentum, then there will be a boost invariance
violation. The reason is that the logarithm will involve a ratio of an
external longitudinal momentum divided by the cutoff momentum.  While
the cutoff dependence can legitimately be subtracted, the finite part
of the logarithm cannot, and so results in a violation of longitudinal
boost invariance.  This violation would be similar to the violation of
scale invariance of canonical equal-time QCD at zero mass.  The scale
invariance violation is never eliminated, it instead results in
asymptotic freedom.  A violation of boost invariance would likewise be
irretrievable in weak coupling, and could only go away at a special
value of $g$. Clearly, this special value would have to be $g_s$.

In addition, if the fourth-order computations turn up such logarithmic
longitudinal divergences, then there would have to be finite subtractions
accompanying the subtractions of the logarithmic cutoff-dependent divergences.
These finite subtractions would involve arbitrary functions of transverse
momentum --- functions that could conceivably be the source of transverse
linear potentials.  Certainly, these finite subtractions would be direct
reflections of vacuum effects, which means linear potentials would not be a
surprise.

However, the most challenging problem to resolve is the computation
of the effective Hamiltonian $H_{\sigma}$ beyond fourth-order.  The
major concern is the computation of higher-order corrections caused by
removal of ``wee parton'' states.  By ``wee parton'' states we mean states
of high mass due to the presence of constituents in the infrared
region and without large transverse momentum.  The simplest example
is a state with two equal mass constituents but one having  a far
smaller longitudinal momentum (smaller $x$ in Feynman's language)
than the other.  Their transverse momentum will be set to $p_{\bot}$
and $-p_{\bot}$,  respectively.  If one constituent has longitudinal
momentum $p^+$, the other has momentum $xp^+$, with $x$ small, then the
states mass squared is about $\frac{m^2+ p_{\bot}^2}{x}$, where $m$
is the constituent mass.  The problem is that when $p_{\bot}$ is
small, masses count, which means there can be major corrections associated
with actual values for bound state masses.  This is a problem once the binding
energies are themselves of order the constituent
masses, namely, when $g$ is near its relativistic value $g_s$.

This challenge will not be resolved easily.  It is unlikely that
it can even be seriously addressed until after the fourth-order
computations are complete, and one can start estimating higher-order
terms and determine the extent to which strong binding energies
result in large higher-order terms in $H_{\sigma}$.
This challenge is further enhanced by the problem of avoiding instabilities in
$H_\sigma$ itself for $g$ near $g_s$.

%%%%%%%%%%%%%%%%%%%%%%%%%%%%%%%%%%%%%%%%%%%%%%%%%%%
%%%%%	concl.tex  (put into QCD paper) 	%%%
%%%%% 		last modified 1/24/94         	%%%
%%%%%%%%%%%%%%%%%%%%%%%%%%%%%%%%%%%%%%%%%%%%%%%%%%%

%%%%%%%%%%%%%%%%%%%%%%%%%%%%%%%%%%%%%%%%%%%%%%%%%%%
\section{Conclusion}
%%%%%%%%%%%%%%%%%%%%%%%%%%%%%%%%%%%%%%%%%%%%%%%%%%%

We have presented a framework for the computation of bound states in
QCD that is based on Hamiltonian methods. The essential ingredients of
our formulation are the use of light-front coordinates, nonzero quark
and gluon masses, severe cutoffs on constituent momenta, and the
similarity renormalization scheme.  We have argued that this
formulation of the theory removes all barriers to a treatment of
hadrons in QCD which is analogous to that of bound states in QED. In
particular, we have shown that in our formulation it is natural to
choose a starting Hamiltonian which contains many characteristics of
the Constituent Quark Model, and therefore we expect that we may
fashion this unperturbed Hamiltonian to model the basic physics well
already at the outset.  We take the view that the general principles
of the CQM are true and propose to proceed from there. To achieve our
perturbative starting point we must introduce artificial stabilizing
and confining potentials to reproduce the necessary physics at small
values of the coupling $g$, and we must extrapolate to the
relativistic value of the renormalized coupling $g_s$.  At $g_s$, the
artificial potential must vanish, and the true dynamics will appear.
The essential point to realize is that our formulation allows the
possibility that this true dynamics may be discovered perturbatively.

Thus the price we pay for our new formulation --- renormalization
problems from severe light-front infrared divergences and the breaking
of both gauge and Lorentz covariance --- may eventually turn out to be
relatively small in comparison to what we gain: both confinement and
chiral symmetry breaking may be studied with bound state perturbation
theory.  Moreover, the eventual restoration of Lorentz covariance ---
which requires the calculation of physical observables so as to
complete the renormalization process --- will be made easier by the
expectation that bound states will be well-approximated by
nonrelativistic, few-body wave functions. The most daring aspect of
this approach is the assignment of zero-mode contributions to
counterterms for infrared divergences.  There is no precedent for our
analysis, and its worth can only be proven by obtaining satisfactory
and relativistic results as $g \to g_s$.

The possibility of combining phenomenology and field theory at weak
coupling opens a wide range of interesting computations. One
feature essential for the feasibility of the entire enterprise is our
contention that a massive gluon prevents the unlimited growth of the
running coupling constant in the infrared domain. For this to happen
all pure light-front infrared divergences (due to $k^+$ getting small)
need to be cancelled in the coupling constant renormalization
calculation. This needs to be demonstrated.  Then there is the
computation of bound states and hadronic observables.  The use
of an artificial potential and massive quarks and gluons makes this
possible to leading order in a perturbative expansion,
with more and more of the true physics being included as one
moves to higher orders. As discussed in the context of Lattice Gauge
Theory by Lepage and Mackenzie \cite{LM 93}, we must extrapolate to
the relativistic value of the renormalized coupling in order to make
the perturbative expansion legitimate. One can also readily conceive
other instructive calculations which would not involve all the
complications of low-energy QCD, such as the application of the
framework developed here to the study of scalar field theory, QED, and
especially heavy quark systems.

The possibility of constructing low-energy bound states in QCD through
a combination of relativistic perturbation theory and many-body
quantum mechanics defines a start-up phase with a wide range of new
and tractable calculations.  We have tried to emphasize those
computations which are most immediately relevant and to define how
these calculations can be carried out.  The final demonstration of the
feasibility of our approach awaits these computations.

\acknowledgements
We acknowledge many helpful discussions over several years with D.
Mustaki, S. Pinsky, and J. Shigemitsu.  We would also like to
acknowledge our useful interactions with S. Brodsky, M. Burkardt, D.
H. Feng, R. Furnstahl, X. Ji,  P. Lepage, and G.A. Miller. This
research is supported in part by the National Science Foundation under
Grant Nos. PHY-8858250, PHY-91029022, PHY-9203145, and PHY-9207889,
and by Komitet Bada{\'n} Naukowych under Grant No. KBN 2 P302 031 05.
Some of the figures in this paper were constructed using the package
``FEYNMAN: A {\LaTeX} Routine for Generating Feynman Diagrams,'' by
Michael Levine, Cavendish-HEP 88/11.

\appendix
%%%%%%%%%%%%%%%%%%%%%%%%%%%%%%%%%%%%%%%%%%%%%%%%%%%%%%%%%%%%%%%
%%%%%%%%%%  zerom.tex     (input to QCD paper)      %%%%%%%%%%%
%%%%%%%%%%%%%%%%%%%%%%%%%%%%%%%%%%%%%%%%%%%%%%%%%%%%%%%%%%%%%%%
%%%%%%%%%%  last modified: 1/24/94      %%%%%%%%%%%%%%%%%%%%%%
%%%%%%%%%%%%%%%%%%%%%%%%%%%%%%%%%%%%%%%%%%%%%%%%%%%%%%%%%%%%%%%

%%%%%%%%%%%%%%%%%%%%%%%%%%%%%%%%%%%%%%%%%%%%%%%%%%%%%%%%%%%%%%%
\section{Sigma model with and without zero modes}
%%%%%%%%%%%%%%%%%%%%%%%%%%%%%%%%%%%%%%%%%%%%%%%%%%%%%%%%%%%%%%%

We will work only at the canonical or tree level here.  The
Lagrangian  for the sigma model is
\begin{eqnarray} 	{\cal L} = {1 \over 2}(\partial_\mu \sigma)^2 + {1 \over 2}
	(\partial_\mu \pi)^2 - V(\sigma^2 + \pi^2), \end{eqnarray}
where
\begin{eqnarray} 	V(\sigma^2 + \pi^2)
 = - {1 \over 2} \mu^2 (\sigma^2 + \pi^2) + { \lambda \over 4} (\sigma^2
	+ \pi^2)^2 . \end{eqnarray}
This Lagrangian has a continuous symmetry corresponding to the transformation
\begin{eqnarray} 	\sigma \rightarrow \sigma' && = \sigma \; \cos \alpha \; + \;
	\pi \; \sin \alpha,
	\nonumber \\
	\pi \rightarrow \pi' && = - \sigma \;
	\sin \; \alpha \; + \; \pi \; \cos \;
	\alpha  \label{cs1}. \end{eqnarray}
The extremum of the potential $V$ is determined by
\begin{eqnarray} 	{\delta V \over \delta \sigma} && = \sigma [ - \mu^2 +
\lambda (\sigma^2 +
	\pi^2) ] = 0 , \nonumber \\
	{\delta V \over \delta \pi} && = \pi [ - \mu^2 + \lambda (\sigma^2 +
	\pi^2) ] = 0 . \end{eqnarray}
For $\mu^2 > 0$, the minimum of the potential is at $ \sigma^2 + \pi^2 =
\sigma_v^2 = {\mu^2 \over \lambda}$. There are an infinite number of degenerate
vacuua. We can pick any one of them to be the true vacuum. For example, we can
pick
\begin{eqnarray} 	<0 \mid \sigma \mid 0> = \sigma_v \neq 0, \; <0\mid \pi \mid
0> =0. \end{eqnarray}
Then the symmetry is said to be spontaneously broken.

The conserved current associated with the symmetry of the Lagrangian is
\begin{eqnarray} 	J^\mu = \sigma \partial^\mu \pi - \pi \partial^\mu \sigma .
\end{eqnarray}
The associated charge is
\begin{eqnarray} 	Q = \frac{1}{2}\int dx^- d^2 x_\perp [ \sigma \partial^+ \pi
		- \pi \partial^+ \sigma ]. \end{eqnarray}
{}From the canonical commutation relations for $\sigma$ and $\pi$ fields,
namely,
\begin{eqnarray} 	\big [ \sigma(x),\sigma(y)\big]\mid_{x^+=y^+} = - {i \over 4}
	\epsilon(x^-- y^-) \delta^2 (x^\perp - y^\perp), \end{eqnarray}
\begin{eqnarray} 	\big [ \pi(x),\pi(y)\big]\mid_{x^+=y^+} = - {i \over 4}
\epsilon(x^-
	- y^-) \delta^2 (x^\perp - y^\perp), \end{eqnarray}
we have
\begin{eqnarray}
	\big [ Q, \pi(0) \big ] && = - i \sigma(0), \nonumber \\
	\big [Q, \sigma(0)\big ] && = i \pi(0). \label{tl1} \end{eqnarray}
Thus we have
\begin{eqnarray} 	<0 \mid [ Q, \pi(0)] \mid 0> = -i \sigma_v  \neq 0.
\label{apeq1} \end{eqnarray}
Starting from $\partial^\mu J_\mu =0$, we have,
\begin{eqnarray} 	{d Q \over dx^+} = 0. \end{eqnarray}
{}From (\ref{apeq1}) it is also easy to verify Goldstone's
theorem and assert the existence of a massless pion.

In the theory we have considered, zero modes are present; and as a result
\begin{itemize}
\item there is spontaneous symmetry breaking,
\item the vacuum is nontrivial, $<0\mid \sigma(0) \mid 0 > \neq 0 $,
\item a massless pion is predicted.
\end{itemize}

Next we discuss the $\sigma$ model with the zero modes removed. To
construct the effective Hamiltonian we cannot use the symmetry
(\ref{cs1}) as a guide  but can rely on power counting and locality.
The Hamiltonian is
\begin{eqnarray}  	P^- = \int dx^- d^2 x_\perp [ {1 \over 2} (\partial_\perp
\phi \cdot
	\partial_\perp \phi + \partial_\perp \pi \cdot
        \partial_\perp \pi) + V] . \end{eqnarray}
Since the symmetry is broken only in the $\sigma $ sector, we insist that $V$
is even in the $\pi$-field so that the symmetry under $\pi\to -\pi$ remains.
Thus, since no inverse powers of mass are allowed, we can write
\begin{eqnarray} 	V = {1 \over 2} m_\phi^2 \phi^2 + {1 \over 2} m_\pi^2 \pi^2 +
\lambda_1
	\phi^4 + \lambda_2 \pi^4 + \lambda_3 \phi^2 \pi^2 + \lambda_4 \phi^3 +
	\lambda_5 \pi^2 \phi. \end{eqnarray}
Since there are no zero modes,
a term of the form $\lambda_6 \phi$ vanishes upon integration in (A13) and
so is not included.

The equations of motion are
\begin{eqnarray} 	-\partial^+  \partial^- \pi + \partial_\perp^2 \pi  && =
	m_\pi^2 \pi + 4 \lambda_2 \pi^3 + 2 \lambda_3 \phi^2 \pi +
           2 \lambda_5  \pi
	\phi , \nonumber \\
	-\partial^+  \partial^- \phi + \partial_\perp^2 \phi
 	&& =
m_\phi^2 \phi + 4 \lambda_1 \phi^3 + 2 \lambda_3 \pi^2 \phi + 3 \lambda_4
\phi^2 + \lambda_5 \pi^2.  \label{em1} \end{eqnarray}

Next we construct the covariant current operator $J^\mu$.
First we determine the canonical dimensions of the components $J^+, J^-$ and $
J^\perp$. Since the charge $ Q \;(= { 1 \over 2} \int dx^- d^2 x_\perp J^+$) is
dimensionless, the
canonical dimension of $J^+$ is $ J^+ \sim {1 \over x^-} { 1 \over
(x_\perp)^2}$. Since each term in $\partial^\mu J_\mu \;(= { 1 \over 2}
(\partial^- J^+ + \partial^+ J^-) - \partial^\perp\cdot
 J^\perp)$ should have the
same dimensions, the canonical dimensions of the other components are
\begin{eqnarray} J^\perp \sim { 1 \over (x_\perp)^3},
 \; \; J^- \sim {x^- \over (x_\perp)^4}. \end{eqnarray}
The components of $J^\mu$ are to be constructed from the operators $\pi, \phi,
\partial^+, \partial^-, $ and $\partial^\perp$ and constants. The constants
are allowed to have dimensions of negative power of $x_\perp$
(as masses do) but no power of $x^-$. Since the operator ${ 1 \over
\partial^+}$ does not appear in the canonical scalar field theory, we do not
allow the operator ${ 1 \over \partial^+}$ to appear in the canonical current
operator. We also implement the symmetry that $J^\mu$ has to change
sign under $\pi \rightarrow - \pi$. Then the allowed structure of components
of $J^\mu$ is
\begin{eqnarray}
J^+ && = a_1 \partial^+ \pi + a_2 (\partial^+ \pi) \phi + a_3
(\partial^+ \phi) \pi, \nonumber \\
J_\perp && = b_1 \partial_\perp \pi +
b_2 (\partial_\perp \phi) \pi + b_3 (\partial_\perp \pi) \phi, \nonumber \\
J^- && = c_1 \partial^- \pi + c_2 (\partial^- \pi) \phi + c_3
(\partial^- \phi) \pi. \end{eqnarray}

Next we compute $\partial^\mu J_\mu$ and set the coefficients of terms
involving derivatives which cannot be replaced by the equation of motion to
be zero. Then we get
\begin{eqnarray}
a_2 = - c_3, \; a_3 = - c_2, \; b_2 = - b_3.
\end{eqnarray}
Further, from the structure of the equations of motion, we also have
\begin{eqnarray}
b_1 = {a_1+ c_1 \over 2}, \; b_2 = - {a_2 + c_2 \over 2}.
\end{eqnarray}
Since the zero modes are dropped, we have the charge
\begin{eqnarray}
Q = { 1 \over 2 } \int dx^- d^2 x_\perp [  a_2 \phi \partial^+ \pi
+ a_3 \pi \partial^+
	\phi ]. \end{eqnarray}
Insisting that $Q$ generates the correct transformation laws (\ref{tl1}),
we arrive at
\begin{eqnarray} a_3 = - a_2 . \end{eqnarray}
Without loss of generality we set $a_2 = 1$, so
\begin{eqnarray}
J^+ && = a_1 \partial^+ \pi + \phi \partial^+ \pi - \pi \partial^+ \phi, \\
J^- && = c_1 \partial^- \pi + \phi \partial^- \pi - \pi \partial^- \phi, \\
J_\perp && = { 1 \over 2} (a_1+ c_1) \partial_\perp \pi
+ \phi \partial_\perp \pi - \pi \partial_\perp \phi.
\end{eqnarray}
Further, if $J^+$ and $J^-$ are to transform as components of a covariant
four-vector, we also have to set $c_1 = a_1$. Then

\begin{eqnarray} J^\mu = \phi \partial^\mu \pi - \pi \partial^\mu \phi + a_1
\partial^\mu \pi. \end{eqnarray}
Since our effective Hamiltonian (even with explicit symmetry breaking)
is supposed to give rise to the same physics as that of spontaneous symmetry
breaking, we demand that the current is conserved. Using the equations of
motion (\ref{em1}),
\begin{eqnarray}  \partial^\mu J_\mu && = -(\phi + a_1)
(m_\pi^2 \pi + 4 \lambda_2 \pi^3 + 2 \lambda_3 \phi^2 \pi + 2 \lambda_5  \pi
\phi) \nonumber \\
&& \qquad + \pi
(m_\phi^2 \phi + 4 \lambda_1 \phi^3 + 2 \lambda_3 \pi^2 \phi + 3 \lambda_4
\phi^2 + \lambda_5 \pi^2) = 0 . \end{eqnarray}

Setting the coefficients of the different combinations of field operators
separately to zero, we have
\begin{eqnarray} m_\pi^2 a_1 =0, \;  m_\pi^2 - m_\phi^2 + 2 \lambda_5  a_1 =0,
\; 4
\lambda_2 - 2 \lambda_3 = 0, \; 4 \lambda_1 - 2 \lambda_3 = 0, \;
\nonumber \\
 \qquad 2 \lambda_5
- 3 \lambda_4 + 2 a_1 \lambda_3 =0, \; - \lambda_5 + 4 \lambda_2 a_1 =0.
\end{eqnarray}
The solutions are
\begin{eqnarray}  	\lambda_1 = \lambda_2 = {1 \over 2} \lambda_3 \equiv
{\lambda\over 4},
\quad \lambda_5 = \lambda a_1, \quad \lambda_4 = \lambda a_1,
\quad m_\phi^2 = m_{\pi}^2 + 2\lambda a_1^2,
\end{eqnarray}
and
\begin{eqnarray}
a_1 = 0
\quad {\rm or }\quad
m_{\pi} = 0.
\end{eqnarray}
If we take $a_1 = 0$, then the potential is reduced to the canonical form
(A2), and the current is also of the canonical form.  This corresponds to
the full canonical theory with a symmetry preserving vacuum and a doublet
in the spectrum.
If we choose $m_{\pi} =0$, we have
\begin{eqnarray} V &=& { 1 \over 2} m_\phi^2 \phi^2 + {\lambda \over 4} (\phi^2
+ \pi^2)^2 +
\sqrt{\lambda m_\phi^2\over 2} (\phi^2 + \pi^2)\phi, \nonumber \\
J^\mu &=& \phi \partial^\mu  \pi - \pi \partial^\mu \phi + \sqrt{{m_\phi^2
\over
2 \lambda}} \; \partial^\mu \pi
 . \end{eqnarray}
Now the potential explicitly breaks the symmetry, and the current
is different from the canonical one.  The charge ${Q}$ does not
commute with the Hamiltonian, and
\begin{eqnarray} {d {Q} \over dx^+} \neq 0. \end{eqnarray}
This corresponds to full canonical theory after a shift of $\sigma$ to remove
its vacuum expectation value. Thus we see that
the power counting rules
allow us to reconstruct the theory without reference to the zero modes.

Thus in the theory with zero modes dropped and the zero pion mass, the
Hamiltonian explicitly breaks the symmetry, the vacuum is trivial, and
there is no longer the notion of spontaneous symmetry breaking.
Current conservation is preserved and vastly reduces the number of
free parameters present in the effective Hamiltonian constructed from
power counting and forces the pion to remain massless.

Of course, we should emphasize that the sigma model we have discussed
is only at the tree level.  For a complete understanding, we need to
consider radiative corrections and the subsequent renormalization
effects.

%%%%%%%%%%%%%%%%%%%%%%%%%%%%%%%%%%%%%%%%%%%%%%%%%%%
%%% appcs.tex 					%%%
%%%%%%%%%%%%%%%%%%%%%%%%%%%%%%%%%%%%%%%%%%%%%%%%%%%
%%         Last modified    1/24/93       	%%%
%%%%%%%%%%%%%%%%%%%%%%%%%%%%%%%%%%%%%%%%%%%%%%%%%%%

%%%%%%%%%%%%%%%%%%%%%%%%%%%%%%%%%%%%%%%%%%%%%%%%%%%%%%%
\section{More about chiral symmetry on the light front}
%%%%%%%%%%%%%%%%%%%%%%%%%%%%%%%%%%%%%%%%%%%%%%%%%%%%%%%

In this appendix, we show that the normal chiral transformation on the
quark field is inconsistent with the light-front constraint equation,
where the latter is a result of the light-front equations of motion.
We also show that chirality is the same as helicity on the light-front.

Recall that in light-front coordinates, the quark
field can be separated into the plus- and minus-components:
$\psi = \psi_+ + \psi_-$, where  $\psi_{\pm} = \frac{1}{2}
\gamma^0 \gamma^\pm \psi$, and only $\psi_+$ is a dynamical
variable.  The minus-component $\psi_-$ is determined by the
light-front constraint
\begin{equation}
	\psi_- = { 1 \over i \partial^+} \big(  \alpha_{\perp}\cdot
	(i \partial_{\perp} + g A_\perp) + \gamma^{0} m_F \big ) \psi_{+}.
\end{equation}

The normal chiral transformation is defined by
\begin{equation}
	\psi \longrightarrow \psi + \delta \psi ~~ {\rm with}~~
		\delta \psi = - i \theta \gamma_5 \psi.
\end{equation}
However, in light-front coordinates, only $\psi_+$ is a dynamical
variable, so that the chiral transformation acts only on $\psi_+$
\begin{equation}
	\psi_+ \longrightarrow \psi_+ + \delta \psi_+ ~~, ~~~ \delta
		\psi_+ = - i \theta \gamma_5 \psi_+,
\end{equation}
and the transformation on $\psi_-$ is given by the equation of
constraint
\begin{eqnarray}
	\delta \psi_- && = { 1 \over i \partial^+}
\big (  \alpha_\perp\cdot(i  \partial_\perp + g A_\perp) + \gamma^{0} m_F \big)
		\delta \psi_{+} \nonumber \\
	&& = - i \theta  \gamma_5 { 1 \over i \partial^+}
\alpha_\perp \cdot \big( i \partial_\perp+ g A_\perp \big) \psi_{+}
		- i \theta m_F \gamma^0 \gamma_5
		{1 \over i \partial^+} \psi_{+}  .
\end{eqnarray}
Thus, the chiral transformation on $\psi$ on the light-front is
\begin{eqnarray}
	\delta  \psi && = \delta { \psi}_+ + \delta
		{\tilde \psi}_-  \nonumber \\
	&& = -i \theta \gamma_5 \psi_+ - i \theta \gamma_5
		{ 1 \over i \partial^+} \alpha_\perp \cdot
\big ( i \partial_\perp + g A_\perp \big) \psi_{+} - i \theta m_F \gamma^0
		\gamma_5 {1 \over i \partial^+} \psi_{+} . \label{eq.lfct}
\end{eqnarray}
If we naively use the form $ \delta \psi = - i \theta \gamma_5 \psi
= -i \theta \gamma_5 (\psi_+ + \psi_-) $,
\begin{equation}
	\delta { \psi} = -i \theta \gamma_5 \psi_+
  		- i \theta \gamma_5 { 1 \over i \partial^+}
	 \alpha_\perp \cdot \big ( i \partial_\perp+ g A_\perp \big ) \psi_{+}
		- i \theta m_F \gamma_5 \gamma^0 {1 \over i
		\partial^+} \psi_{+}  .
\end{equation}
This is obviously inconsistent with (\ref{eq.lfct}), which
is a consequence of the light-front equation of constraint.

For the massive quark theory,
\begin{eqnarray} \partial_{\mu} j^{\mu}_{5}  = 2 i m_{F} {\bar \psi} \gamma_{5}
\psi , \end{eqnarray}
where $j_5^{\mu} =
\bar{\psi} \gamma^{\mu} \gamma_5 \psi$ is the axial
vector current.
The axial vector charge
on the light-front is then
\begin{eqnarray} Q^{5}_{LF}  =  \int dx^- d^2 x_\perp  j^{+}_{5}(x).
\end{eqnarray}
Explicit calculation using the field expansions and normal ordering leads to
\begin{eqnarray}
	Q^{5}_{LF} = && \int  {dk^+ d^2 k_\perp \over 2(2 \pi)^3}
		\sum_{\lambda} \lambda \big [ b_{\lambda}^{\dagger}(k)
		b_{\lambda}(k)  + d_{\lambda}^{\dagger}(k)
		d_{\lambda}(k)  \big ] . \nonumber \\
&&  \end{eqnarray}
Thus $Q^{5}_{LF}$ measures the
helicity.

In the canonical theory the chiral current is conserved for zero
fermion mass. In the theory with the renormalized effective
Hamiltonian we expect to observe the consequences of spontaneous
symmetry breaking, signals of which are a conserved axial current and
a zero mass pion. However, we do not expect to generate a massless
pion unless the coupling $g$ exactly matches its renormalized value
$g_s$. Hence the issues of a massless pion and a conserved axial
vector current (which we expect to be non-canonical in form) can be
addressed meaningfully only after we reach the relativistic limit of
the theory.

%%%%%%%%%%%%%%%%%%%%%%%%%%%%%%%%%%%%%%%%%%%%%%%%%%%%%%%%%%%%%
%%%%%%%  nrpot.tex  (input to QCD paper)         %%%%%%%%%%%%
%%%%%%%%%%%%%%%%%%%%%%%%%%%%%%%%%%%%%%%%%%%%%%%%%%%%%%%%%%%%%
%%%%%%%  last modified 1/24/93    %%%%%%%%%%%%%%%%%%%%%%%%%%
%%%%%%%%%%%%%%%%%%%%%%%%%%%%%%%%%%%%%%%%%%%%%%%%%%%%%%%%%%%%%

%%%%%%%%%%%%%%%%%%%%%%%%%%%%%%%%%%%%%%%%%%%%%%%%%%%%%%%%%%%%%
\section{Nonrelativistic limit of massless gluon exchange}
%%%%%%%%%%%%%%%%%%%%%%%%%%%%%%%%%%%%%%%%%%%%%%%%%%%%%%%%%%%%%

In this Appendix we discuss the nonrelativistic limit of one gluon
exchange; however, we insist on maintaining the separate dimensional
assignments for longitudinal and transverse coordinates when taking
this limit, and we arrive at results that differ substantially from
other analyses which allow longitudinal and transverse dimensions to
mix.  Our entire program is based on separate power-counting analyses
for these dimensions, and if they mix our power counting fails.

In second-order perturbation theory, the effective interaction Hamiltonian in
the $q {\bar q}$ sector due to {\it massless} gluon exchange is
\begin{eqnarray}
	H_{I\sigma ij}^{(2)} = H_{I\sigma ij 1}^{(2)} + H_{I\sigma ij 2}^{(2)}
\end{eqnarray}
where
\begin{eqnarray}
	H_{I\sigma ij 1}^{(2)} && = - 4 g^2 T_{\alpha_3 \alpha_1}^a
		T_{\alpha_4 \alpha_2}^a \delta_{s_1 s_3} \delta_{s_2 s_4}
		{1 \over 2}{ 1 \over (p_1^+ - p_3^+)^2} \nonumber \\
	&& \qquad - g^2 T_{\alpha_3 \alpha_1}^a T_{\alpha_4 \alpha_2}^a
		{\cal M}_{2ij} { 1 \over p_1^+ - p_3^+} {1 \over 2}
	\Bigg[ {1 \over  {m_F^2 + p_{1 \perp}^2 \over p_1^+} -
                {m_F^2 + p_{3 \perp}^2 \over p_3^+} - { (p_{1 \perp}
		- p_{3 \perp})^2 \over p_1^+ - p_3^+} } \Bigg], \\
	H_{I\sigma ij 2}^{(2)} && = - 4 g^2 T_{\alpha_3 \alpha_1}^a
		T_{\alpha_4 \alpha_2}^a \delta_{s_1 s_3} \delta_{s_2 s_4}
		{1 \over 2}{ 1 \over (p_1^+ - p_3^+)^2} \nonumber \\
	&& \qquad - g^2 T_{\alpha_3 \alpha_1}^a T_{\alpha_4 \alpha_2}^a
		{\cal M}_{2ij} { 1 \over p_1^+ - p_3^+} {1 \over 2}
	\Bigg[ {1 \over  {m_F^2 + p_{4 \perp}^2 \over p_4^+} -
                {m_F^2 + p_{2 \perp}^2 \over p_2^+} - { (p_{1 \perp}
		- p_{3 \perp})^2 \over p_1^+ - p_3^+} } \Bigg] \end{eqnarray}
and
\begin{eqnarray}
	{\cal M}_{2ij} && = \chi_{s_{3}}^{\dagger} \Big[ 2 { p_1^{i_1}
		- p_3^{i_1} \over p_1^+ - p_3^+} - i m_F \Big({1 \over
		p_1^+} - { 1 \over p_3^+} \Big) \sigma^{i_1} - \Big(
  		\sigma^{i_1} {\sigma_\perp \cdot p_{1 \perp} \over p_1^+}
		+ {\sigma_\perp \cdot p_{3 \perp} \over p_3^+}
		\sigma^{i_1} \Big) \Big ] \chi_{s_{1}} \nonumber \\
	&& \qquad \chi_{-s_{2}}^{\dagger} \Big [ 2 {p_1^{i_1} -
		p_3^{i_1} \over p_1^+ - p_3^+} - i m_F \Big(
		{1 \over p_4^+} - { 1 \over p_2^+} \Big) \sigma^{i_1}
		- \Big( \sigma^{i_1} {\sigma_\perp \cdot p_{4 \perp}
		\over p_4^+} + {\sigma_\perp \cdot p_{2 \perp} \over p_2^+}
		\sigma^{i_1} \Big ) \Big ] \chi_{-s_{4}} . \end{eqnarray}

Let us denote the total momenta by $(P^+, P_\perp)$.  For simplicity, we
consider the case $P_\perp = 0$, that is, $p_{2 \perp} = - p_{1 \perp}, \;
p_{4 \perp} = - p_{3 \perp}$.
To get the form of the effective Hamiltonian in the nonrelativistic
limit, as an example, we consider the matrix element for $ s_1 = s_2
= s_3 = s_4 = \uparrow$. Then
\begin{eqnarray} {\cal M}_{2ij} = 4 { (p_{1 \perp} - p_{ 3 \perp})^2 \over
(p_1^+ - p_3^+)^2}
- 2 {p_{1 \perp}^2 (p_2^+ - p_1^+) \over p_1^+ p_2^+ (p_1^+ - p_3^+)}
- 2 {p_{3 \perp}^2 (p_3^+ - p_4^+) \over p_3^+ p_4^+ (p_1^+ - p_3^+)}
- 2 { (P^+)^2 \over p_1^+ p_2^+ p_3^+ p_4^+} p_1^L p_3^R \end{eqnarray}
where $ k^L = k^1 - i k^2$ and $k^R = k^1 + i k^2$.
This gives, after simplification,
\begin{eqnarray}
	H_{I\sigma ij}^{(2)} && = - g^2 T_{\alpha_3 \alpha_1}^a
		T_{\alpha_4 \alpha_2}^a \Bigg[ { 1 \over 2} \Big(
		{4 m_F^2 \over p_1^+ p_3^+} + 2 {(P^+)^2 \over
		p_1^+ p_2^+ p_3^+ p_4^+} p_1^L p_3^R \Big ) { 1 \over E_1}
		\nonumber \\
	&& \qquad ~~~~~~~~~~~~~~~~~~~~ + { 1 \over 2} \Big ( {4 m_F^2 \over
		p_2^+ p_4^+} + 2 {(P^+)^2 \over p_1^+ p_2^+ p_3^+
		p_4^+} p_1^L p_3^R \Big ) { 1 \over E_2} \nonumber \\
	&& \qquad ~~~~~~~~~~~~~~~~~~~~ - { 1 \over E_1 E_2} (P^+)^2 \Big( {p_{1
	\perp}^2 \over p_1^+ p_2^+} - { p_{3 \perp}^2 \over p_3^+ p_4^+}
		\Big)^2 \Bigg] \end{eqnarray}
where
\begin{eqnarray}
	E_1 = (p_{1 \perp} - p_{3 \perp})^2 + {m_F^2 \over p_1^+
		p_3^+}(p_1^+ - p_3^+)^2 - (p_1^+ - p_3^+) \Big(
		{p_{1 \perp}^2 \over p_1^+} - {p_{3 \perp}^2
		\over p_3^+} \Big), \end{eqnarray}
and
\begin{eqnarray}
	E_2 = (p_{1 \perp} - p_{3 \perp})^2 + {m_F^2 \over p_2^+
		p_4^+}(p_1^+ - p_3^+)^2 - (p_1^+ - p_3^+) \Big(
		{p_{3 \perp}^2 \over p_4^+} - {p_{1 \perp}^2
		\over p_2^+} \Big). \end{eqnarray}

In the nonrelativistic limit, the term involving $ p_1^L p_3^R$ is negligible
compared to the mass term; in both $E_1$ and $E_2$ the third term is
negligible compared to the first two terms;
furthermore, ${m_F^2 \over p_1^+ p_3^+} \;
\rightarrow \; ({m_F \over p_1^+})^2 $ and
${m_F^2 \over p_2^+ p_4^+} \;
\rightarrow \; ({m_F \over p_2^+})^2 $.
Thus in the nonrelativistic limit, the effective interaction in the $q {\bar
q}$ sector due to {\it massless} one-gluon exchange reduces to
\begin{eqnarray}
	{\cal H}_{2ij}^{NR} && = - 4 g^2 T^a T^a { 1 \over 2}
		\Bigg [ \Big({m_F \over p_1^+}\Big)^2 { 1 \over
		(p_{1 \perp} - p_{3 \perp})^2 + ({m_F
		\over p_1^+})^2 (p_1^+ - p_3^+)^2 } \nonumber \\
	&& \qquad \qquad ~~~~~~~~ + \Big({m_F \over p_2^+}\Big)^2
		{ 1 \over (p_{2 \perp} - p_{4 \perp})^2 + ({m_F
		\over p_2^+})^2 (p_2^+ - p_4^+)^2 }\Bigg ].  \end{eqnarray}

The Coulomb interaction has the form $ { 1 \over p^2}$
in momentum space. If we further make the approximation for (C9)
that $p_1^+ \approx m_F + p_1^z$ and $ p_3^+ \approx m_F + p_3^z $,
we arrive at the familiar Coulomb interaction in momentum space
\begin{eqnarray}
	{\cal H}_{cij} =
	- 4 g^2 T^a T^a { 1 \over ( {\vec p_1} - {\vec p_3})^2}
\label {ci} \end{eqnarray}
with $  ( {\vec p_1} - {\vec p_3})^2 \; = \; (p_{1 \perp} - p_{3 \perp})^2 +
(p_1^z - p_3^z)^2$. The Coulomb potential in position space is then obtained
by the Fourier transformation,
\begin{eqnarray}
	{ 1 \over ( {\vec p_1} - {\vec p_3})^2} = {1 \over 4\pi}
		\int d^3r e^{i( {\vec p_1} -
		{\vec p_3}) \cdot {\vec r} } { 1 \over r}, \end{eqnarray}
with $ r = |\vec x - \vec x^\prime|$ the separation between the two
fermions.
Obviously, $H_c = -\int d^3x d^3x' j^a_q(\vec x)
{ g^2 \over 4 \pi r} j^a(\vec x')$ has lost boost
invariance but has recovered rotational invariance.

We wish to represent the interaction ${\cal H}_2^{NR}$
in a form where longitudinal boost invariance is kept.
To do so, we use an interpolating Fourier
transformation:
\begin{eqnarray}
	\Big({m_F \over p_1^+}\Big)^2 && { 1 \over (p_{1 \perp}
		- p_{3 \perp})^2 + ({m_F \over p_1^+})^2 (p_1^+
		- p_3^+)^2 } \nonumber \\
	&& ~~~~ \sim {1\over 4\pi} \int d \Big( {p_1^+ \over m_F} y^- \Big)
		d^2 y_{\bot} e^{ i\big\{ \big( {m_F \over p_1^+}
		(p_1^+ -p_3^+)\big) \big({p_1^+ \over m_F } y^- \big)
		- (p_{1\bot} - p_{3\bot}) \cdot y_{\bot} \big\} }
		\nonumber \\
	&& ~~~~~~~~~~~~~~~~~~~~~~~~~~~~~~~~~~~ \times \Big( { m_F \over p_1^+ }
		\Big)^2 { 1 \over \sqrt{y_\perp^2 + ({ p_1^+
		\over m_F} y^-)^2 } } \nonumber \\
	&& = {1\over 4\pi} \int dy^- d^2 y_{\bot}~ e^{i \big\{
		(p_1^+ -p_3^+)y^- - (p_{1\bot} - p_{3\bot})
		\cdot y_{\bot} \big\} } { m_F \over p_1^+ }
		{ 1 \over \sqrt{y_\perp^2 + ({ p_1^+ \over m_F})^2
		(y^-)^2 } } , \end{eqnarray}
where $y_\bot =x_{1\bot} - x_{2\bot} \equiv \delta x_\bot$ is the
transverse separation of the two constituents, but the
longitudinal separation $y^- \equiv \delta x$ is more complicated since
the above expression
is not a complete Fourier transformation to light-front coordinate space.
But for qualitative purposes we may define
a light-front ``radial'' coordinate $ r_1 = \sqrt{(\delta x_\perp)^2
+ ({ p_1^+ \over m_F})^2 (\delta x^-)^2}$, and we see from (C12) that
$H_2^{NR}$ has the form of the light-front Coulomb potential
(2.6) which we have introduced in the artificial potential in Section
II.E.  This light-front radial coordinate is invariant
under longitudinal boosts.  However, it is not invariant under
transverse boosts.  For more physical remarks, see
Section II.E.

Relativistically, the spinor matrix elements are different in different
helicity sectors. In the nonrelativistic limit this helicity dependence
vanishes, and we get the same interaction in all helicity sectors.

%%%%%%%%%%%%%%%%%%%%%%%%%%%%%%%%%%%%%%%%%%%%%%%%%%%%%
%%%%   etpsc.tex  (input to QCD paper)        %%%%%%%
%%%%%%%%%%%%%%%%%%%%%%%%%%%%%%%%%%%%%%%%%%%%%%%%%%%%%
%%%%   last modified: 1/24/93          %%%%%%%%%%%%%
%%%%%%%%%%%%%%%%%%%%%%%%%%%%%%%%%%%%%%%%%%%%%%%%%%%%%

%%%%%%%%%%%%%%%%%%%%%%%%%%%%%%%%%%%%%%%%%%%%%%%%%%%%%%%%%%%
\section{Phase space cell division: Equal-time case}
%%%%%%%%%%%%%%%%%%%%%%%%%%%%%%%%%%%%%%%%%%%%%%%%%%%%%%%%%%%

Let us try to motivate why a phase space cell analysis is useful for the
study of Hamiltonians. Recall free field theory in equal-time
coordinates. A free field is
diagonalized in terms of momentum eigenstates. Thus it is customary to express
free fields in terms of creation and annihilation operators for particles in
plane wave states.
The free Hamiltonian is
\begin{eqnarray} H_0 = \int \;  d^3k \; \omega_k \; a^\dagger_k a_k,
\end{eqnarray}
with $\omega_k = \sqrt{\mu^2 + k^2}$.

 Next consider interactions. For illustrative purposes,
we will choose a
$\phi^3$-type interaction,
\begin{eqnarray} H_{int} = \int d^3x \int d^3k_1 \int d^3k_2 \int d^3k_3
a^{\dagger}(k_1)
a(k_2) a(k_3) \, e^{-i(k_2+k_3- k_1)\cdot x} . \end{eqnarray}
Before trying to solve the
Hamiltonian quantitatively, it is important to make qualitative estimates of
various terms in the Hamiltonian. For this purpose, the creation and
annihilation operators $a_k$ and $a^{\dagger}_k$ are quite inappropriate. For
example, the operator $a^\dagger_k$ creates a particle in a plane wave state
and hence creates a state of infinite norm. Thus when we try to estimate the
$``$order of magnitude'' of $a^{\dagger}_k$ we get infinity! What can be
done to avoid this catastrophe?

It has been suggested \cite{Wi 65} (also see \cite{Wi 71a}) some time ago
that the
creation and annihilation operators which depend on a continuous variable $k$
be expanded in terms of a discrete, complete, orthonormal set of functions
$u_i(k)$.
Specifically,
\begin{eqnarray}  a_k = \sum_i u_i(k) a_i \; ,\end{eqnarray}
\begin{eqnarray}  a^\dagger_k = \sum_i u_i^*(k) a^\dagger_i \; . \end{eqnarray}
The coefficients of this expansion are themselves creation and
annihilation operators for particles in normalizable $``$wave packet'' states.
Now
\begin{eqnarray} H_0 = \sum_{i_1i_2} \; C_{i_1i_2} \; a^\dagger_{i_1} a_{i_2},
\end{eqnarray}
where
\begin{eqnarray} C_{i_1i_2} = \int \; d^3 k \; \omega_k \; u^*_{i_1}(k)
u_{i_2}(k)\; . \end{eqnarray}

$H_0$ is now the Hamiltonian of an infinite number of oscillators coupled to
each other (no longer diagonal).
Now making $H_0$ diagonal means making self-interactions of
the individual oscillators more important than interactions between different
oscillators. Consider the matrix $C$. If the
$u_l$'s are orthogonal, $C$ would be
diagonal except for the factor $\omega_k$. Even with the factor $\omega_k$, one
should be able to keep the off-diagonal elements small. If the
$u_i$'s are
$``$properly chosen'' (that is, as
localized in momentum space as possible)
then for distinct momentum shells $i_1$ and $i_2$, the
functions $u_{i_1}$ and $u_{i_2}$ do not overlap very much.
If $u_{i_1}$ and $u_{i_2}$ are in
the same momentum shell but different spatial cells, this fact will be
reflected in a rapid phase variation of $u_{i_1}^* u_{i_2}$
as a function of $k$, which
again makes the integral small. Thus with a properly chosen set of $u_i$'s, for
order of magnitude estimates we need to consider only diagonal elements
$C_{i_1i_1}$ of $C$.

The interaction Hamiltonian in terms of wave packet states is
\begin{eqnarray} H_{int} = \sum_{i_1i_2i_3} a^{\dagger}_{i_1} a_{i_2} a_{i_3}
D_{i_1 i_2 i_3} \end{eqnarray}
with
\begin{eqnarray} D_{i_1 i_2 i_3} && = \int d^3x \int d^3k_1 \int d^3k_2 \int
d^3k_3
u^*_{i_1}(k_1) u_{i_2}(k_2) u_{i_3}(k_3)
\, e^{-i(k_2 + k_3  - k_1).x} \nonumber \\
&& = \int d^3x v_{i_1}^*(x) v_{i_2}(x) v_{i_3}(x) . \end{eqnarray}
If the $u$'s are as localized
as possible in momentum space, then the $v$'s are as
spread out as possible in coordinate space and hence the off-diagonal elements
of $D$ are as big as the diagonal elements of $D$. Thus if we want to
approximate $D$ by its diagonal elements for order of magnitude estimates, we
need $v$'s to be as localized in coordinate space as possible. Thus we clearly
have to satisfy almost mutually exclusive requirements, which means that we
have to settle for a compromise.

A qualitative implementation of the compromise is as follows. We
choose $u_{i}(k)$ such that if $\Delta k_i$ is the momentum width of the
function $u_i(k)$ and $\Delta x_i$ is the width of the Fourier transform
$v_i(x)$ of $u_i(k)$, then the product $\Delta k_i \Delta x_i$ is near the
lower limit set by the uncertainty principle, that is,
\begin{eqnarray}
\Delta k_i \Delta x_i \approx {1 \over 2} \; . \end{eqnarray}
The set of functions $\{u_i\}$ is chosen to be complete and orthonormal. What
does this mean? Think of $u_i$ as occupying a cell of unit volume in phase
space. Completeness means that the total volume occupied by the
$u_i$'s must fill all space. Orthogonality means that the regions occupied by
different $u_i$'s do not overlap.

How are we going to divide the phase space into cells?
Consider one dimensional example for visualization.
Divide the momentum space into an infinite number of cells, the
$i^{th}$ cell being
\begin{eqnarray}  2^{-i} < k <  2^{-i+1} \; . \end{eqnarray}
For each momentum cell separately, divide the position space
linearly into cells of
the appropriate size.
The position space coordinate is chosen
to be $ l  \le 2^{-i} x \le l + 1 $.

Let us give some examples. Consider $i=0$. Then $ 1 < k < 2 $, and
the length of the momentum cell is $L_k = 1$.
The position cell division is
$l  \le  x \le
l + 1 $. Thus the position space coordinates are, for example,
\begin{eqnarray} l = -1 && \qquad - 1 < x <  0, \\ \nonumber
    l =  0 && \qquad 0 < x <  1, \\ \nonumber
    l = +1 && \qquad 1  < x <  2, \end{eqnarray}
and so on,
so that each position space cell has length $L_x = 1$. Thus the $``$ volume''
of each phase space cell is $L_k L_x = 1$.

Consider $i=1$. Then $ {1 \over 2}  < k < 1 $, and the
length of the momentum cell is $L_k = {1 \over 2}$.
The position cell division is
$l  \le {1 \over 2}  x \le
l + 1 $. Thus the position space coordinates are, for example,
\begin{eqnarray} l = -1 && \qquad - 2 < x < 0,  \\ \nonumber
    l =  0 && \qquad 0 < x < 2,  \\ \nonumber
    l = +1 && \qquad 2 < x < 4, \end{eqnarray}
and so on, so that each position space cell has length $L_x = 2$.
Thus the $``$volume''
of each phase space cell is again $L_k L_x = 1$.

One can label each phase space cell by $(l,i)$.
An illustration of the division of phase space into cells is given in Fig. 13.

Why have we chosen a logarithmic scale for momentum cells? One would
like to have each momentum cell correspond to a distinct energy scale
so that order of magnitude estimates become meaningful. A linear
choice obviously fails for this purpose, while a logarithmic choice is
the most obvious one that fulfills it. Note that there is nothing
sacred about the factor of 2; for a specific example where the factor
of 2 appears as the optimum choice, see Ref. {\cite{Wi 71a}.

%%%%%%%%%%%%%%%%%%%%%%%%%%%%%%%%%%%%%%%%%%%%%%%%%%%%%%%%%%%%
%%%%%%% refer.tex       (input to QCD paper)           %%%%%
%%%%%%%%%%%%%%%%%%%%%%%%%%%%%%%%%%%%%%%%%%%%%%%%%%%%%%%%%%%%
%%%% Last modified	12/15/93			%%%%
%%%%%%%%%%%%%%%%%%%%%%%%%%%%%%%%%%%%%%%%%%%%%%%%%%%%%%%%%%%%

%%%%%%%%%%%%%%%%%%%%%%%%%%%%%%%%%%%%%%%%%%%%%%%%%%%%%%%%%%%
%%%%%%%% Figures.tex (input to QCD paper)          %%%%%%%%
%%%%%%%%%%%%%%%%%%%%%%%%%%%%%%%%%%%%%%%%%%%%%%%%%%%%%%%%%%%
\vskip .5in
\centerline{List of Figures}
\vskip .5in
\begin{enumerate}
\item Light-front coordinate system.
\item Examples of instantaneous interactions sensitive to fermion zero modes.
\item Instantaneous four-fermion and two-fermion--two-gluon interactions.
\item  Cutoff region of center of mass momenta.
\item  Cutoff region of constituent momenta.
\item Triangle of renormalization.
\item Gluon mass correction from two-gluon intermediate states.
\item Contribution to quark--antiquark scattering
  amplitude from gluon exchange.
\item Contribution to quark--gluon scattering
  amplitude from gluon exchange.
\item Contribution to quark-gluon scattering amplitude from quark exchange.
\item Quark-antiquark annihilation and contribution to qqg vertex
      renormalization.
\item Higher-order multifermion processes requiring counterterms.
\item Example of phase space cell division in one dimension.
\end{enumerate}
\end{document}